\documentclass{jfm}
\usepackage{graphicx}
\usepackage{epstopdf, epsfig}

\usepackage{amsmath}
\usepackage{mathrsfs}
\usepackage{bm}

\newcommand{\olra}{\overleftrightarrow}

\shorttitle{Theory for the rheology of dense non-Brownian suspensions}
\shortauthor{Koshiro Suzuki and Hisao Hayakawa}

\title{Theory for the rheology of dense non-Brownian suspensions:
divergence of viscosities and $\mu$-$J$ rheology}

\author{Koshiro Suzuki\aff{1}
  \corresp{\email{suzuki.koshiro@mail.canon}},
 \and Hisao Hayakawa\aff{2}}

\affiliation{\aff{1}Simulation \& Analysis R\&D Center, Canon Inc., 30-2
Shimomaruko 3-chome, Ohta-ku, Tokyo 146-8501, Japan 
\aff{2}Yukawa Institute for Theoretical Physics, Kyoto University,
Kitashirakawaoiwake-cho, Sakyo-ku, Kyoto 606-8502, Japan}

\usepackage{version}	
\begin{document}

\maketitle

\begin{abstract}
A systematic microscopic theory for the rheology of dense non-Brownian
suspensions characterized by the volume fraction $\varphi$ is developed.
The theory successfully derives the critical behavior in the vicinity of
the jamming point (volume fraction $\varphi_{J}$), for both the pressure
$P$ and the shear stress $\sigma_{xy}$, i.e. $P \sim \sigma_{xy} \sim
\dot\gamma \eta_0 \delta\varphi^{-2}$, where $\dot\gamma$ is the shear
rate, $\eta_0$ is the shear viscosity of the solvent, and $\delta\varphi
= \varphi_J - \varphi > 0$ is the distance from the jamming point.
It also successfully describes the behavior of the stress ratio $\mu =
\sigma_{xy}/P$ with respect to the viscous number
$J=\dot\gamma\eta_{0}/P$.
\end{abstract}


\section{Introduction}
\label{sec:intro}
The physics of the rheology of suspensions begins with the seminal work
by Einstein~\citep{MW, Einstein1905}.
He has shown that the effective shear viscosity $\eta_s(\varphi)$
defined by the ratio of the shear stress $\sigma_{xy}(\varphi)$ to the
shear rate $\dot\gamma$ as
$\eta_{s}(\varphi)
=
\sigma_{xy}(\varphi)/\dot\gamma$
is enhanced as $\eta_s(\varphi)/\eta_0=1+5\varphi/2+O(\varphi^2)$ in
dilute suspensions, where $\varphi$ is the volume fraction of the
suspended particles and $\eta_0$ is the viscosity of the solvent.
On the other hand, it has been empirically shown that $\eta_s(\varphi)$
behaves as $\eta_s(\varphi)/\eta_0 \sim (\varphi_m-\varphi)^{-2}$ near a
critical volume fraction $\varphi_m$ in dense
suspensions~\citep{CCD1971, Krieger1972, Quemada1977, ZHL2000}.

Recently, the divergence of the shear stress $\sigma_{xy}$ has been well
studied in the context of the jamming transition, which is an athermal
phase transition of dense disordered materials such as
suspensions~\citep{Pusey91}, emulsions, foams~\citep{DW1994}, and granular
materials~\citep{OLLN2002, OSLN2003, OH2014, CSD2014}.
It is well established that the shear viscosity of non-Brownian
suspensions which are insensitive to thermal fluctuations near the
jamming point behaves as $\eta_s(\varphi)/\eta_{0} \sim
(\varphi_{J}-\varphi)^{-\lambda}$ with $\lambda \approx 2$ and
$\varphi_J$ the jamming volume fraction~\citep{BGP2011, BDCL2010},
although numerical simulations for soft spheres exhibit $\lambda \approx
2.2$~\citep{ABH2012} or $\lambda \approx 1.67 - 2.55$~\citep{KCIB2015},
and a theoretical approach by DeGiuli et al. asserts $\lambda\approx
2.83$~\citep{GDLW2015}.

On the other hand, the pressure of suspensions $P$ has been less
investigated.
Experimentally, it has been shown that the pressure viscosity defined by
$\eta_{n}(\varphi)
=
P(\varphi)/\dot\gamma$
exhibits $\eta_{n}(\varphi)/\eta_0 \sim
(\varphi_J-\varphi)^{-2}$~\citep{DGMYM2009, BGP2011, CW2014, DBHGP2015}.
This is non-trivial, since it differs from the pressure at equilibrium
given by $P^{(\rm{eq})}(\varphi) = nT[1 + 4\varphi g_0(\varphi)]$, where
$n=6\varphi/(\pi d^3)$ is the average number density, $d$ is the
diameter of the particle, $T$ is the temperature, and $g_0(\varphi)$ is
the radial distribution function at contact~\citep{HM}.
Together with the relation $g_0(\varphi)\sim
(\varphi_J-\varphi)^{-1}$~\citep{DTS2005}, this leads to
$P^{(\rm{eq})}(\varphi)\sim nT(\varphi_J-\varphi)^{-1}$, which is
inconsistent with the experimental observations for non-Brownian
suspensions.
To be consistent with the experimental expression $P(\varphi)\sim
\eta_0\dot\gamma (\varphi_J-\varphi)^{-2}$, we need to explain two
non-trivial relations, i.e. $P\propto\dot\gamma\eta_0$ and
$P\propto(\varphi_J-\varphi)^{-2}$.
The former one, $P\propto\dot\gamma\eta_0$, has been argued by
phenomenological considerations~\citep{JM1990,NB1994} or by
microstructural and structure-property analyses~\citep{BM1997}.
The latter one, $P\propto(\varphi_J-\varphi)^{-2}$, is more
non-trivial.
Several phenomenological models are proposed to explain this
property~\citep{ZHL2000, MS2009}, but practically it is merely given as
an empirical law without a theoretical basis~\citep{MB1999}.

Another rheological property of our interest is the stress ratio, 
$\mu = \sigma_{xy}/P$.
It is known that $\mu$ converges to a constant in approaching the
jamming point, while it varies on departure from the point, by
experiments and simulations~\citep{GDRMidi2004, BGP2011, KH2011, ICH2014,
DBHGP2015, KCIB2015}.
In fact, a constitutive equation for $\mu(J) = \sigma_{xy}/P$ together
with $\varphi = \varphi(J)$, where $J=\dot\gamma\eta_0/P$ is the viscous
number, is proposed and confirmed by experiments conducted with a
pressure-imposed cell ($\mu$-$J$ rheology)~\citep{BGP2011, DBHGP2015}.
The reported result exhibits
$\mu(J) = \mu_0 + C J^{1/2}$,
where $C$ is a constant and $\mu_0$ is its value in the jamming limit,
$J\to 0$.
However, there exists no theory to explain this law so far.

Derivation of the rheological properties of suspensions from a
microscopic theory is difficult even for the shear viscosity.
It has been shown by Brady and his coworkers that the effective
self-diffusion constant satisfies $D(\varphi)\propto
D_{0}(\varphi_m-\varphi)$, where $D_0 = T_{\mathrm{s}}/(3\pi d\eta_0)$
with $T_{\mathrm{s}}$ the solvent temperature, which is crucial to
obtain $\eta_s(\varphi)/\eta_0 \sim (\varphi_m-\varphi)^{-2}$ for
Brownian suspensions~\citep{Brady1993, BM1997, FB2000}.
However, this theory is not applicable to non-Brownian suspensions,
because $D(\varphi)$ is an increasing function of $\varphi$ in
non-Brownian suspensions~\citep{LA1987-1, LA1987-2, BETA1998, BEBJM2002,
HBB2010, Olsson2010}.
Hence, an alternative framework is necessary for dense non-Brownian
suspensions.
In this paper, we attempt to derive the divergent behavior of the shear
and pressure viscosities, $\eta_s/\eta_0 \sim \eta_n/\eta_0 \sim
(\varphi_J-\varphi)^{-2}$, and the $\mu$-$J$ rheology, $\mu(J) = \mu_0 +
CJ^{1/2}$, by means of a microscopic theory for an idealistic model of
non-Brownian suspensions.

\section{Basic equations and exact equations for the stress}

\subsection{Microscopic basic equations}
\label{sec:Microscopic}
We consider an assembly of $N$ frictionless monodisperse spherical
particles of diameter $d$ contained in a box of volume $V$ and immersed
in a liquid of viscosity $\eta_0$.
A simple steady shear with shear rate $\dot\gamma$ is applied to the
system.
The coordinate is chosen such that the flow is in the $x$-direction and
the velocity gradient is in the $y$-direction.
We consider the overdamped equation of motion
\begin{eqnarray}
\sum_{j=1}^{N}\zeta^{(N)}_{ij}
\left(
\dot{\boldsymbol{r}}_{j}
-
\dot\gamma y_j \boldsymbol{e}_{x}
\right)
=
\boldsymbol{F}_{i}^{(\mathrm{p})}
\hspace{1em}
(i=1,\cdots,N),
\label{eq:eom}
\end{eqnarray}
where $\boldsymbol{r}_{i}$ and $\dot{\bm{r}}_{i}$ are the position and
velocity of particle $i$, respectively, $\boldsymbol{e}_x$ is the unit
vector in the $x$-direction, $\boldsymbol{F}_i^{(\mathrm{p})}$ is the
interparticle force exerted on particle $i$ from other particles, and
$\{\zeta_{ij}^{(N)}\}_{i,j=1}^N$ is the resistance matrix of the suspension,
which depends on the configuration of the particles, $\{
\bm{r}_{i}\}_{i=1}^{N}$.
Note that $\{\zeta_{ij}^{(N)}\}_{i,j=1}^N$ is a $3N\times 3N$ matrix,
where each component $\zeta_{ij}^{(N)}$ is a $3\times 3$ matrix.
In particle suspensions, the inertia of the particles is absorbed by the
background fluid and hence insignificant.
Thus we neglect it in Eq.~(\ref{eq:eom}).
We also neglect the rotation of the particles and the thermal
fluctuating force exerted on the particles from the solvent in
Eq.~(\ref{eq:eom}).
%

The time evolution of an arbitrary observable $A(\boldsymbol{\Gamma})$
is determined by the Liouville equation
\begin{eqnarray}
\dot{A}(\boldsymbol{\Gamma}(t)) 
=
\dot{\boldsymbol{\Gamma}} \cdot 
\frac{\partial}{\partial \boldsymbol{\Gamma}}
A(\boldsymbol{\Gamma}(t)) 
:=
i {\cal L} A(\boldsymbol{\Gamma}(t)),
\end{eqnarray}
where $i{\cal L}$ is the Liouvillian.
In simple shear flows, $\boldsymbol{\Gamma}$ is given by
$\boldsymbol{\Gamma}=\{\boldsymbol{r}_{i}, \boldsymbol{v}_{i}
\}_{i=1}^{N}$, where 
\begin{eqnarray}
\boldsymbol{v}_{i} 
:=
\dot{\boldsymbol{r}}_{i} -
\dot\gamma y_{i}\boldsymbol{e}_{x} 
=
\sum_{j=1}^{N}\zeta_{ij}^{(N)-1} \boldsymbol{F}_{j}^{(\mathrm{p})}
\label{eq:peculiar_velocity}
\end{eqnarray}
is the peculiar velocity, which is the velocity in the sheared frame.
For Eq.~(\ref{eq:eom}), $i\mathcal{L}$ is given by
\begin{eqnarray}
i\mathcal{L}
:=
\sum_{i=1}^{N}
\left(
\bm{v}_{i}
\cdot \frac{\partial}{\partial \boldsymbol{r}_{i}}
+
\dot{\bm{v}}_{i} \cdot
\frac{\partial}{\partial \bm{v}_{i}}
\right)
=
\sum_{i=1}^{N}
\left(
\sum_{j=1}^{N} \zeta_{ij}^{(N)-1}
\boldsymbol{F}_{j}^{(\mathrm{p})}
\!\cdot\! \frac{\partial}{\partial \boldsymbol{r}_{i}}
+
\dot{\bm{v}}_{i} \cdot
\frac{\partial}{\partial \bm{v}_{i}}
\right),
\label{eq:Liouvillian_0}
\end{eqnarray}
where $\dot{\bm{v}}_{i}$ is evaluated as
\begin{eqnarray}
\dot{\boldsymbol{v}}_{i}
=
\ddot{\bm{r}}_{i}-\dot\gamma \dot{y}_{i} \boldsymbol{e}_{x} 
=
\ddot{\bm{r}}_{i}
-\dot\gamma \sum_{j=1}^{N} \zeta_{ij}^{(N)-1} F_{j,y}^{(\mathrm{p})}
\boldsymbol{e}_{x}.
\label{eq:vdot}
\end{eqnarray}
Note that Eq.~(\ref{eq:peculiar_velocity}) is utilized in the second
equality of Eq.~(\ref{eq:vdot}).
In the formulation of overdamped Liouville equation, we neglect
$\ddot{\bm{r}}_{i}$ in Eq.~(\ref{eq:vdot}), because $\bm{v}_{i}$ rather
than $\ddot{\bm{r}}_{i}$ resides in the equation of motion,
Eq.~(\ref{eq:peculiar_velocity}).
This leads us to the expression
\begin{eqnarray}
i\mathcal{L}
=
\sum_{i=1}^{N}
\left(
\sum_{j=1}^{N} \zeta_{ij}^{(N)-1}
\boldsymbol{F}_{j}^{(\mathrm{p})}
\!\cdot\! \frac{\partial}{\partial \boldsymbol{r}_{i}}
-
\dot\gamma
F_{i,y}^{(\mathrm{p})}
\frac{\partial}{\partial F_{i,x}^{(\mathrm{p})}}
\right).
\label{eq:Liouvillian}
\end{eqnarray}
Note, however, that neglecting $\ddot{\bm{r}}_{i}$ in
Eq.~(\ref{eq:vdot}) does not mean the absence of $\dot{\bm{v}}_i$ as can
be seen in Eq.~(\ref{eq:vdot}).
In fact, because of the existence of $\bm{v}_i$, trajectories
of the particles can be curved.

The Liouville equation of the microscopic stress tensor
$\tilde{\sigma}_{\alpha\beta}(\boldsymbol{\Gamma})$ ($\alpha, \beta = x,
y, z$) reads
\begin{eqnarray}
\frac{d}{dt}
\tilde{\sigma}_{\alpha\beta} (\boldsymbol{\Gamma})
\!=\!
\sum_{i=1}^{N}
\!\left(
\sum_{j=1}^{N} \zeta_{ij}^{(N)-1}
\boldsymbol{F}_{j}^{(\mathrm{p})}
\!\!\cdot\!
\frac{\partial \tilde{\sigma}_{\alpha\beta} (\boldsymbol{\Gamma})}{\partial \boldsymbol{r}_{i}}
-
\dot\gamma F_{i,y}^{(\mathrm{p})} 
\frac{\partial \tilde{\sigma}_{\alpha\beta} (\boldsymbol{\Gamma})}{\partial F_{i,x}^{(\mathrm{p})}}
\!\right)\!\!,
\hspace{1.5em}
\label{eq:micro_stress_eom}
\end{eqnarray}
where $\tilde{\sigma}_{\alpha\beta}(\boldsymbol{\Gamma})$ is given by
\begin{eqnarray}
\tilde{\sigma}_{\alpha\beta}(\boldsymbol{\Gamma})
:=
-\frac{1}{2V}
\sum_{i=1}^N
\left(
r_{i,\beta}
F_{i,\alpha}^{(\mathrm{p})}
+
r_{i,\alpha}
F_{i,\beta}^{(\mathrm{p})}
\right).
\label{eq:micro_stress}
\end{eqnarray}
In simple shear flows, the only non-zero components of
$\tilde{\sigma}_{\alpha\beta}(\bm{\Gamma})$ are
$\tilde{\sigma}_{xy}(\bm{\Gamma})$ and the diagonal ones, from which the
microscopic pressure is given by
$\tilde{P}(\boldsymbol{\Gamma})
=
-
(
\tilde{\sigma}_{xx}(\boldsymbol{\Gamma}) 
+ 
\tilde{\sigma}_{yy}(\boldsymbol{\Gamma}) 
+ 
\tilde{\sigma}_{zz}(\boldsymbol{\Gamma})
)/3$.
Note that we mainly consider the Cauchy stress, which contributes to the
divergence at $\varphi\approx\varphi_{J}$, but it is possible to define
the kinetic stress by the peculiar velocity.

\subsection{Exact equations for the stress}
\label{sec:EqContinuity}
Macroscopic equation of continuity of the stress tensor is obtained by
multiplying Eq.~(\ref{eq:micro_stress_eom}) by the nonequilibrium
distribution function $f(\boldsymbol{\Gamma},t)$ and integrating over
$\boldsymbol{\Gamma}$,
\begin{eqnarray}
\frac{d}{dt}
\sigma_{\alpha\beta}
\!+\!
\frac{1}{2}\dot\gamma 
\left(
\delta_{\alpha x} \sigma_{y\beta}
+
\delta_{\beta x} \sigma_{y\alpha}
\right)
&=&
-
\frac{1}{2V}
\!\!\sum_{i}
\!\left\langle\!
\sum_{j}
\zeta_{ij}^{(N)-1}
F_{j,\beta}^{(\mathrm{p})}F_{i,\alpha}^{(\mathrm{p})}
\!\right\rangle
+
(\alpha \leftrightarrow \beta)
\nonumber \\
&&
-
\frac{1}{2V}
\!\!\sum_{i,j}{}'
\!\left\langle \!
\sum_{k}
\zeta_{ik}^{(N)-1}
F_{k,\lambda}^{(\mathrm{p})}
r_{j,\beta}
\frac{\partial F_{j,\alpha}^{(\mathrm{p})}}{\partial r_{i,\lambda}}
\!\right\rangle
+
(\alpha \leftrightarrow \beta),
\hspace{1.5em}
\label{eq:dsdt}
\end{eqnarray}
where
\begin{eqnarray}
\left\langle \cdots \right\rangle 
:=
\int d\boldsymbol{\Gamma}\,
f(\boldsymbol{\Gamma},t)\cdots
\end{eqnarray}
is the macroscopic average, $\sum_{i,j}'$ denotes the summation over $i$
and $j$ with $i\ne j$, and the macroscopic stress tensor denotes
\begin{eqnarray}
\sigma_{\alpha\beta}
:=
\int d\boldsymbol{\Gamma}\,
f(\boldsymbol{\Gamma},t) 
\tilde{\sigma}_{\alpha\beta}(\boldsymbol{\Gamma}).
\end{eqnarray}
It might be noteworthy that the equation of continuity of the stress
tensor, Eq.~(\ref{eq:dsdt}), is consistent with that for the Enskog
theory of moderately dense inertial suspensions~\citep{HTG2017}.
In fact, the two terms on the right-hand side (r.h.s.) of
Eq.~(\ref{eq:dsdt}), which are proportional to $\zeta_{ij}^{(N)-1}$ and
originate from particle contacts, correspond to the collision integral
terms in the Enskog theory.

\section{Approximate expression of the interparticle force for dense
 frictionless hard spheres}
\label{sec:ApproximateInterparticleForce}

\begin{figure}
\centerline{
\includegraphics[width=3.5cm]{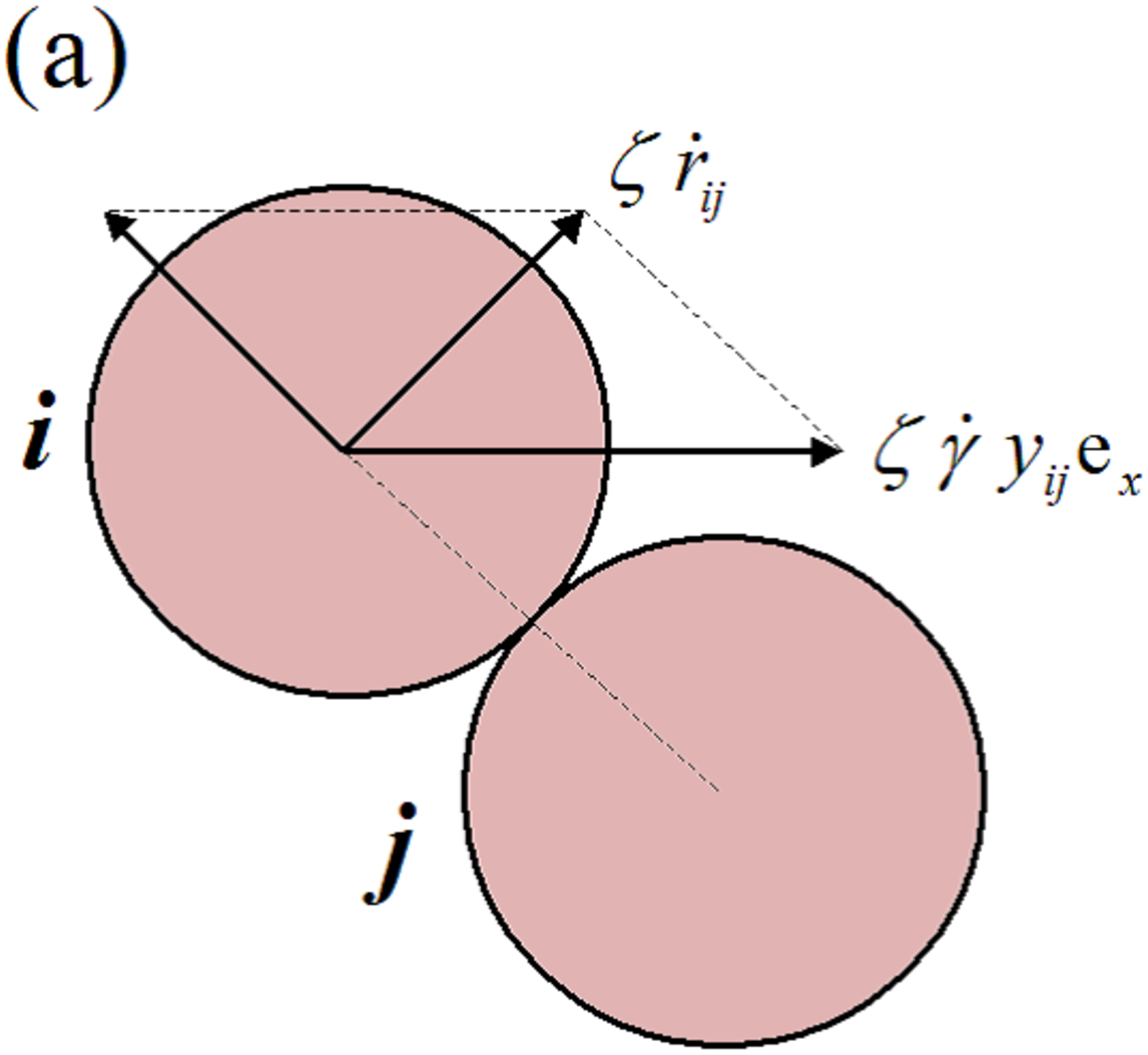}
\hspace{2em}
\includegraphics[width=3.5cm]{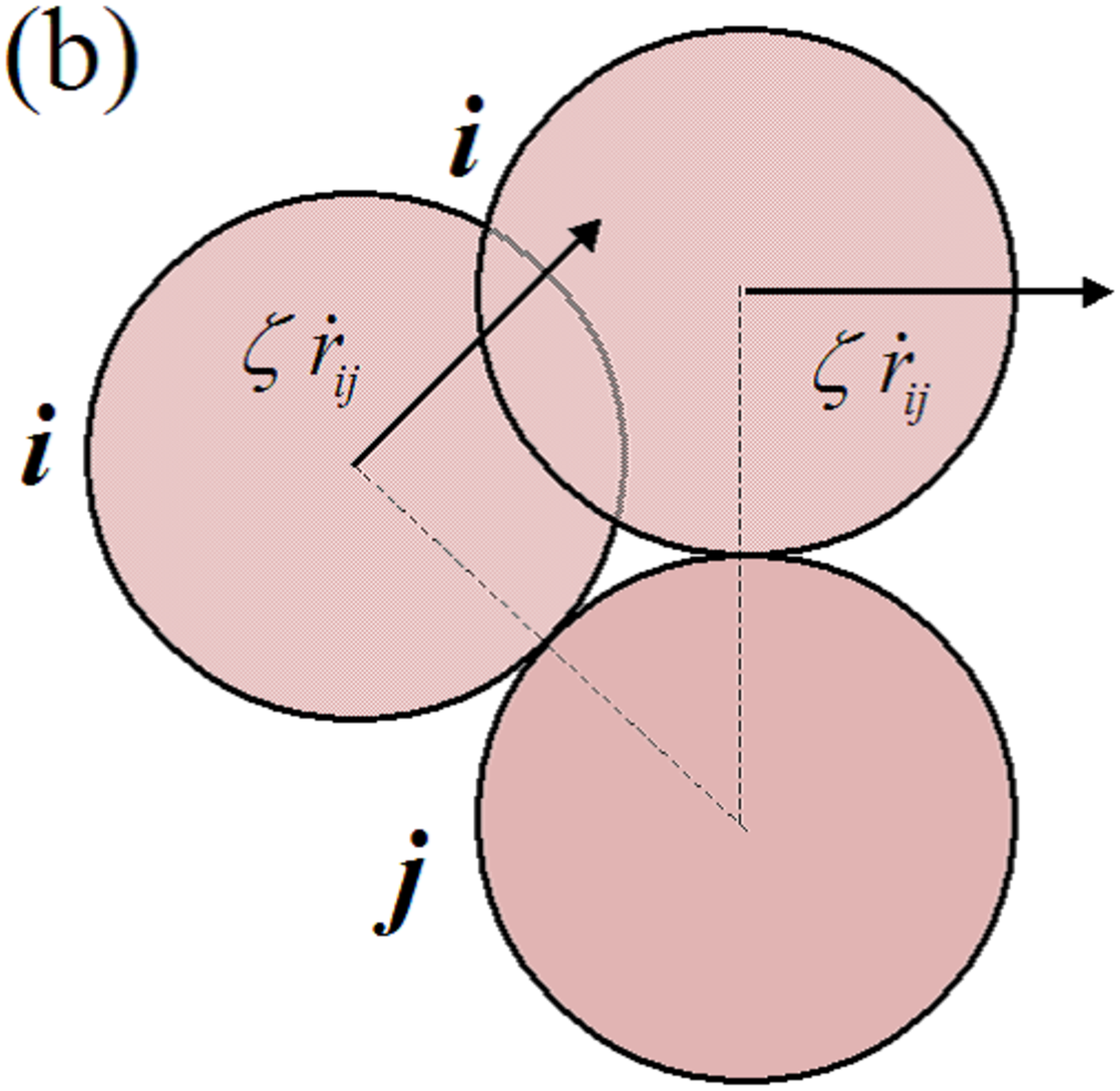}
\hspace{2em}
\includegraphics[width=3.5cm]{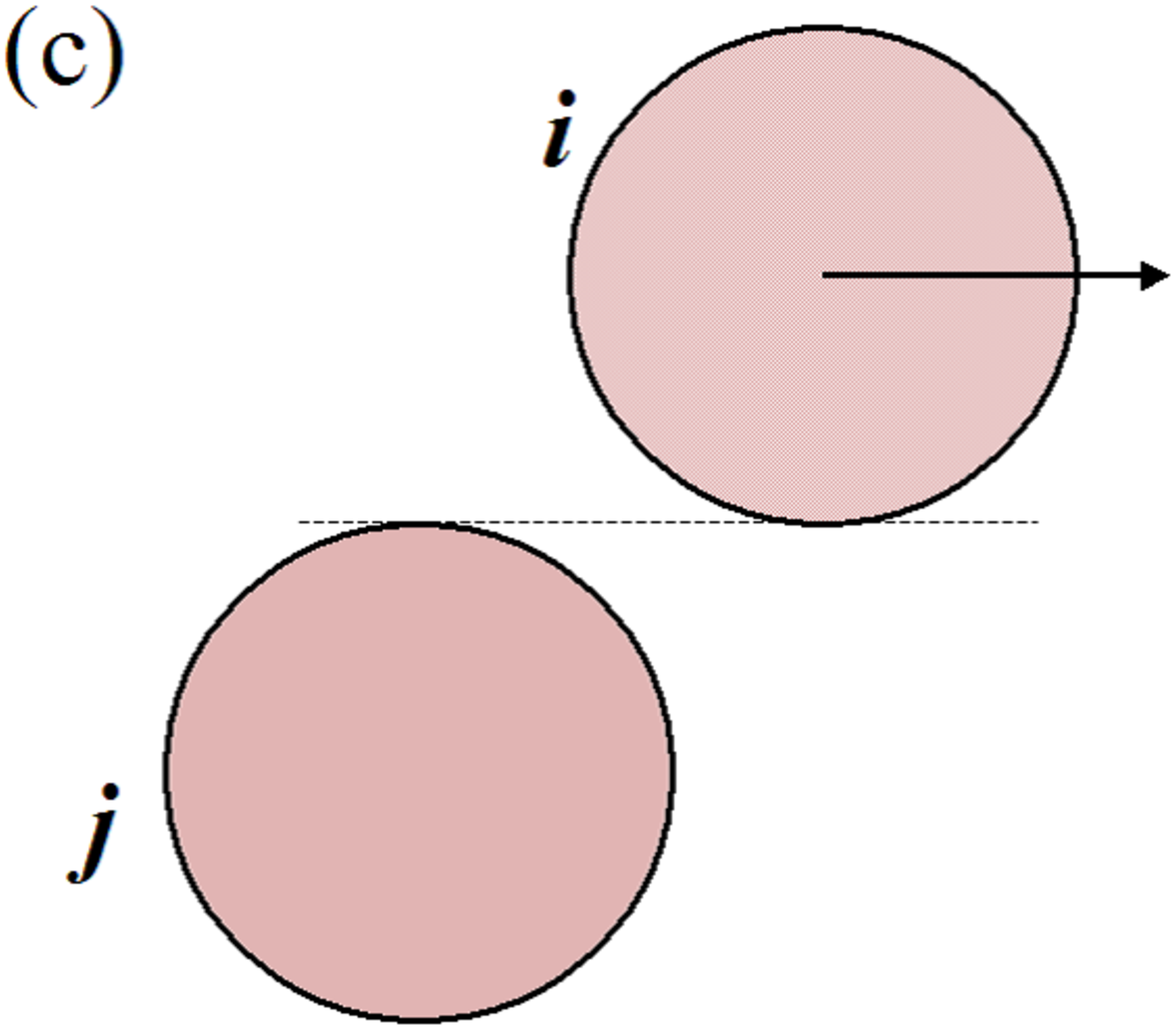}
} 
\caption {Dynamics of two spheres in contact: (a) instance of contact,
(b) relative motion; two spheres in contact relatively move in the
tangential direction until their departure, which occurs when they are
aligned in the $y$-direction, (c) after departure.}
\label{Fig:dynamics}
\end{figure}

To proceed, let us derive the specific form of the interparticle force
$\boldsymbol{F}_{i}^{(\mathrm{p})}(\{\bm{r}_{j}\}_{j=1}^N)$ for dense
frictionless hard spheres.
For dense spheres where all of them are at or close to contact, we can
expect that the far-field part of
$\olra{\zeta^{(N)}}(\{\bm{r}_{i}\}_{i=1}^N)$ does not contribute and
only the lubrication part, which is well approximated by the sum of the
two-body terms $\olra{\zeta^{(2)}_{\rm{lub}}}(\bm{r}_{ij})$, is
significant~\citep{KK, SMMD2013, MSMD2014}.
Thus, we approximate the resistance matrix $\olra{\zeta^{(N)}}$ as
\begin{eqnarray}
\olra{\zeta^{(N)}}(\{ \bm{r}_{i}\}_{i=1}^{N})
\approx
\zeta_0  \bm{I}
+
\olra{\zeta^{(2)}_{\rm{lub}}}(\bm{r}_{ij}),
\label{eq:resistance_approx}
\end{eqnarray}
where the first term on the r.h.s. with $\zeta_0 := 3\pi \eta_0 d$ and
$\bm{I}$ the unit matrix is the Stokesean one-body drag force.
This leads to the approximate equation of motion
\begin{eqnarray}
\zeta_0 
(\dot{\bm{r}}_{i} - \dot\gamma y_i \bm{e}_x)
+
\sum_{j\neq i} \zeta^{(2)}_{{\rm lub},ij}
(\dot{\bm{r}}_{j} - \dot\gamma y_j \bm{e}_x)
\approx
\bm{F}_{i}^{(\rm{p})}.
\end{eqnarray}
Then, accordingly, the interparticle force should be well approximated
by a sum of two-body forces, 
\begin{eqnarray}
\bm{F}_{i}^{(\rm{p})}
\approx 
\sum_{j\neq i} \bm{F}_{ij}^{(\rm{p})},
\label{eq:Fip_2body}
\end{eqnarray}
where $\bm{F}_{ij}^{(\rm{p})}$ is the
two-body force exerted on particle $i$ from $j$.
The dynamics of the interacting two spheres is schematically described
in Fig.~\ref{Fig:dynamics}.
When the approaching two spheres come into contact (a), they slide in
the tangential direction until they are aligned in the velocity-gradient
direction (b), and then depart (c).
In general, there are not only two but multiple of particles in contact,
but every pair slides mutually in the tangential direction until their
departure, so it is reasonable to consider the dynamics as a
superposition of the two-body counterpart.
Indeed, the simulation in terms of Stokesean dynamics is performed in
terms of the superposition of the two-body interactions~\citep{SMMD2013,
MSMD2014}.
Hence, it is sufficient to consider the two-body dynamics to determine
$\bm{F}_{ij}^{(\rm{p})}$.

Let us consider two spheres $i$ and $j$ in contact.
The equation of motion of the two spheres is given by
\begin{eqnarray}
\zeta_0
\left[
\begin{array}{c}
\dot{\bm{r}}_{i} - \dot\gamma\, y_{i} \bm{e}_{x} \\
\dot{\bm{r}}_{j} - \dot\gamma\, y_{j} \bm{e}_{x} 
\end{array}
\right]
+
\olra{\zeta^{(2)}_{\rm{lub}}}(\bm{r}_{ij})
\left[
\begin{array}{c}
\dot{\bm{r}}_{i} - \dot\gamma\, y_{i} \bm{e}_{x} \\
\dot{\bm{r}}_{j} - \dot\gamma\, y_{j} \bm{e}_{x} 
\end{array}
\right]
&=&
\left[
\begin{array}{c}
\bm{F}_{ij}^{(\rm{p})} \\
-\bm{F}_{ij}^{(\rm{p})} 
\end{array}
\right],
\label{eq:eom_2body_1}
\end{eqnarray}
where we have utilized $\bm{F}_{ji}^{(\rm{p})}=-\bm{F}_{ij}^{(\rm{p})}$.
The matrix $\olra{\zeta^{(2)}_{\rm{lub}}}$ is explicitly given 
by~\citep{JO1984}
\begin{eqnarray}
\olra{\zeta^{(2)}_{\rm{lub}}}(\bm{r}_{ij})
&=&
\zeta_0
\left[
\begin{array}{cc}
\Pi(\bm{r}_{ij}) & -\Pi(\bm{r}_{ij})
 \\
-\Pi(\bm{r}_{ij}) & \Pi(\bm{r}_{ij})
\end{array}
\right],
\\
\Pi(\bm{r}_{ij})
&:=&
\frac{1}{8}\frac{1}{\delta r_{ij}+\epsilon}P_{ij} + \frac{1}{12}\ln
\frac{1}{\delta r_{ij}+\epsilon}P_{ij}',
\end{eqnarray}
where $\zeta_0 := 3\pi \eta_0 d$, $\delta r_{ij}:=r_{ij}-d$, and
$P_{ij}:=\hat{\bm{r}}_{ij}\hat{\bm{r}}_{ij}$,
$P_{ij}':=\bm{I}-\hat{\bm{r}}_{ij}\hat{\bm{r}}_{ij}$ are projection
operators.
At contact, i.e. $\delta r_{ij}=0$, $\olra{\zeta^{(2)}_{\rm{lub}}}$
exhibits singularities of the form $\epsilon^{-1}$ and $\ln \epsilon^{-1}$,
where $\epsilon$ is a cut off which is physically interpreted as
e.g. surface roughness.
In this work we keep $\epsilon$ finite and do not consider these
singularities.
Then, Eq.~(\ref{eq:eom_2body_1}) is given by
\begin{eqnarray}
&&
\zeta_0
\left[
\begin{array}{c}
(\dot{\bm{r}}_{i} - \dot\gamma y_i \bm{e}_x)
+
\frac{1}{8\epsilon}P_{ij}\cdot (\dot{\bm{r}}_{ij} - \dot\gamma y_{ij} \bm{e}_{x})
+ \frac{1}{12}\ln \epsilon^{-1} P_{ij}'\cdot (\dot{\bm{r}}_{ij} - \dot\gamma
y_{ij} \bm{e}_{x}) \\
(\dot{\bm{r}}_{j} - \dot\gamma y_j \bm{e}_x)
-
\frac{1}{8\epsilon}P_{ij}\cdot (\dot{\bm{r}}_{ij} - \dot\gamma y_{ij} \bm{e}_{x})
- \frac{1}{12}\ln \epsilon^{-1} P_{ij}'\cdot (\dot{\bm{r}}_{ij} - \dot\gamma
y_{ij} \bm{e}_{x})
\end{array}
\right] 
\nonumber \\
&&
\hspace{2em}
=
\left[
\begin{array}{c}
\bm{F}_{ij}^{(\rm{p})} \\
-\bm{F}_{ij}^{(\rm{p})}
\end{array}
\right].
\label{eq:eom_2body_2}
\end{eqnarray}
By subtracting the two equations, Eq.~(\ref{eq:eom_2body_2}) reduces
to
\begin{eqnarray}
\bm{F}_{ij}^{(\rm{p})}
=
\zeta_0
\left[
\frac{1}{2}
(\dot{\bm{r}}_{ij} - \dot\gamma y_{ij} \bm{e}_{x})
+
\frac{1}{8\epsilon}P_{ij}\cdot (\dot{\bm{r}}_{ij} - \dot\gamma y_{ij} \bm{e}_{x})
+ \frac{1}{12}\ln \epsilon^{-1} P_{ij}'\cdot (\dot{\bm{r}}_{ij} - \dot\gamma
y_{ij} \bm{e}_{x})
\right],
\hspace{1.5em}
\label{eq:F_ij_p}
\end{eqnarray}
where the r.h.s. consists of the Stokesean drag force (first term), the
normal lubrication force (second term), and the tangential lubrication
force (third term).
These three terms are of the order of 1, $\epsilon^{-1}$, and $\ln
\epsilon^{-1}$, respectively.
The second term is dominant for $\epsilon \ll 1$, so
Eq.~(\ref{eq:F_ij_p}) is reduced to
\begin{eqnarray}
\bm{F}_{ij}^{(\rm{p})}
\approx
\frac{\zeta_0}{8\epsilon}
P_{ij}\cdot (\dot{\bm{r}}_{ij} - \dot\gamma y_{ij} \bm{e}_{x})
=
\frac{\zeta_0}{8\epsilon}
(\hat{\bm{r}}_{ij}\cdot\dot{\bm{r}}_{ij} - \dot\gamma r_{ij} \hat{y}_{ij} \hat{x}_{ij}) \hat{\bm{r}}_{ij}.
\label{eq:F_ij_p_2}
\end{eqnarray}
%

For hard spheres, the relative velocity of $i$ and $j$ is in the
direction perpendicular to $\hat{\bm{r}}_{ij}$ in order not to overlap,
i.e. $\hat{\bm{r}}_{ij}\cdot\dot{\bm{r}}_{ij}=0$ or
$P_{ij}\cdot\dot{\bm{r}}_{ij}=0$ (cf. Fig.~\ref{Fig:dynamics}(a)).
Then we obtain
\begin{eqnarray}
\bm{F}_{ij}^{(\rm{p})}
=
-\frac{1}{2}\zeta_{e}\dot\gamma r_{ij} \hat{x}_{ij} \hat{y}_{ij}
\hat{\bm{r}}_{ij},
\label{eq:F_ij_p_3}
\end{eqnarray}
where we have defined
\begin{eqnarray}
\zeta_{e} 
:= 
\frac{\zeta_0}{4\epsilon} = \frac{3\pi \eta_0 d }{4\epsilon}.
\end{eqnarray}
Note that Eq.~(\ref{eq:F_ij_p_3}) is valid only at contact, i.e. $r_{ij}
= d$, and $\bm{F}_{ij}^{(\rm{p})}=0$ for $r_{ij}>d$ for hard spheres.
That is, $\bm{F}_{ij}^{(\rm{p})}\propto \delta(r_{ij}-d)$.
Thus we modify Eq.~(\ref{eq:F_ij_p_3}) by replacing $r_{ij}$ with $d^2
\delta(r_{ij}-d)$,
\begin{eqnarray}
\bm{F}_{ij}^{(\rm{p})}
=
-\frac{1}{2}\zeta_{e}\dot\gamma d^2
\delta(r_{ij}-d) \hat{x}_{ij} \hat{y}_{ij}
\hat{\bm{r}}_{ij}.
\end{eqnarray}
Note that $\boldsymbol{F}_{ij}^{(\mathrm{p})}\propto \delta (r_{ij}-d)$
results in an important feature that the spatial correlations are
expressed solely by~\citep{DTS2005}
\begin{eqnarray}
g_0(\varphi)\sim (\varphi_{J}-\varphi) ^{-1},
\label{eq:g0}
\end{eqnarray}
where there is no dependence on its spatial derivative, $g'(r)$, because
our dynamics inhibits the overlap of the contacting particles.

\begin{figure}
\centerline{\includegraphics[width=8.5cm]{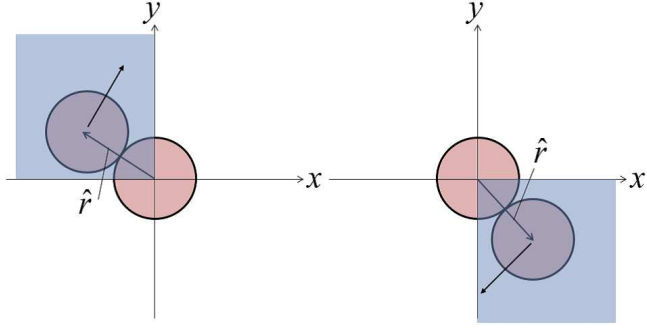}}
\caption{Dominant relative position of spheres in contact.
The arrows perpendicular to $\hat{\bm{r}}$ show the direction of the
velocity of the spheres.}
\label{Fig:config}
\end{figure}

Furthermore, in order for $\bm{F}_{ij}^{(\rm{p})}$ to be a repulsive
force, $\hat{x}_{ij}\hat{y}_{ij}<0$ is necessary.
Hence, we introduce a projection operator
\begin{eqnarray}
\mathcal{P}(\hat{x},\hat{y})
:=
-\hat{x}\hat{y}\, \Theta(-\hat{x}\hat{y}) > 0
\label{eq:Projection}
\end{eqnarray}
to assure this property,
\begin{eqnarray}
\bm{F}_{ij}^{(\rm{p})} 
=
\frac{1}{2} \zeta_{e} \dot\gamma d^2 \delta(r_{ij}-d)
\mathcal{P}(\hat{x}_{ij}, \hat{y}_{ij})
\hat{\bm{r}}_{ij}.
\label{eq:F_ij_p_final}
\end{eqnarray}
Here, $\Theta (x)$ is Heaviside's step function, i.e. $\Theta (x)=1$ for
$x>0$ and $\Theta (x)=0$ otherwise.
The projection operator in Eq.~(\ref{eq:F_ij_p_final}) implies that
$\bm{F}_{ij}^{(\rm{p})}$ is non-zero only when the separation vector of
the contacting two spheres $\bm{r}_{ij}:=\bm{r}_i-\bm{r}_j$ is in the
compression quadrant (cf. Fig.~\ref{Fig:config}).
This results from the approximation where we have neglected the first
and third terms in the r.h.s. of Eq.~(\ref{eq:F_ij_p}).
In fact, these two terms are in general non-zero, irrespective of the
direction of $\bm{r}_{ij}$.
Hence, although the direction of $\bm{r}_{ij}$ can be in any direction
in dense suspensions, the dominant contribution of the interparticle
force comes from configurations where $\bm{r}_{ij}$ is in the
compression quadrant.

To summarize, the equation of motion for dense hard-sphere suspensions
is reduced to
\begin{eqnarray}
\zeta_0 
(\dot{\bm{r}}_{i} - \dot\gamma y_i \bm{e}_x)
+
\sum_{j\neq i} \zeta^{(2)}_{{\rm lub},ij}
(\dot{\bm{r}}_{j} - \dot\gamma y_j \bm{e}_x)
=
\sum_{j\neq i} \bm{F}_{ij}^{(\rm{p})},
\label{eq:eom_approx_final}
\end{eqnarray}
where the two-body interparticle force $\bm{F}_{ij}^{(\rm{p})}$ is given
by Eq.~(\ref{eq:F_ij_p_final}), and the summation is over the contacting
particles.
Note that Eq.~(\ref{eq:eom_approx_final}) is exact, under the assumption
that the interparticle force is expressed as a superposition of two-body
forces, Eq.~(\ref{eq:Fip_2body}), and hydrodynamic forces other than the lubrication force are
neglected, Eq.~(\ref{eq:resistance_approx}).
%
%
%
%
%
%

In Eq.~(\ref{eq:dsdt}), not only the interparticle force
$\bm{F}_{i}^{(\rm{p})}$ but also the inverse of the resistance matrix
$\zeta_{ij}^{(N)-1}$ should be evaluated.
We assume that $\zeta_{ij}^{(N)-1}$ can be approximated by the sum of
the two-body mobility matrix $\olra{\mathcal{M}^{(2)}}$ as
\begin{eqnarray}
\bm{v}_{i}
=
\dot{\bm{r}}_{i} - \dot\gamma y_{i} \bm{e}_{x}
=
\sum_{j=1}^N \zeta_{ij}^{(N)-1} \bm{F}_{j}^{(\rm{p})}
\approx
\sum_{j=1}^N \mathcal{M}_{ij}^{(2)} \bm{F}_{j}^{(\rm{p})}
\label{eq:v_mobility}
\end{eqnarray}
for dense suspensions.
Let us consider the two-body dynamics to evaluate the r.h.s. of
Eq.~(\ref{eq:v_mobility}),
\begin{eqnarray}
\left[
\begin{array}{c}
\bm{v}_{i} \\
\bm{v}_{j}
\end{array}
\right]
=
\olra{\mathcal{M}^{(2)}}(\bm{r}_{ij})
\left[
\begin{array}{c}
\bm{F}_{i}^{(\rm{p})} \\
\bm{F}_{j}^{(\rm{p})} 
\end{array}
\right]
=
\olra{\mathcal{M}^{(2)}}(\bm{r}_{ij})
\left[
\begin{array}{c}
\bm{F}_{ij}^{(\rm{p})} \\
-\bm{F}_{ij}^{(\rm{p})} 
\end{array}
\right],
\label{eq:eom_2body}
\end{eqnarray}
where $\olra{\mathcal{M}}^{(2)}$ is explicitly given by~\citep{RP1969}
\begin{eqnarray}
\olra{\mathcal{M}^{(2)}} (\bm{r}_{ij})
&=&
\frac{1}{3\pi \eta_0 d}
\left[
\begin{array}{cc}
\bm{I} & \Xi(\bm{r}_{ij}) \\
\Xi(\bm{r}_{ij}) & \bm{I}
\end{array}
\right],
\\
\Xi(\bm{r}_{ij})
&:=&
\left[
\frac{3}{4} \frac{d}{r_{ij}} - \left( \frac{d}{2r_{ij}}\right)^3
\right]
P_{ij}
+
\left[
\frac{3}{8} \frac{d}{r_{ij}} + \frac{1}{2} \left(
\frac{d}{2r_{ij}} \right)^3
\right]
P_{ij}'.
\end{eqnarray}
Here, $P_{ij}:=\hat{\bm{r}}_{ij} \hat{\bm{r}}_{ij}$ and
$P_{ij}':=\bm{I}-\hat{\bm{r}}_{ij}\hat{\bm{r}}_{ij}$ are projection
operators, as defined before.
At contact, Eq.~(\ref{eq:eom_2body}) is explicitly written by
\begin{eqnarray}
\left[
\begin{array}{c}
\bm{v}_{i} \\
\bm{v}_{j}
\end{array}
\right]
=
\frac{1}{3\pi \eta_0 d}
\left[
\begin{array}{c}
\bm{F}_{ij}^{(\rm{p})} 
- \left[
\frac{3}{4} \frac{d}{r_{ij}} - \left( \frac{d}{2r_{ij}}\right)^3
\right] \bm{F}_{ij}^{(\rm{p})} \\
-\bm{F}_{ij}^{(\rm{p})} 
+ \left[
\frac{3}{4} \frac{d}{r_{ij}} - \left( \frac{d}{2r_{ij}}\right)^3
\right] \bm{F}_{ij}^{(\rm{p})}
\end{array}
\right]
\approx
\frac{1}{8\pi \eta_0 d}
\left[
\begin{array}{c}
\bm{F}_{ij}^{(\rm{p})} \\
- \bm{F}_{ij}^{(\rm{p})} 
\end{array}
\right],
\end{eqnarray}
where we have utilized the projection properties,
$P_{ij}\cdot\bm{F}_{ij}^{(\rm{p})}=\bm{F}_{ij}^{(\rm{p})}$ and
$P_{ij}'\cdot\bm{F}_{ij}^{(\rm{p})}=0$ which follow from
$\bm{F}_{ij}^{(\rm{p})}\propto \hat{\bm{r}}_{ij}$, and $r_{ij}\approx
d$.
Hence, Eq.~(\ref{eq:v_mobility}) is evaluated as
\begin{eqnarray}
\bm{v}_{i}
=
\sum_{j=1}^N \zeta_{ij}^{(N)-1} \bm{F}_{j}^{(\rm{p})}
\approx
\frac{1}{8\pi \eta_0 d}
\sum_{j\neq i}
\bm{F}_{ij}^{(\rm{p})},
\label{eq:v_mobility_2}
\end{eqnarray}
where the summation is over the contacting particles.

\section{Approximate formulas for the viscosities and $\mu$-$J$ rheology}
\label{sec:StressTensorFormulas}

In Sec.~\ref{sec:ApproximateInterparticleForce}, we have derived
approximate expressions for the the interparticle force and the inverse
of the resistance matrix, which appear in the r.h.s. of the exact
equation of the stress tensor, Eq.~(\ref{eq:dsdt}).
Using these expressions, i.e. Eqs.~(\ref{eq:F_ij_p_final}) and
(\ref{eq:v_mobility_2}), an approximate equation for the stress tensor
can be derived.
The first term on the r.h.s. of Eq.~(\ref{eq:dsdt}) is evaluated as
\begin{eqnarray}
\sum_{i=1}^N
\left\langle
\sum_{j} \zeta_{ij}^{(N)-1} F_{j,\beta}^{(\rm{p})}
F_{i,\alpha}^{(\rm{p})}
\right\rangle
&\approx&
\frac{1}{8\pi \eta_0 d}
\sum_{i=1}^N
\left\langle
\sum_{j\neq i} F_{ij,\beta}^{(\rm{p})}
\sum_{k\neq i} F_{ik,\alpha}^{(\rm{p})}
\right\rangle
\nonumber \\
&=&
\frac{1}{4}
\frac{\zeta_e^2}{8\pi \eta_0 d}
\dot\gamma^2 d^4
\sum_{i=1}^{N}
\left\langle
\sum_{j\neq i}
\sum_{k\neq i}
\Delta_{ij}^{xy} 
\hat{r}_{ij,\beta}\,
\Delta_{ik}^{xy}
\hat{r}_{ik,\alpha}
\right\rangle
\nonumber \\
&=&
\frac{1}{4}
\zeta
\dot\gamma^2 d^4
\sum_{i=1}^{N}
\left\langle
\sum_{j\neq i}
\sum_{k\neq i}
\Delta_{ij}^{xy} 
\hat{r}_{ij,\beta}\,
\Delta_{ik}^{xy}
\hat{r}_{ik,\alpha}
\right\rangle\!,
\label{eq:FpalphaFpbeta_ss}
\end{eqnarray}
where we have introduced 
\begin{eqnarray}
\Delta_{ij}^{\alpha\beta}
:=
- \delta(r_{ij}-d)
\hat{r}_{ij,\alpha}\hat{r}_{ij,\beta}\Theta(-\hat{x}_{ij}\hat{y}_{ij})
\label{eq:Delta}
\end{eqnarray}
and
\begin{eqnarray}
\zeta
:=
\frac{\zeta_e^2}{8\pi \eta_0 d}
=
\frac{3}{32}\frac{\zeta_e}{\epsilon} 
=
\frac{3}{128} \frac{\zeta_0}{\epsilon^2}
\label{eq:zeta_def}
\end{eqnarray}
for abbreviation.
For the second term on the r.h.s. of Eq.~(\ref{eq:dsdt}) which includes
a derivative of the interparticle force, special attention should be
paid.
Here, we only show the results (see
Appendix~\ref{app:sec:ForceCorrelation} for the detailed derivation):
\begin{eqnarray}
\sum_{i,j}{}'
\!\left\langle
\sum_{k}
\zeta_{ik}^{(N)-1}
F_{k,\lambda}^{(\mathrm{p})}
r_{j,\beta}
\frac{\partial F_{j,\alpha}^{(\mathrm{p})}}{\partial r_{i,\lambda}}
\!\right\rangle
&\approx&
\frac{1}{4}
\zeta \dot\gamma^2 d^4
\!\sum_{i=1}^N
\!\left\langle
\sum_{j\neq i}
\!\sum_{k\neq i}
\!\Delta_{ij}^{xy} 
\hat{r}_{ij,\alpha}\,
\Delta_{ik}^{xy}
\hat{r}_{ik,\beta}
\right.
\nonumber \\
&&
\left.
+
\Delta_{ik}^{xy} 
\Delta_{ij}^{\alpha\beta}
\!\left( \!
\hat{x}_{ik} 
\hat{y}_{ij} 
\!+\!
\hat{y}_{ik} 
\hat{x}_{ij} 
\!\right)
\!\right\rangle\!.
\label{eq:FrdFdr_ss}
\end{eqnarray}
From Eqs.~(\ref{eq:dsdt}), (\ref{eq:FpalphaFpbeta_ss}), and
(\ref{eq:FrdFdr_ss}), we obtain an approximate equation for the stress
evolution:
\begin{eqnarray}
\frac{d}{dt}
\sigma_{\alpha\beta}
\!+\!
\frac{1}{2} \dot\gamma \left(
\delta_{\alpha x} \sigma_{y\beta}
\!+\!
\delta_{\beta x} \sigma_{y\alpha}
\right)
\!
&\approx&
- \zeta \dot\gamma^2\frac{d^4}{4V}
\!\sum_{i=1}^{N}
\!\sum_{j\neq i}
\!\sum_{k\neq i}
\left\langle
\Delta_{ij}^{xy} 
\Delta_{ik}^{xy}
\hat{r}_{ij,\alpha}
\hat{r}_{ik,\beta}
\right\rangle
+
(\alpha\leftrightarrow\beta)
\nonumber \\
&&
\hspace{-7em}
-
\zeta \dot\gamma^2 \frac{d^4}{4V}
\!\sum_{i=1}^{N}
\!\sum_{j\neq i}
\!\sum_{k\neq i}
\left\langle
\Delta_{ij}^{xy}
\Delta_{ik}^{\alpha\beta} 
\left( 
\hat{x}_{ij} 
\hat{y}_{ik} 
+
\hat{y}_{ij}
\hat{x}_{ik} 
\right)
\right\rangle.
\label{eq:ss2}
\end{eqnarray}

\subsection{Grad's 13-moment-like expansion}
\label{subsec:Grad}
To obtain the stress tensor from Eq.~(\ref{eq:ss2}), it is still
necessary to have the distribution function $f(\bm{\Gamma},t)$ at hand
to evaluate the statistical averages on the r.h.s.
However, the exact expression of $f(\bm{\Gamma},t)$ for many-body
problems is unknown and thus we should resort to approximations.
Here we adopt Grad's 13-moment-like expansion for $f(\bm{\Gamma},t)$.
This method is well established to approximate the distribution function
in the kinetic theory of dilute or moderately dense
gases~\citep{Grad1949, HH1982, JR1985-1, JR1985-2, TK1995, SMTK1996,
Garzo2002, SGD2004, Kremer, Garzo2013, CRG2015, HT2017,
HT2016-2, HTG2017}.
It is an expansion in terms of the heat and stress currents in addition
to the five conserved currents for the collisional invariants.
For simple shear without spatial inhomogeneity, the current for the heat
and the conserved quantities are negligible, and hence the velocity
distribution function is dominated by the stress current,
\begin{eqnarray}
f(\boldsymbol{v})
\approx
f_{\mathrm{eq}}(\boldsymbol{v})
\left[
1 + \frac{V}{2T_K} \Pi_{\alpha\beta}^{(K)}
\tilde{\sigma}_{\alpha\beta}^{(K)}(\boldsymbol{v}) 
\right],
\label{eq:f_Grad_kinetic}
\end{eqnarray}
where summation over repeated indices $\alpha, \beta$ is taken,
e.g. $\sigma_{\alpha\alpha} := {\rm tr} (\sigma_{\alpha\beta}) =
\sigma_{xx}+\sigma_{yy}+\sigma_{zz}$.
Here, $\tilde{\sigma}_{\alpha\beta}^{(K)}(\boldsymbol{v}) := -m
v_{\alpha}v_{\beta}/V$ is the microscopic kinetic stress, where $m$ and
$\boldsymbol{v}$ are the mass and velocity of the particle,
$\Pi_{\alpha\beta}^{(K)} :=
\sigma_{\alpha\beta}^{(K)}/P^{(K)}+\delta_{\alpha\beta}$ is the
normalized deviatoric stress, where $\sigma_{\alpha\beta}^{(K)}:=\int
d\boldsymbol{v}
f(\boldsymbol{v})\tilde{\sigma}_{\alpha\beta}^{(K)}(\boldsymbol{v})$ and
$P^{(K)}:=-\sigma_{\alpha\alpha}^{(K)}/3$ are the macroscopic kinetic
stress and the pressure, $T_K :=-\sigma_{\alpha\alpha}^{(K)}/(3n)$ is
the kinetic temperature, where $n$ is the average number density, and
$f_{\mathrm{eq}}(\boldsymbol{v})$ is the equilibrium distribution
function.
Note that the kinetic pressure satisfies the relation $P^{(K)}=nT_K$.
This distribution function gives reasonably precise description of
nonequilibrium gases, e.g. continuous as well as discontinuous shear
thickening~\citep{CRG2015, HT2017, HT2016-2, HTG2017}.
It is also notable that Eq.~(\ref{eq:f_Grad_kinetic}) satisfies the
Green-Kubo formula within the BGK (Bhatnagar-Gross-Krook)
approximation~\citep{HT2016-2}, while it further incorporates the normal
stress differences, which is not the case for the Green-Kubo formula.

The distribution function Eq.~(\ref{eq:f_Grad_kinetic}) cannot be
directly applied to dense non-Brownian suspensions, where the Cauchy
stress dominates the kinetic stress.
A possible extension of this expansion for non-Brownian suspensions
would be
\begin{eqnarray}
f(\boldsymbol{\Gamma},t) 
\approx
f_{\mathrm{eq}}(\boldsymbol{\Gamma}_v)
f_{\mathrm{eq}}(\boldsymbol{\Gamma}_r)
\left[
1 + \frac{V}{2T} \Pi_{\alpha\beta}(t)
\tilde{\sigma}_{\alpha\beta}(\boldsymbol{\Gamma}_r)
\right],
\label{eq:f_Grad}
\end{eqnarray}
where $\bm{\Gamma}:=\{\bm{\Gamma}_{r}, \bm{\Gamma}_{v}\}$ with
$\bm{\Gamma}_r := \{\bm{r}_i\}_{i=1}^N$ and $\bm{\Gamma}_v :=
\{\bm{v}_i\}_{i=1}^N$, and the kinetic stress is replaced by the Cauchy
stress.
Here, 
\begin{eqnarray}
\Pi_{\alpha\beta} 
:= 
\frac{\sigma_{\alpha\beta}}{P} + \delta_{\alpha\beta}
\label{eq:Pi}
\end{eqnarray}
is the normalized deviatoric stress and the appropriate definition of
the temperature $T$ for non-Brownian suspensions will be discussed in
Sec.~\ref{subsec:temperature}.
Here it is postulated that the distribution function is factorized into
peculiar velocity-dependent and position-dependent parts, where the
peculiar velocity-dependent part can be approximated by Gaussian,
$f_{\mathrm{eq}}(\boldsymbol{\Gamma}_v)$, and the position-dependent
part is approximated by an expansion around equilibrium,
$f_{\mathrm{eq}}(\boldsymbol{\Gamma}_r)$, with the stress current.
This expansion around equilibrium is non-trivial for non-Brownian
suspensions, where the equilibrium state is absent.
Nonetheless, we will show in Sec.~\ref{sec:validation} that the velocity
distribution is nearly Gaussian and thus the factorization of
Eq.~(\ref{eq:f_Grad}) seems to be valid, and the expansion of the
position distribution with the stress current is applicable for the
evaluation of the stress.

From Eq.~(\ref{eq:f_Grad}), the macroscopic average of an arbitrary
observable $A(\bm{\Gamma}_r)$ which depends on $\bm{\Gamma}_r$ is given
by
\begin{equation}
\langle A(\bm{\Gamma}_r(t)) \rangle
\approx
\langle A(\bm{\Gamma}_r) \rangle_{\mathrm{eq}}
+
\frac{V}{2T}
\Pi_{\alpha\beta}(t)
\langle 
A(\bm{\Gamma}_r)
\tilde{\sigma}_{\alpha\beta}(\bm{\Gamma}_r) \rangle_{\mathrm{eq}},
\label{eq:ss_formula}
\end{equation}
where the first term on the r.h.s. is the canonical term with
\begin{eqnarray}
\langle\cdots\rangle_{\mathrm{eq}}
:=
\int d\boldsymbol{\Gamma}_{r}\,
f_{\mathrm{eq}}(\boldsymbol{\Gamma}_{r})\cdots
\end{eqnarray}
and the second term is its non-canonical correction.
Note that Eq.~(\ref{eq:ss_formula}) is formally equivalent to the
multiple relaxation-time approximation of the Green-Kubo
formula~\citep{SH2015}, where the dimensionless tensor
$\Pi_{\alpha\beta}$ plays the role of the multiple relaxation times.

\subsection{Approximate expressions}
\label{sec:ApproximateExpressions}

Let us evaluate the two averages on the r.h.s. of Eq.~(\ref{eq:ss2})
via the approximate formula Eq.~(\ref{eq:ss_formula}).
Detailed evaluation of the averages is shown in
Appendix~\ref{app:sec:ApplyGrad}. 
These two averages are written as sums of terms of the form
\begin{eqnarray}
\sum_{i=1}^{N} \sum_{j\neq i} \sum_{k\neq i}
\langle
X(\bm{r}_{ij}) Y(\bm{r}_{ik})
\rangle,
\label{eq:XY}
\end{eqnarray}
where $X(\bm{r}_{ij})$ or $Y(\bm{r}_{ik})$ abbreviates term which
depends on $\bm{r}_{ij}$ or $\bm{r}_{ik}$, respectively.
For instance, for the first term on the r.h.s. of Eq.~(\ref{eq:ss2}),
$X(\bm{r}_{ij})$ and $Y(\bm{r}_{ik})$ are given by
$
X(\bm{r}_{ij})
=
\hat{r}_{ij,\alpha}
\Delta_{ij}^{xy},
Y(\bm{r}_{ik})
=
\hat{r}_{ik,\beta}
\Delta_{ik}^{xy}
$,
or
$X(\bm{r}_{ij})
=
\hat{r}_{ij,\beta}
\Delta_{ij}^{xy},
Y(\bm{r}_{ik})
=
\hat{r}_{ik,\alpha}
\Delta_{ik}^{xy}$.
Terms of the form Eq.~(\ref{eq:XY}) are evaluated by
Eq.~(\ref{eq:ss_formula}) as 
\begin{eqnarray}
\sum_{i=1}^{N} \!\sum_{j\neq i} \!\sum_{k\neq i}
\langle 
X(\bm{r}_{ij})
Y(\bm{r}_{ik})
\rangle 
\!= \!
\sum_{i=1}^{N} \!\sum_{j\neq i} \!\sum_{k\neq i}
\!\left\{
\!\langle
X(\bm{r}_{ij})
Y(\bm{r}_{ik})
\rangle_{\mathrm{eq}} 
\!+\!
\frac{V}{2T}\Pi_{\rho\sigma}
\langle 
X(\bm{r}_{ij})
Y(\bm{r}_{ik})\tilde{\sigma}_{\rho\sigma}
\rangle_{\mathrm{eq}}
\!\right\} \!\!.
\hspace{2.em}
\end{eqnarray}
Note that $\tilde{\sigma}_{\rho\sigma}$ depends on the
relative coordinate, e.g. $\bm{r}_{lm}$, but either $l$ or $m$ must be
identical to e.g. $i$; otherwise the correlation decouples and vanishes.
Hence, the nonequilibrium term is given by
\begin{eqnarray}
\sum_{i=1}^{N} \sum_{j\neq i} \sum_{k\neq i} \sum_{l\neq i}
\langle 
X(\bm{r}_{ij})
Y(\bm{r}_{ik})
\tilde{\sigma}_{\rho\sigma}(\bm{r}_{il})
\rangle_{\mathrm{eq}},
\end{eqnarray}
which can be decomposed into four-, three-, and
two-body correlations as
\begin{eqnarray}
&&
\hspace{-2em}
\sum_{i=1}^{N} \sum_{j\neq i} \sum_{k\neq i} \sum_{l\neq i}
\langle 
X(\bm{r}_{ij})
Y(\bm{r}_{ik})
\tilde{\sigma}_{\rho\sigma}(\bm{r}_{il})
\rangle_{\mathrm{eq}}
=
\sum_{i,j,k,l}\!\!{}'''
\langle 
X(\bm{r}_{ij})
Y(\bm{r}_{ik})
\tilde{\sigma}_{\rho\sigma}(\bm{r}_{il})
\rangle_{\mathrm{eq}}
\nonumber \\
&&
+
\sum_{i,j,k}\!{}''
\langle 
X(\bm{r}_{ij})
Y(\bm{r}_{ik})
\tilde{\sigma}_{\rho\sigma}(\bm{r}_{ik})
\rangle_{\mathrm{eq}}
+
\sum_{i,j}{}'
\langle 
X(\bm{r}_{ij})
Y(\bm{r}_{ij})
\tilde{\sigma}_{\rho\sigma}(\bm{r}_{ij})
\rangle_{\mathrm{eq}}.
\end{eqnarray}
Here, the four-, three-, and two-body terms are given by
\begin{eqnarray}
\sum_{i,j,k,l}\!\!{}''' \!
\left\langle 
X(\bm{r}_{ij}) 
Y(\bm{r}_{ik}) 
\tilde{\sigma}_{\rho\sigma}(\bm{r}_{il})
\right\rangle_{\mathrm{eq}}
\!
&=&
\!
N n^3 \!
\int \!\! d^3\bm{r} \!\! 
\int \!\! d^3 \bm{r}' \!\! 
\int \!\! d^3 \bm{r}''
g^{(4)}(\bm{r},\bm{r}',\bm{r}'')
X(\bm{r}) 
Y(\bm{r}') 
\tilde{\sigma}_{\rho\sigma}(\bm{r}''),
\hspace{2.5em}
\label{eq:4body}
\\
\sum_{i,j,k}\!{}'' \!
\left\langle 
X(\bm{r}_{ij}) 
Y(\bm{r}_{ik}) 
\tilde{\sigma}_{\rho\sigma}(\bm{r}_{ik})
\right\rangle_{\mathrm{eq}}
&=&
N n^3 \!
\int \! d^3\bm{r} \! 
\int \! d^3 \bm{r}' 
g^{(3)}(\bm{r},\bm{r}')
X(\bm{r}) 
Y(\bm{r}') 
\tilde{\sigma}_{\rho\sigma}(\bm{r}'),
\label{eq:3body}
\\
\sum_{i,j (i\neq j)}{}' \!
\left\langle 
X(\bm{r}_{ij}) 
Y(\bm{r}_{ij}) 
\tilde{\sigma}_{\rho\sigma}(\bm{r}_{ij})
\right\rangle_{\mathrm{eq}}
&=&
N n^3 \!
\int \! d^3\bm{r} 
g(r)
X(\bm{r}) 
Y(\bm{r}) 
\tilde{\sigma}_{\rho\sigma}(\bm{r}),
\label{eq:2body}
\end{eqnarray}
respectively, where $g^{(4)}(\bm{r},\bm{r}',\bm{r}'')$ and
$g^{(3)}(\bm{r},\bm{r}')$ are the quadruplet-correlation
function~\citep{HM}
\begin{equation}
g^{(4)}(\boldsymbol{r},\boldsymbol{r}',\boldsymbol{r}'')
:=
\frac{1}{Nn^3}
\sum_{i,j,k,l}\!\!{}'''
\langle
\delta(\boldsymbol{r}-\boldsymbol{r}_{ij})\delta(\boldsymbol{r}'-\boldsymbol{r}_{ik})\delta(\boldsymbol{r}''-\boldsymbol{r}_{il})\rangle_{\mathrm{eq}}
\label{eq:g4}
\end{equation}
and the triplet-correlation function~\citep{HM}
\begin{equation}
g^{(3)}(\boldsymbol{r},\boldsymbol{r}')
:=
\frac{1}{Nn^2}
\sum_{i,j,k}\!{}''\langle
\delta(\boldsymbol{r}-\boldsymbol{r}_{ij})\delta(\boldsymbol{r}'-\boldsymbol{r}_{ik})\rangle_{\mathrm{eq}},
\label{eq:g3}
\end{equation}
and the summations $\sum_{i,j,k}''$, and $\sum_{i,j,k,l}'''$ are
performed over different particles.
For instance, $\sum_{i,j,k}''$ is performed for $i,j,k$ with $i\neq j$,
$j\neq k$, and $k\neq i$.
In the spatial integrations over $g(r)$, $g^{(3)}(\boldsymbol{r},\boldsymbol{r}')$, or
$g^{(4)}(\boldsymbol{r},\boldsymbol{r}',\boldsymbol{r}'')$, it is crucial that these correlation
functions are accompanied by the delta functions $\delta(r-d)$,
$\delta(r-d)\delta(r'-d)$, or $\delta(r-d)\delta(r'-d)\delta(r''-d)$,
respectively.
This feature can be explicitly traced back in Eq.~(\ref{eq:ss2}), and is
a consequence of the hard-core collision of the particles,
Eq.~(\ref{eq:F_ij_p_final}).
This implies that only the contact values of the correlation functions
contribute.
By virtue of this feature, we can conveniently approximate
$g^{(3)}(\boldsymbol{r},\boldsymbol{r}')$ and $g^{(4)}(\boldsymbol{r},\boldsymbol{r}',\boldsymbol{r}'')$.

Let us consider $g^{(3)}(\boldsymbol{r},\boldsymbol{r}')$ for illustration.
First of all, we adopt the factorization
approximation~\citep{Kirkwood1935},
$g^{(3)}(\boldsymbol{r},\boldsymbol{r}')\approx
g(r)g(r')g(|\boldsymbol{r}-\boldsymbol{r}'|)$.
Although this approximation is not accurate in general, it has been
argued that it is valid at contacts, where $r, r',
|\boldsymbol{r}-\boldsymbol{r}'| \approx d$~\citep{Alder1964,GZS2004}.
Furthermore, we have shown in \cite{SH2015} that only the radial
contacts contribute to the divergence in the vicinity of the jamming
point, i.e. 
\begin{eqnarray}
g^{(3)}(\boldsymbol{r},\boldsymbol{r}')\approx g(r)g(r')
\label{eq:3body_approx}
\end{eqnarray}
for $r, r'\approx d$.
Similarly, $g^{(4)}(\boldsymbol{r},\boldsymbol{r}',\boldsymbol{r}'')$
can be approximated as
\begin{eqnarray}
g^{(4)}(\boldsymbol{r},\boldsymbol{r}',\boldsymbol{r}'')\approx
g(r)g(r')g(r'')g(|\boldsymbol{r}-\boldsymbol{r}'|)g(|\boldsymbol{r}'-\boldsymbol{r}''|)g(|\boldsymbol{r}''-\boldsymbol{r}|)\approx
g(r)g(r')g(r'')
\hspace{1.5em}
\label{eq:4body_approx}
\end{eqnarray} 
for $r, r', r'' \approx d$, as far as divergence is concerned.
%
%
We will examine the validity of the factorization approximation for the
evaluation of the stress in Sec.~\ref{sec:validation}.

These approximations, together with
$g(r)\delta(r-d)=g_0(\varphi)\delta(r-d)$, enable us to express the two
averages on the r.h.s. of Eq.~(\ref{eq:ss2}) in terms of polynomials of
the radial distribution function at contact, $g_0(\varphi)$.
From tedious but straightforward calculation as shown in
Appendix~\ref{app:sec:ApplyGrad} and \ref{app:sec:Factorization}, we
reach the approximate equation of the stress
\begin{eqnarray}
&&
\frac{d}{dt}
\sigma_{\alpha\beta}
+
\frac{1}{2}\dot\gamma 
\left(
\delta_{\alpha x}\sigma_{y\beta}
\!+\!
\delta_{\beta x}\sigma_{y\alpha}
\right)
\approx
\frac{\zeta\dot\gamma^2}{4d}
\sum_{\ell =1}^{2}
\left\{
-\varphi^{*3} g_{0}(\varphi)^2 \mathcal{S}_{\alpha\beta}^{(\ell :\mathrm{c2})}
\!-\!
\varphi^{*2} g_{0}(\varphi) \mathcal{S}_{\alpha\beta}^{(\ell :\mathrm{c1})}
\right.
\nonumber \\
&&
\hspace{5em}
\left.
+
\Lambda
\Pi_{\rho\lambda}
\!\left[
\varphi^{*4}
g_{0}(\varphi)^3 \mathcal{S}_{\alpha\beta\rho\lambda}^{(\ell :\mathrm{nc3})} 
\!+\!
\varphi^{*3}
g_{0}(\varphi)^2 \mathcal{S}_{\alpha\beta\rho\lambda}^{(\ell :\mathrm{nc2})} 
\!+\!
\varphi^{*2}
g_{0}(\varphi) \mathcal{S}_{\alpha\beta\rho\lambda}^{(\ell :\mathrm{nc1})} 
\right]
\right\},
\hspace{2.5em}
\label{eq:ss3}
\end{eqnarray}
where the two terms with coefficients $\mathcal{S}_{\alpha\beta}^{(\ell
:\mathrm{c2})}$ and $\mathcal{S}_{\alpha\beta}^{(\ell :\mathrm{c1})}$
are the canonical contributions, and the three terms with coefficients
$\mathcal{S}_{\alpha\beta\rho\lambda}^{(\ell :\mathrm{nc3})}$,
$\mathcal{S}_{\alpha\beta\rho\lambda}^{(\ell :\mathrm{nc2})}$, and
$\mathcal{S}_{\alpha\beta\rho\lambda}^{(\ell :\mathrm{nc1})}$ are the
nonequilibrium corrections.
The coefficients $\mathcal{S}_{\alpha\beta}^{(\ell :\mathrm{c1})}$,
$\mathcal{S}_{\alpha\beta}^{(\ell :\mathrm{c2})}$ and
$\mathcal{S}_{\alpha\beta\rho\lambda}^{(\ell :\mathrm{nc3})} $,
$\mathcal{S}_{\alpha\beta\rho\lambda}^{(\ell :\mathrm{nc2})} $, 
$\mathcal{S}_{\alpha\beta\rho\lambda}^{(\ell :\mathrm{nc1})} $ are
numbers which arise from angular integrals.
Here, we have introduced a dimensionless scalar
\begin{eqnarray}
\Lambda
:=
\frac{\zeta\dot\gamma d^2}{4T}, 
\label{eq:Lambda}
\end{eqnarray}
and $\varphi^{*}= 6\varphi/\pi = nd^3$ denotes the dimensionless number
density.

\subsection{Implications of the symmetry}
\label{subsec:ImplicationsSymmetry}
A specific feature of non-Brownian suspensions under simple shear is the
symmetry under the parity ``$\hat{x}\to -\hat{x}$ and $\hat{y}\to
-\hat{y}$'', which follows from $\mathcal{P}(\hat{x},\hat{y})$
introduced in Eq.~(\ref{eq:Projection}).
This implies that only the parity-even terms in Eq.~(\ref{eq:ss3})
survive.
Furthermore, even though parity even, terms odd with respect to
$\hat{z}$ vanish.
These features can be summarized as follows,
\begin{eqnarray}
\int d\mathcal{S}\,
\Theta(-\hat{x}\hat{y}) \hat{x}^{i}\hat{y}^{j}\hat{z}^{k}
\neq 0
\hspace{1em}
\mbox{if and only if}
\hspace{1em}
i+j = \mbox{even and } k=\mbox{even},
\label{eq:sym1}
\end{eqnarray}
where $\int d\mathcal{S}\,\cdots$ expresses an angular integral with
respect to $\hat{\boldsymbol{r}}$.
As a consequence of this property we have
\begin{eqnarray}
\int d\mathcal{S}\,
\Theta(-\hat{x}\hat{y}) \hat{x}^{i}\hat{y}^{j}\hat{z}^{k}
= 0
\hspace{1em}
\mbox{if}
\hspace{1em}
i+j+k = \mbox{odd}.
\label{eq:sym2}
\end{eqnarray}
The terms proportional to $g_0(\varphi)^3$ in Eq.~(\ref{eq:ss3}), which
are cubic with respect to $\hat{\boldsymbol{r}}$, vanish because of
Eq.~(\ref{eq:sym2}),
\begin{eqnarray}
\mathcal{S}_{\alpha\beta\rho\lambda} ^{(1:\mathrm{nc3})}
=
\mathcal{S}_{\alpha\beta\rho\lambda} ^{(2:\mathrm{nc3})}
=
0. 
\label{eq:S_12nc3}
\end{eqnarray}
The same observation holds for the canonical terms proportional to
$g_0(\varphi)^2$, which are also cubic in $\hat{\boldsymbol{r}}$,
\begin{eqnarray}
\mathcal{S}_{\alpha\beta}^{(1:\mathrm{c2})} 
=
\mathcal{S}_{\alpha\beta}^{(2:\mathrm{c2})} 
=
0.
\label{eq:S_12c2}
\end{eqnarray}
From Eqs.~(\ref{eq:S_12nc3}) and (\ref{eq:S_12c2}), Eq.~(\ref{eq:ss3})
reduces to
\begin{eqnarray}
&&
\frac{d}{dt}
\sigma_{\alpha\beta}
+
\frac{1}{2}\dot\gamma 
\left(
\delta_{\alpha x}\sigma_{y\beta}
\!+\!
\delta_{\beta x}\sigma_{y\alpha}
\right)
\approx
\frac{\zeta\dot\gamma^2}{4d}
\sum_{\ell =1}^{2}
\left\{
-\varphi^{*2} g_{0}(\varphi) \mathcal{S}_{\alpha\beta}^{(\ell
:\mathrm{c1})}
\right.
\nonumber \\
&&
\hspace{5em}
\left.
+
\Lambda
\Pi_{\rho\lambda}
\!\left[
\varphi^{*3}
g_{0}(\varphi)^2 \mathcal{S}_{\alpha\beta\rho\lambda}^{(\ell :\mathrm{nc2})} 
\!+\!
\varphi^{*2}
g_{0}(\varphi) \mathcal{S}_{\alpha\beta\rho\lambda}^{(\ell :\mathrm{nc1})} 
\right]
\right\},
\label{eq:ss4}
\end{eqnarray}
or, explicitly in components,
\begin{eqnarray}
\frac{d}{dt} \!
\left(
\!\!\!
\begin{array}{c}
\sigma_{xy} \\
-3P \\
\sigma_{xx} \\
\sigma_{yy}
\end{array}
\!\!\!
\right)
\!
+
\!\left(
\!\!
\begin{array}{c}
\frac{1}{2}\sigma_{yy} \\
\sigma_{xy} \\
\sigma_{xy} \\
0
\end{array}
\!\!
\right)\!
\!
\approx
\!\frac{\zeta\dot\gamma}{4d}
\varphi^{*2}
g_{0}(\varphi)
\!\left\{ \!\!
-\bm{\mathcal{B}}
+\!
\Lambda
\left( \varphi^* g_0(\varphi) \bm{\mathcal{A}}^{(2)} 
+ \bm{\mathcal{A}}^{(1)}\right)
\!\!
\left(\!\!
\begin{array}{c}
\Pi_{xy} \\
\Pi_{xx} \\
\Pi_{yy} 
\end{array}
\!\!\right)
\!\! \right\}\!,
\hspace{2.5em}
\label{eq:ss4_components}
\end{eqnarray}
where the matrices $\bm{\mathcal{A}}^{(m)}$ ($m=1,2$) and the vector
$\bm{\mathcal{B}}$ are given by Eqs.~(C64)--(C66).
Note that $\Pi_{\alpha\beta}$ is given by Eq.~(\ref{eq:Pi}) and hence
Eq.~(\ref{eq:ss4_components}) is a closed set of equations for $P$,
$\sigma_{xy}$, $\sigma_{xx}$, and $\sigma_{yy}$.

Note that Eq.~(\ref{eq:ss4}) includes terms proportional to
$g_0(\varphi)^2$ or $g_0(\varphi)$.
Hence, we decompose the stress tensor as $\sigma_{\alpha\beta} =
\sigma_{\alpha\beta}^{(2)}+\sigma_{\alpha\beta}^{(1)}$, where
$\sigma_{\alpha\beta}^{(m)}$ is $\mathcal{O}(g_0(\varphi)^m)$ ($m=1,2$),
because Eq.~(\ref{eq:g0}) suggests that it is separable near the jamming
point.
Then, Eq.~(\ref{eq:ss4}) is cast into two equations, each for
$\sigma_{\alpha\beta}^{(2)}$ and $\sigma_{\alpha\beta}^{(1)}$,
respectively,
\begin{eqnarray}
\frac{d}{dt}
\sigma_{\alpha\beta}^{(2)}
\!+\!
\frac{1}{2}\dot\gamma \!\left(
\delta_{\alpha x}\sigma_{\beta y}^{(2)}
\!+\!
\delta_{\beta x}\sigma_{\alpha y}^{(2)}
\right)
\!&=&\!
\frac{\zeta\dot\gamma^2}{4d}
\Lambda
\varphi^{*3}
g_{0}(\varphi)^2 
\Pi_{\rho\lambda}^{(2)}
\sum_{\ell =1}^{2}
\mathcal{S}_{\alpha\beta\rho\lambda}^{(\ell :\mathrm{nc2})}.
\label{eq:ss3_order2}
\\
\frac{d}{dt}
\sigma_{\alpha\beta}^{(1)}
\!+\!
\frac{1}{2}\dot\gamma \!\left(
\delta_{\alpha x}\sigma_{\beta y}^{(1)}
\!+\!
\delta_{\beta x}\sigma_{\alpha y}^{(1)}
\right)
\!&=&\!
\frac{\zeta\dot\gamma^2}{4d}
\varphi^{*2}
g_{0}(\varphi) 
\!\sum_{\ell =1}^{2}
\!\left\{
-\mathcal{S}_{\alpha\beta}^{(\ell :\mathrm{c1})}
\!\!+\!
\Lambda
\Pi_{\rho\lambda}^{(1)}
\mathcal{S}_{\alpha\beta\rho\lambda}^{(\ell :\mathrm{nc1})} 
\right\}\!.
\hspace{1.5em}
\label{eq:ss3_order1}
\end{eqnarray}
Here, we have introduced
\begin{eqnarray}
\Pi_{\alpha\beta}^{(m)}
:=
\frac{\sigma_{\alpha\beta}^{(m)}}{P^{(m)}}
+
\delta_{\alpha\beta}
\hspace{1em}
(m=1,2),
\end{eqnarray}
where $P^{(m)}$ is the component of the pressure of
$\mathcal{O}(g_0(\varphi)^m)$ ($m=1,2$).
The valid range of Eqs.~(\ref{eq:ss3_order2}) and (\ref{eq:ss3_order1})
from Eq.~(\ref{eq:ss4}) is discussed in
Appendix~\ref{app:sec:NumericalSolution}.

Another implication of the symmetry is that the uniform profile of the
velocity distribution is maintained.
This issue is discussed in Appendix~\ref{app:sec:Implication_symmetry}.

\subsection{Interpretation of the temperature}
\label{subsec:temperature}
In contrast to the case of the kinetic theory of dilute and moderately
dense gases, or even dense inertial suspensions near the jamming point,
the determination method of the temperature $T$ is not clear for
non-Brownian suspensions.
As discussed in Sec.~\ref{subsec:Grad}, in the kinetic theory, the
temperature is determined by the equation of state $P^{(K)} = nT_K$,
where $P^{(K)}$ and $T_K$ are the kinetic pressure and the kinetic
temperature, respectively.
We attempt to introduce the temperature by the Cauchy stress, which
dominates the kinetic stress in dense non-Brownian suspensions.
Because $T$ appears in the position-dependent part in
Eq.~(\ref{eq:f_Grad}), $T$ should be defined by the position-dependent
Cauchy stress.
Note that $T$ determines the magnitude of the nonequilibrium correction
of the distribution function, Eq.~(\ref{eq:f_Grad}).
Accordingly, $\Lambda$ introduced in Eq.~(\ref{eq:Lambda}) determines
the nonequilibrium correction of the stress, Eq.~(\ref{eq:ss4}) or
(\ref{eq:ss4_components}).
In this paper, let us introduce $T$ by the equation of state
\begin{eqnarray}
P^{(\rm{eq})} 
= 
nT \left[ 1 +
2\varphi\, g_0(\varphi)\right],
\label{eq:eos}
\end{eqnarray}
where $P^{(\rm{eq})}$ is the equilibrium part of the pressure determined
from Eq.~(\ref{eq:ss4}), (\ref{eq:ss4_components}), or
(\ref{eq:ss3_order1}) with $\Lambda =0$.
In other words, $P^{(\rm{eq})}$ can be estimated only by $f_{\rm
eq}(\bm{\Gamma}_r)$ in Eq.~(\ref{eq:f_Grad}).
The solution of Eq.~(\ref{eq:ss3_order1}) is given in
Eqs.(\ref{eq:ss_final_1}), (\ref{eq:ss_final_12}) in
Sec.~\ref{app:sec:analytic_solution}, from which we obtain
$P^{(\rm{eq})}$ as ($P^{(1)}$ with $\Lambda =0$)
\begin{eqnarray}
P^{(\rm{eq})} 
=
0.0060 \times \frac{\zeta\dot\gamma}{4d} \varphi^{*2} g_0(\varphi)
=
0.022 \times \frac{\zeta\dot\gamma}{4d} \varphi^{2} g_0(\varphi),
\label{eq:P_eq}
\end{eqnarray}
where $\varphi^* = 6\varphi/\pi$.
Thus, from Eqs.~(\ref{eq:eos}) and (\ref{eq:P_eq}), we obtain
\begin{eqnarray}
T
=
0.022 \times \frac{\zeta\dot\gamma}{4nd} 
\frac{\varphi^2 g_0(\varphi)}{1 + 2 \varphi g_0(\varphi)}
\approx
0.022 \times \frac{\zeta\dot\gamma d^2}{8nd^3} \varphi
=
0.0014\, \zeta\dot\gamma d^2,
\label{eq:T_}
\end{eqnarray}
where we have approximated $1 + 2\varphi g_0(\varphi) \approx 2\varphi
g_0(\varphi)$ in the second equality.
This determines $\Lambda$ as
\begin{eqnarray}
\Lambda 
= 
\frac{\zeta\dot\gamma d^2}{4T}
\approx
174,
\label{eq:Lambda2}
\end{eqnarray}
which is independent of dimensional physical variables such as
$\dot\gamma$, $d$, or $\eta_0$, and is merely a number.
This value is larger than the value $\Lambda = 0.04$ determined by
fitting the absolute values of the shear and pressure viscosities to the
result of molecular dynamics (MD) simulation (cf. Secs.~\ref{sec:muJ}
and \ref{sec:Comparison}).
This might be due to the fact that $\zeta$ in Eq.~(\ref{eq:Lambda2}) is
not equivalent to $\zeta_0$ which sets the magnitude of the viscosities
in MD simulation.
In fact, $\zeta$ is given by Eq.~(\ref{eq:zeta_def}) as $\zeta =
3\zeta_0/(128\epsilon^2)$, where $\epsilon \ll 1$ is the magnitude of
the separation of contacting particles.
The ratio $\zeta/\zeta_0 = 174/0.04$ corresponds to $\epsilon \approx
0.002$, but since we cannot determine the magnitude of $\epsilon$ in our
framework, we will leave $\Lambda$ as a fitting parameter.

The important implication of Eq.~(\ref{eq:T_}) is
\begin{eqnarray}
T 
\propto 
\zeta\dot\gamma d^2
\propto
3\pi d^3 \eta_0 \dot\gamma,
\label{eq:Teff}
\end{eqnarray}
which is consistent with the ``effective temperature'' of
suspensions~\citep{OODLLN2002, EKMW2010} defined via the Stokes-Einstein
relation, $D=T_{\mathrm{eff}}/(3\pi d\,\eta_0)$, where $D$ is the
diffusion coefficient.
It has been shown by experiment~\citep{LA1987-1, LA1987-2, BETA1998,
BEBJM2002, EKMW2010} and simulation~\citep{FB1999, Olsson2010, HBB2010}
that $D\sim d^2 \dot\gamma$ holds below the jamming point and $D$
exhibits only a weak dependence on the density, which suggests
$T_{\mathrm{eff}}=3\pi d\,\eta_0 D\sim 3\pi d^3 \eta_0 \dot\gamma$.

\subsection{Viscosities and $\mu$-$J$ rheology
in the steady state} 
\label{sec:muJ}
Let us analyse the rheology in the steady state, i.e. consider
Eq.~(\ref{eq:ss4}) with $d\sigma_{\alpha\beta}/dt = 0$.
Then, Eqs.~(\ref{eq:ss3_order2}) and (\ref{eq:ss3_order1}) can be solved
analytically.
The analytic solutions for $\sigma_{\alpha\beta}^{(2)}$ and
$\sigma_{\alpha\beta}^{(1)}$ are given by
Eqs.~(\ref{eq:ss_final_2})--(\ref{eq:ss_final_13}) in
Appendix~\ref{app:sec:analytic_solution}, together with
Eq.~(\ref{eq:g0}):
\begin{eqnarray}
\sigma_{\alpha\beta}^{(2)} 
&\sim&
\frac{\zeta\dot\gamma}{d}
g_{0}(\varphi)^2
\sim
\frac{\zeta\dot\gamma}{d}
\delta\varphi^{-2},
\\
\sigma_{\alpha\beta}^{(1)} 
&\sim&
\frac{\zeta\dot\gamma}{d}
g_{0}(\varphi)
\sim
\frac{\zeta\dot\gamma}{d}
\delta\varphi^{-1}.
\end{eqnarray}
In particular, the normalized shear viscosity
$\eta_s^* := \eta_s/\eta_0$ and pressure viscosity $\eta_n^* :=
\eta_n/\eta_0$ are given by
\begin{eqnarray}
\eta_s^{*} \!
&\approx&
\!0.0073\, \varphi^{*3} \delta\varphi^{-2} 
\!\!+ 
\!0.080\, \varphi^{*2} \delta\varphi^{-1},
\label{eq:sxy_Lambda0.04}
\\
\eta_n^{*} \!
&\approx&
\!0.045\, \varphi^{*3} \delta\varphi^{-2}
\!\!+
\!0.0005\, \varphi^{*2} \delta\varphi^{-1}
\!\!\approx
0.045\, \varphi^{*3} \delta\varphi^{-2},
\hspace{1.8em}
\label{eq:P_Lambda0.04}
\end{eqnarray}
where $\Lambda = 0.04$ is adopted, which is determined by fitting the
absolute values of $\eta_s^*$ and $\eta_n^*$ to the results of the MD
simulation shown in the next section.
The empirical formula
$g_{0}(\varphi)=g_{{\mathrm{CS}}}(\varphi_{f})(\varphi_J-\varphi_f)/(\varphi_J-\varphi)$
with $\varphi_f = 0.49$ and
$g_{{\mathrm{CS}}}(\varphi):=(1-\varphi/2)/(1-\varphi)^{3}$ valid for
$\varphi_f < \varphi < \varphi_J$~\citep{Torquato1995} is adopted, from
which we obtain $g_0(\varphi)\approx 0.848\, \delta\varphi^{-1}$ for
$\varphi_J = 0.639$.
Although it is widely recognized that $\varphi_{J}\approx 0.64$ for
monodisperse frictionless hard spheres without any solvent, it has been
reported that the value of $\varphi_{J}$ is not uniquely identified and
depends on the protocols used to generate the jammed
configurations~\citep{CCC2010}.
In this work, we adopt $\varphi_J = 0.639$, which is obtained by the
conjugate-gradient protocol~\citep{OSLN2003}.
The numerical coefficients in Eqs.~(\ref{eq:sxy_Lambda0.04}) and
(\ref{eq:P_Lambda0.04}) are determined analytically in rational forms,
but we only show their approximate values in decimals for brevity.

It is evident from Eqs.~(\ref{eq:sxy_Lambda0.04}) and
(\ref{eq:P_Lambda0.04}) that $\eta_s^*$ and $\eta_n^*$ exhibit 
\begin{eqnarray}
\eta_s^*
\sim \eta_n^* \sim \delta\varphi^{-2}
\end{eqnarray}
near the jamming point.
On the other hand, $\mathcal{O}(\delta\varphi^{-1})$ term in $\eta_n^*$
is small for finite $\delta\varphi$, while the corresponding term in
$\eta_s^*$ is significant.
Noting $J=1/\eta_n^{*}$, $J$ and $\delta\varphi$ are uniquely invertible
via Eq.~(\ref{eq:P_Lambda0.04}) as
\begin{eqnarray}
J^{1/2}
\approx
4.74\varphi^{*-3/2}\delta\varphi,
\label{eq:J1/2_dphi}
\end{eqnarray}
or, equivalently,
\begin{eqnarray}
\varphi
=
\varphi_{J}
-
0.211\,\varphi^{*3/2} J^{1/2}.
\end{eqnarray} 
From Eqs.~(\ref{eq:sxy_Lambda0.04}) and (\ref{eq:P_Lambda0.04}), we
obtain the stress ratio $\mu$ as
\begin{eqnarray}
\mu
\approx
0.163 + 
1.78\varphi^{*-1}\, \delta\varphi
=
0.163 + 
0.377\,\varphi^{*1/2} J^{1/2}.
\hspace{1.5em}
\label{eq:mu_theory}
\end{eqnarray}
Note that $\mu$ in the jamming limit, $\mu(J\to 0)=0.163$, is
independent of $\Lambda$.
%
%
\begin{figure*}
\centerline{
\includegraphics[width=7.0cm]{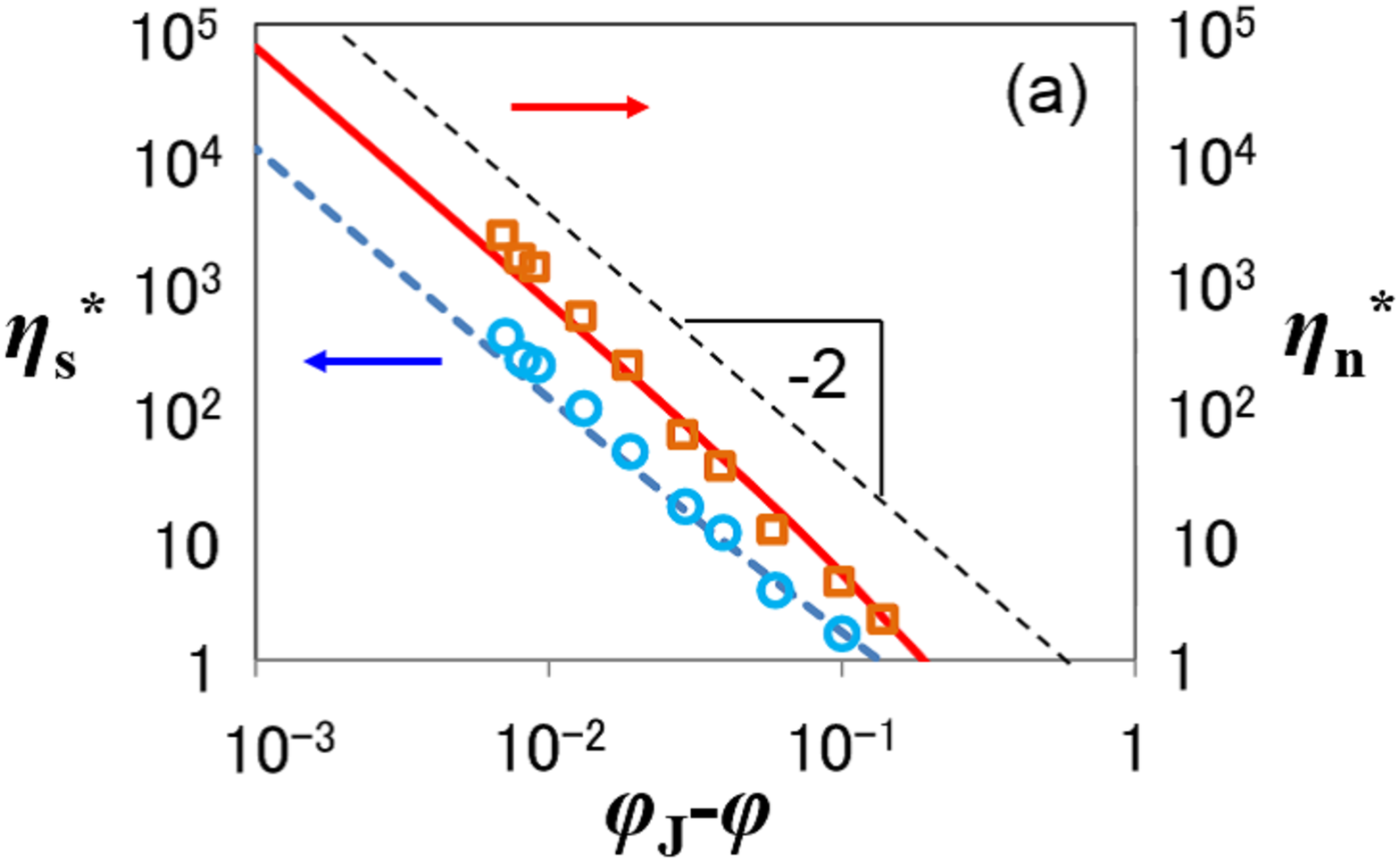}
\hspace{1em}
\includegraphics[width=7.0cm]{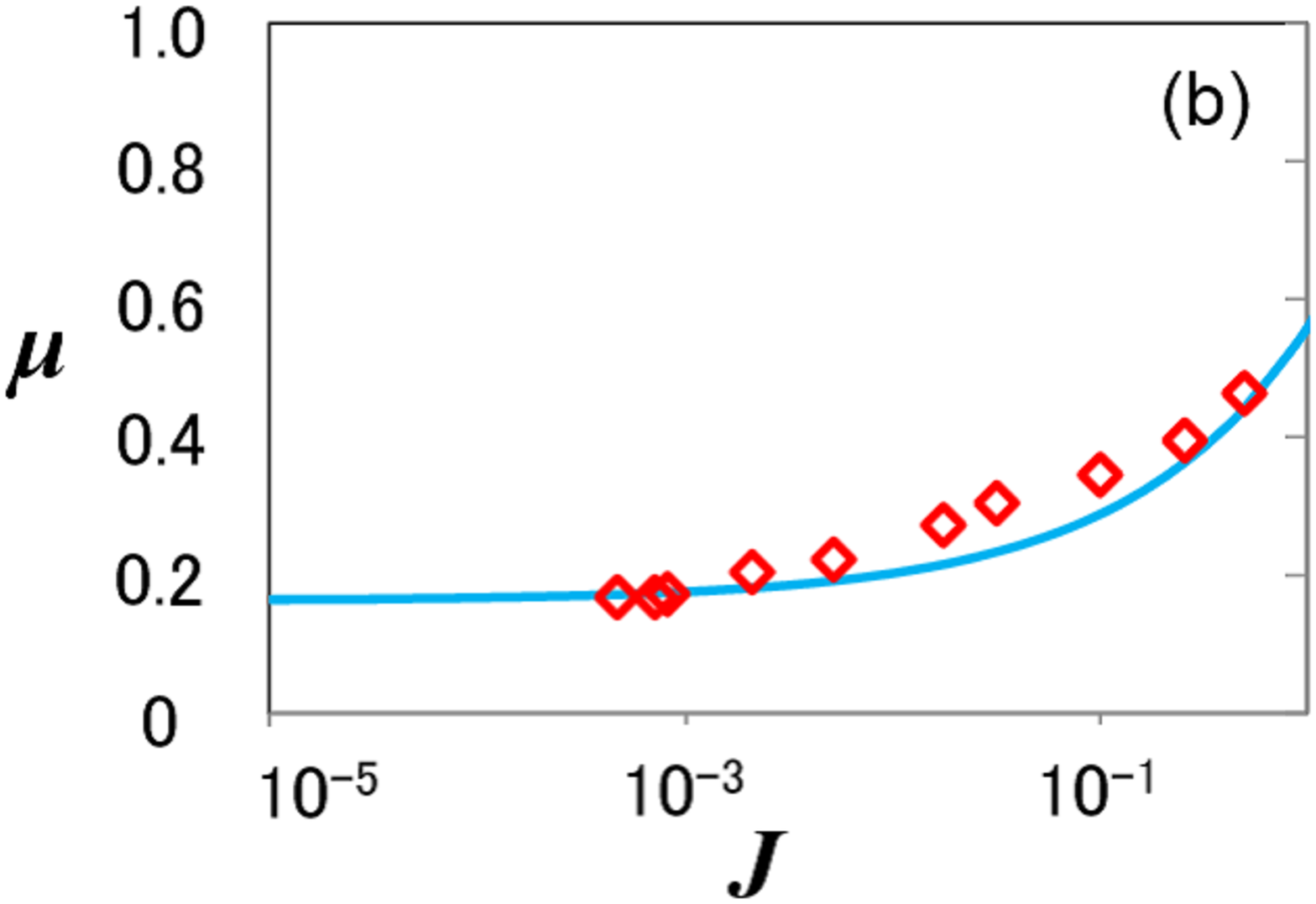}
}
\caption {Comparison of theory with MD simulation. 
(a) The normalized shear viscosity $\eta_s^{*}=\eta_s/\eta_0$ and the
pressure viscosity $\eta_n^{*}=\eta_n/\eta_0$, and (b) the stress ratio
$\mu = \sigma_{xy}/P$.
The results of the theory and MD for the shear (pressure) viscosity are
shown in dashed (solid) lines and open circles (squares) in (a), and
those for the stress ratio are shown in solid line and open diamonds in
(b), respectively.
The theoretical result adopts the fitting parameter $\Lambda = 0.04$. 
The result of MD is for $\varphi=$ 0.632, 0.631, 0.63, 0.626, 0.62,
0.61, 0.6, 0.58, 0.54, and 0.5.  }
\label{Fig:viscosities}
\end{figure*}
%

\section{Comparison with simulation and previous results}
\label{sec:Comparison}
We compare our theory with the MD simulation and previous results.
We adopt the algorithm for Brownian hard spheres~\citep{SVM2007} applied
at zero thermal fluctuations for the MD simulation.
By this choice the only source of the velocity fluctuation is the
interparticle contacts.
The resistance matrix is simplified to $\olra{\zeta^{(N)}} = \zeta_0
\bm{I}$, where $\zeta_0 := 3\pi d\, \eta_0$.
The details of the simulation scheme is presented in
Appendix~\ref{app:sec:EDMD}.

We show the results for the normalized shear viscosity $\eta_s^*$,
Eq.~(\ref{eq:sxy_Lambda0.04}), pressure viscosity $\eta_n^*$,
Eq.~(\ref{eq:P_Lambda0.04}), and the stress ratio $\mu$,
Eq.~(\ref{eq:mu_theory}), in Fig.~\ref{Fig:viscosities}.
As is already mentioned, we adopt $\Lambda=0.04$ as a fitting parameter
for $\eta_s^*$ and $\eta_n^*$.
In the dense region, the theory predicts $\eta_s^{*}\sim\eta_n^{*} \sim
\delta\varphi^{-2}$ in accordance with the MD result, and the stress
ratio approaches $\mu(J\to 0)= 0.163$, which is also in good agreement
with MD.
Because there exists slight difference between $\eta_s^*$ and $\eta_n^*$
for finite $\delta\varphi$, we find that $\mu$ depends on $J$ or
$\delta\varphi$ ($\mu$-$J$ rheology).
A reasonable agreement between our theory and MD is found, although the
applicability for large $J$ in our theory is questionable (see
Appendix~\ref{app:sec:NumericalSolution}).

We compare our theory with previous results.
It is notable that Eqs.~(\ref{eq:J1/2_dphi}) and (\ref{eq:mu_theory})
are consistent with the experimental results $\delta\varphi\propto
J^{1/2}$ and $\mu(J\to 0)\approx 0.32$~\citep{BGP2011}.
%
%
\begin{figure}
\centerline{\includegraphics[width=6.0cm]{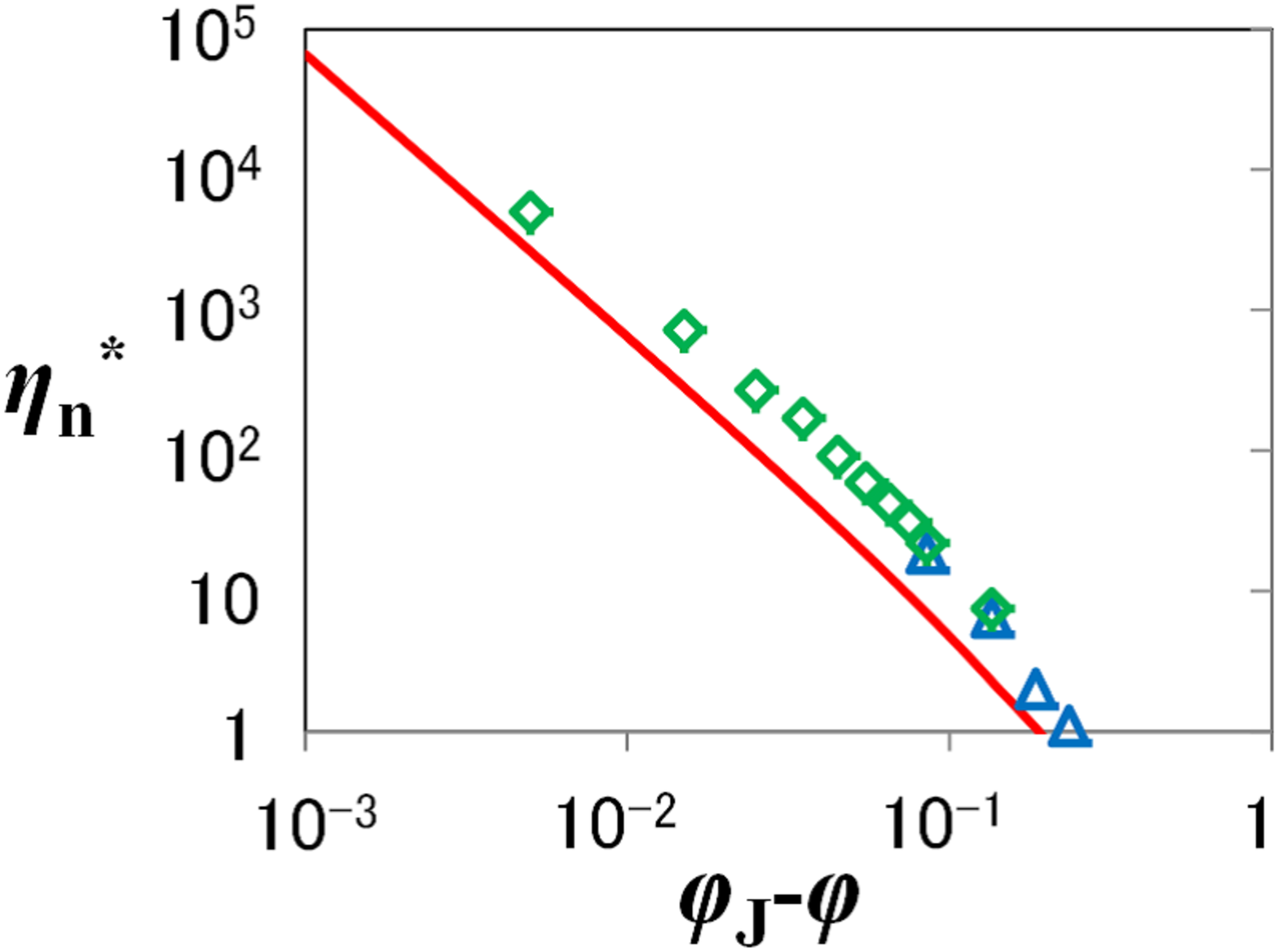}} 
\caption {Comparison with experiments~\cite{DGMYM2009,DBHGP2015}.
The results of the normalized pressure viscosity $\eta_{n}^{*}$ are
shown for our theory (solid line) and the experiments in
\cite{DGMYM2009} (open triangles) and \cite{DBHGP2015} (open
diamonds).
The theoretical result is for $\Lambda = 0.04$.}
\label{Fig:compare_exp}
\end{figure}
%
%
Discrepancy in the value of $\mu(J\to 0)$ might be caused by the
friction or the hydrodynamic interaction between the particles.
%
A comparison of the pressure viscosity with experimental
results~\citep{DGMYM2009,DBHGP2015} shows an agreement within a factor
of 2 (see Fig.~\ref{Fig:compare_exp}), and a comparison with empirical
relations~\citep{MB1999} also shows good agreement (see
Appendix~\ref{app:sec:MorrisBoulay}).

Next we discuss the results for the normal stress differences.
The two normal stress differences, $N_1 = \sigma_{xx}-\sigma_{yy}$ and
$N_2 = \sigma_{yy}-\sigma_{zz}$, are evaluated at
$\mathcal{O}(g_0(\varphi)^2)$ as
\begin{eqnarray}
N_1
&\approx&
-1.11\Lambda\, \Sigma^{(2)},
\\
N_2
&\approx&
0.576\Lambda\, \Sigma^{(2)},
\end{eqnarray}
where $\Sigma^{(2)}:= \frac{\zeta\dot\gamma}{4d} \varphi^{*3}
g_{0}(\varphi)^2$.
From these we see that $N_1 <0$, $N_2 >0$, and 
%
$N_1/N_2 \approx -1.9$, $N_2/P \approx 0.9$, which are independent of
$\Lambda$.
This is consistent with the experimental observation that $N_1 <
0$, $N_2 > 0$, and $N_1/N_2 = -2$~\citep{Laun1994}.
It should be noted, however, that the magnitudes and even the signs of
the normal stress differences are controversial.
For instance, \cite{LDHH2005} report that the sign of $N_{1}$ depends
on the volume fraction, and \cite{MSMD2014} exhibit $N_2 < 0$ and $|N_2|
\gg |N_1|$, while \cite{CW2014} assert $N_1, N_2 <0$ and $|N_1| \approx
|N_2|$.
The pressure and the normal stress differences can be expressed in the
form $P = \dot\gamma \eta_0 \eta_n (1+\lambda_2+\lambda_3)/3$, $N_1 =
-\dot\gamma \eta_0 \eta_n (1-\lambda_2)$, and $N_2 = -\dot\gamma \eta_0
\eta_n (\lambda_2-\lambda_3)$, where $\lambda_2 :=
\sigma_{yy}/\sigma_{xx}$ and $\lambda_3 :=
\sigma_{zz}/\sigma_{xx}$~\citep{MB1999}. 
From Eq.~(\ref{eq:ss_final_2}), we obtain
$\lambda_2
\approx
0.08$,
$\lambda_3
\approx
0.56$,
which are also independent of $\Lambda$.
The value of $\lambda_3$ is close to the value 1/2 determined in
\cite{MB1999}.
The value of $\lambda_2$ is left controversial as in \cite{MB1999}, but
the assumed values such as 0.6, 0.8, or 1.0 are significantly larger
than 0.08 in \cite{MB1999}.

\begin{figure}
\centerline{
\includegraphics[width=3.0cm]{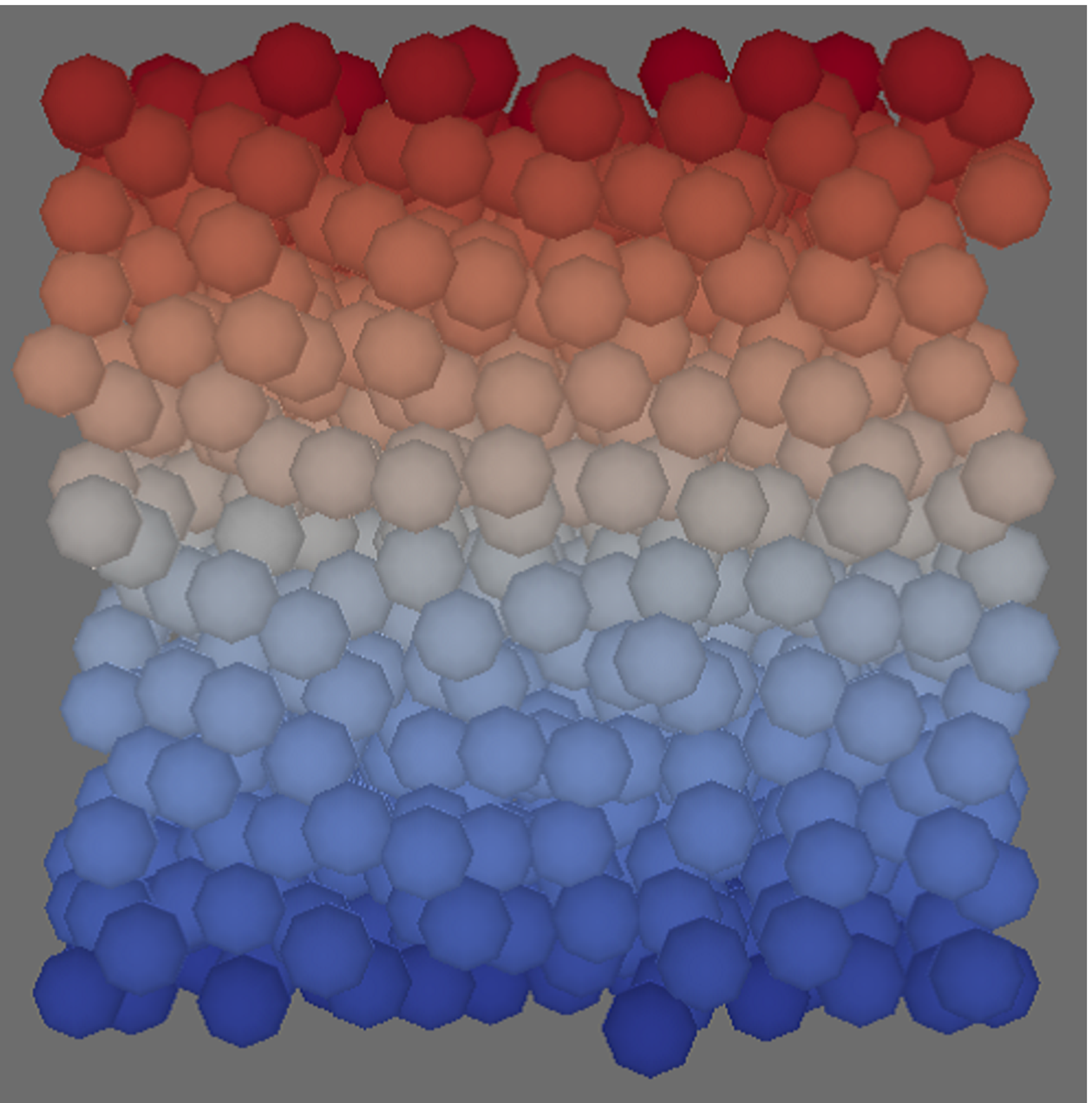} 
\includegraphics[width=3.0cm]{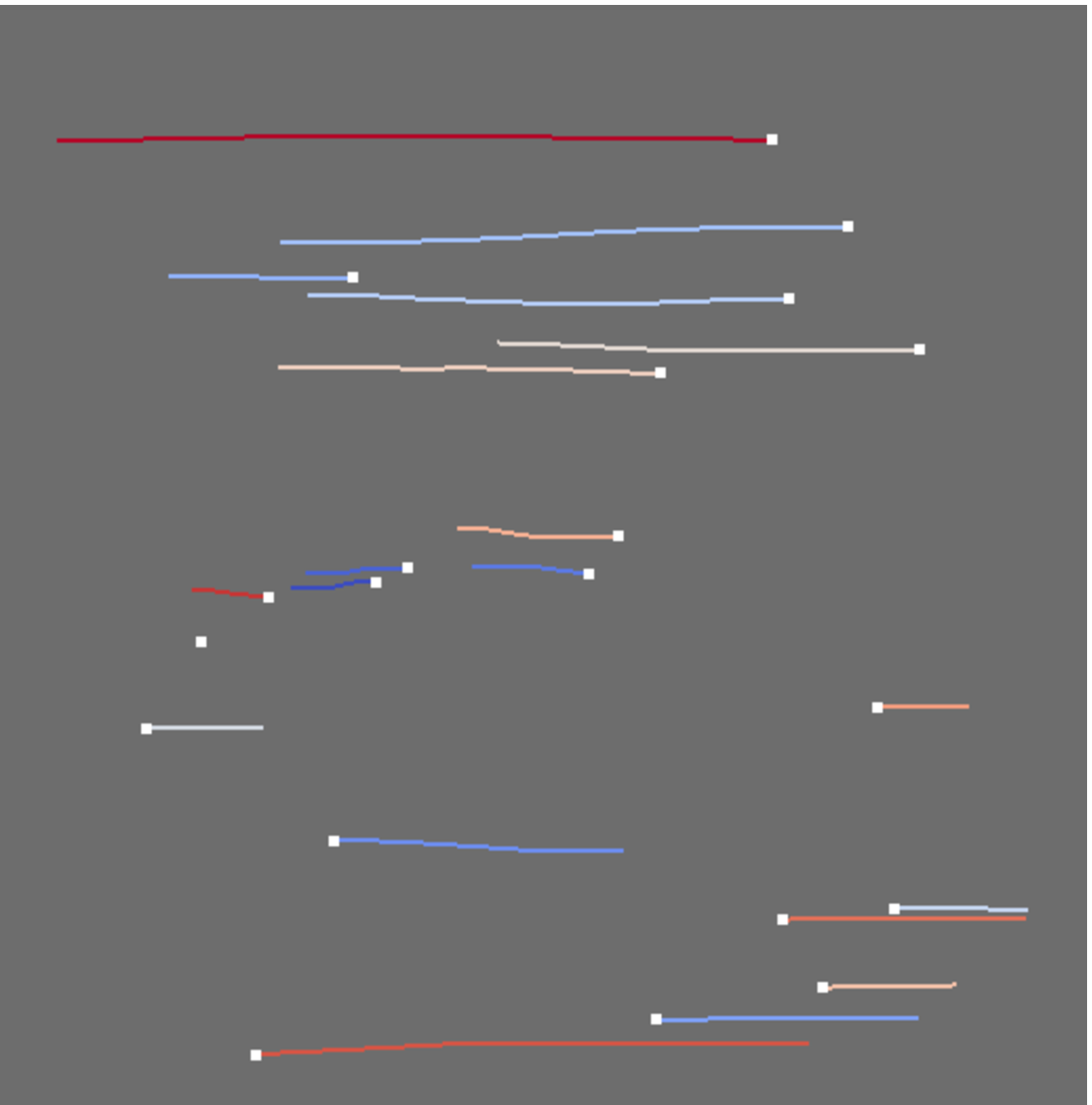} 
\includegraphics[width=3.0cm]{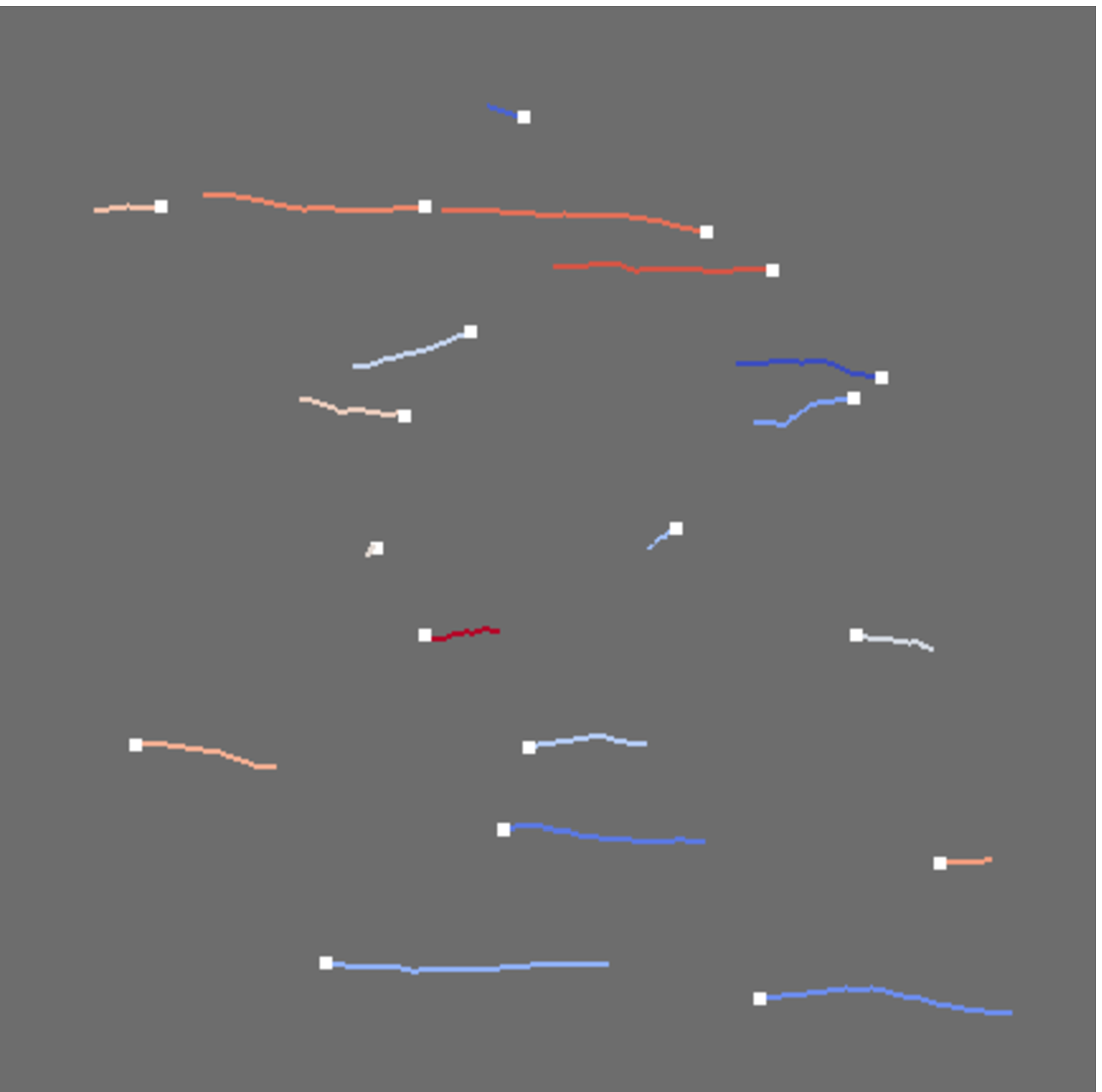} 
\includegraphics[width=3.0cm]{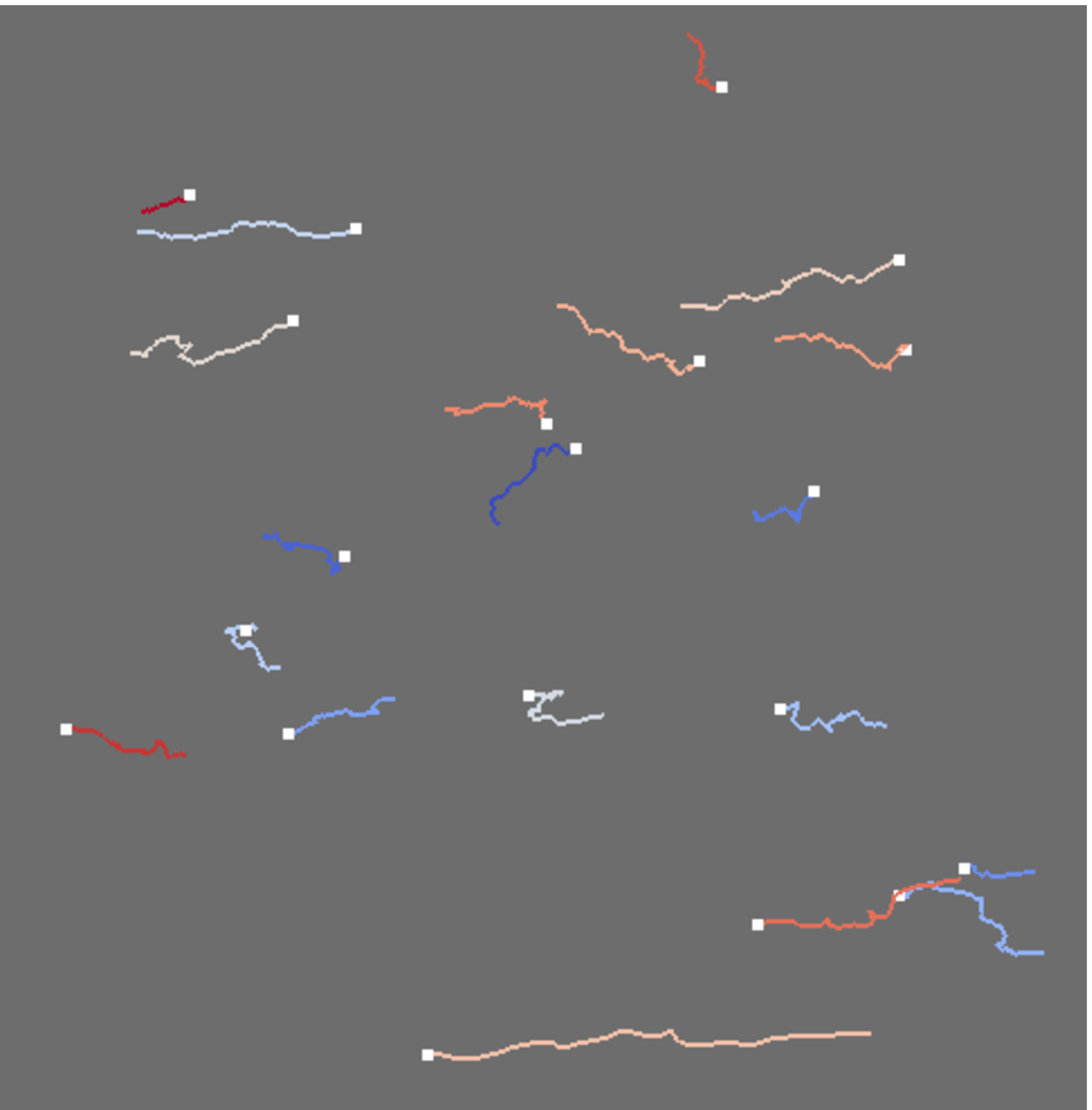} 
}
\caption {Trajectory of randomly sampled particles for
 $\varphi = 0.30, 0.60, 0.635$, from left to right. The trajectory is
 ballistic for $\varphi=0.30$, while it is diffusive for $\varphi =
 0.635$. 
%
}
\label{Fig:Trajectory}
\end{figure}

Finally we clarify the difference between Brownian and non-Brownian
suspensions.
It is crucial in Brownian suspensions that the effective (long-time)
self-diffusion constant $D_{\infty}(\varphi)$ vanishes at the jamming
point, $D_{\infty}(\varphi)\sim\delta\varphi =\varphi_J - \varphi$,
since the shear stress scales as $\sigma_{xy}\sim
g_0(\varphi)/D_{\infty}(\varphi)$~\citep{Brady1993}.
%
%
In contrast, the above argument is not valid for non-Brownian
suspensions.
In fact, it is reported that $D_{\infty}(\varphi)$ increases as the
density is increased and saturates at the jamming point~\citep{LA1987-1,
LA1987-2, BETA1998, BEBJM2002, HBB2010, Olsson2010}.
This feature can be understood from the fact that, in non-Brownian
suspensions, the source of the non-affine displacement is the
contact interaction between the particles, rather than the fluctuating
force from the solvent.
It is obvious that the contact interaction is more significant in dense
suspensions, which results in larger $D_{\infty}(\varphi)$.
This feature can be seen in the trajectory of randomly sampled particles
shown in Fig.~\ref{Fig:Trajectory}.
This clearly suggests that $\sigma_{xy}\sim\delta\varphi^{-2}$ is not
the consequence of the diffusion constant in non-Brownian suspensions.
Although the density dependence is similar, the underlying physics of
the rheology is distinct for Brownian and non-Brownian suspensions.

%
%
%
%

\section{Validation of the postulates}
\label{sec:validation}

In this section we articulate the crucial postulates of our theory and
report the results of the simulations performed for their verification.
The postulates are
\begin{enumerate}
 \item the factorization of the distribution function into the peculiar
       velocity-dependent and position-dependent parts;
 \item Grad's 13-moment-like expansion of the position-dependent
       distribution function, Eq.~({\ref{eq:f_Grad}}), which is an
       expansion around thermal equilibrium;
 \item the factorization approximation of the multi-body correlations
       used in Eqs.~(\ref{eq:3body_approx}) and (\ref{eq:4body_approx}).
\end{enumerate}

\subsection{Peculiar velocity distribution function}

\begin{figure}
\centerline{
\includegraphics[width=5.5cm]{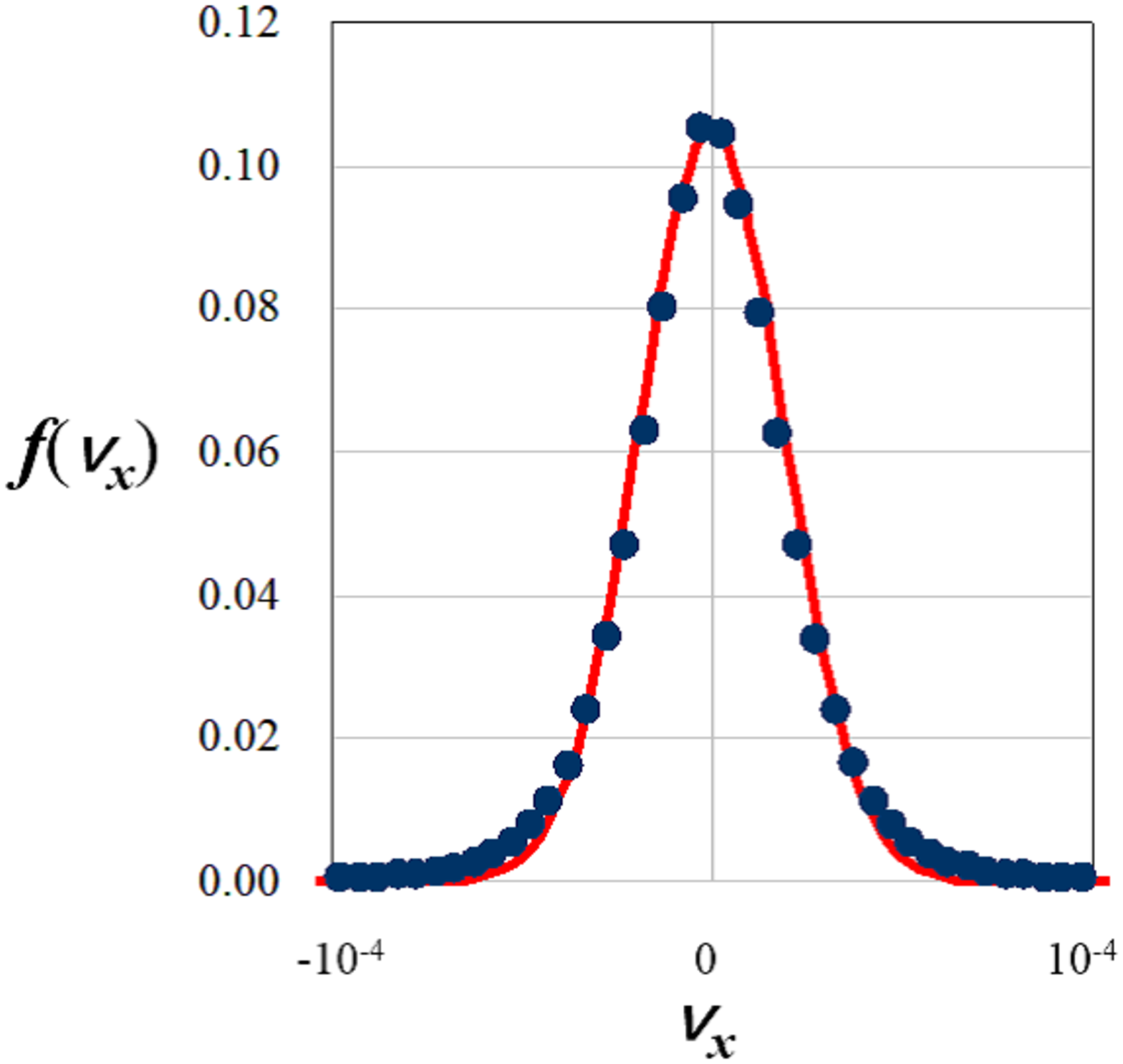}
\includegraphics[width=5.5cm]{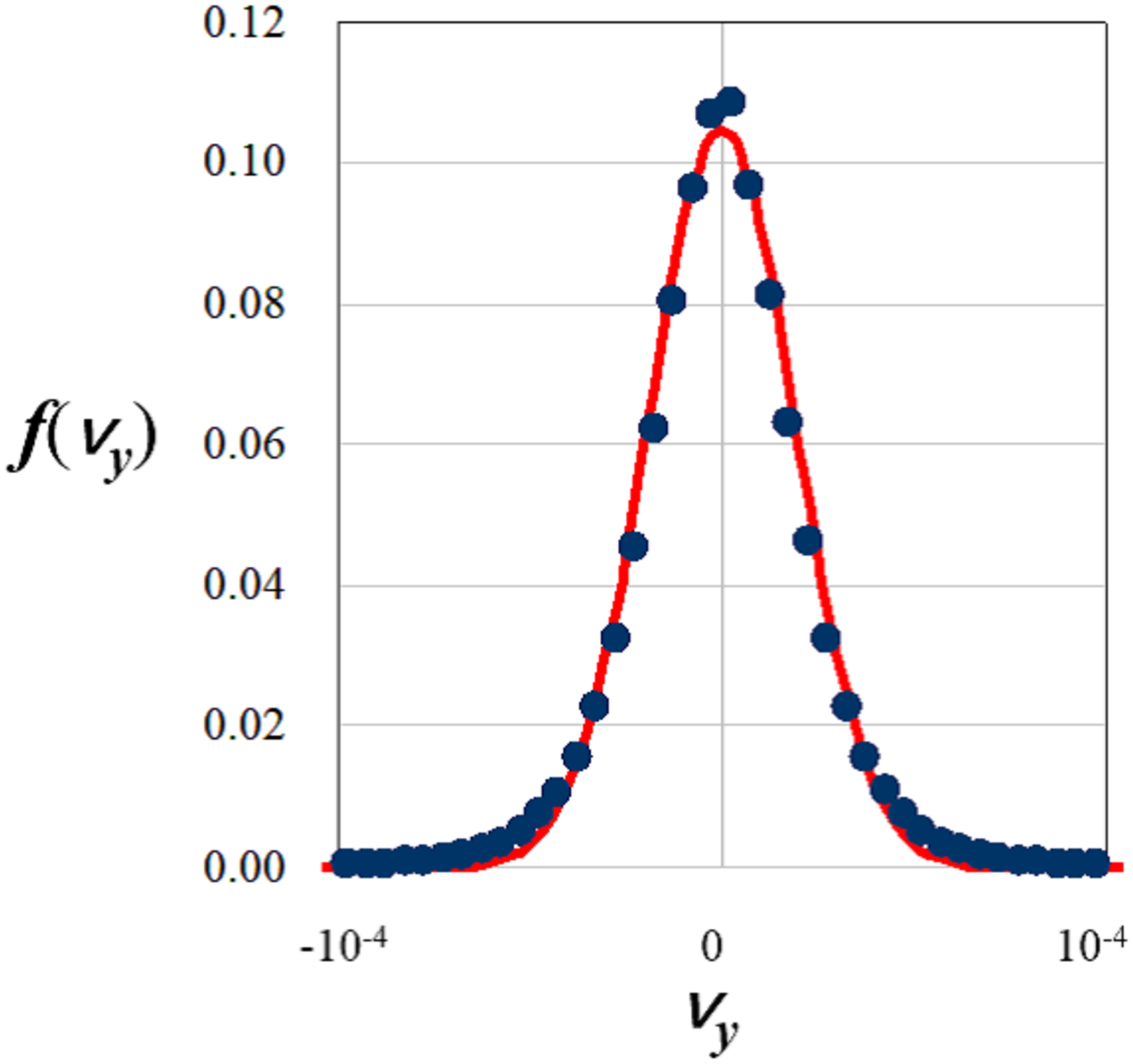}
}
\caption{Distribution of the peculiar velocity measured by MD simulation (circles). 
The conditions are $\varphi=0.63$, $N=1000$, $\dot\gamma^{*} = 1\times 10^{-5}$. 
The number of data averaged for each point is $10^5$. 
The figure shows the $x$- and $y$-components.
Also shown is the Gaussian distribution fitted to the measured data
(solid line).}
\label{Fig:vdf}
\end{figure}

The distribution of the peculiar velocity $\bm{v}_i = \dot{\bm{r}}_i -
\dot\gamma y_i \bm{e}_x$ is measured by MD simulation.
As shown in Fig.~\ref{Fig:vdf}, the peculiar velocity distribution
function is nearly Gaussian in spite of the absence of any inertial
effect in the dynamics.
This result implies that the factorization of the distribution function
assumed in Eq.~(\ref{eq:f_Grad}) is valid.
The result of our simulation supports another theoretical assumption
that the base state is nearly equilibrium.

\subsection{Grad's 13-moment-like expansion and 
factorization of multi-body correlations}

We validate the expansion of the distribution of the position with the
stress current, Eq.~(\ref{eq:f_Grad}).
For this purpose, we consider the radial distribution function at the
steady state, $g_{\dot\gamma}(\bm{r})$, which is anisotropic under shear.
It is given by
\begin{eqnarray}
g_{\dot\gamma}(\bm{r}) 
:=
\frac{1}{Nn}
\left\langle
\sum_i \sum_{j\neq i}
\delta(\bm{r}-\bm{r}_{ij})
\right\rangle
=
\frac{1}{Nn}
\int d\bm{\Gamma} f(\bm{\Gamma})
\sum_i \sum_{j\neq i}
\delta(\bm{r}-\bm{r}_{ij}).
\label{eq:g_gamma}
\end{eqnarray}
From Eq.~(\ref{eq:f_Grad}), after trivially integrating out
$f_{\rm{eq}}(\bm{\Gamma}_v)$, we obtain
\begin{eqnarray}
g_{\dot\gamma}(\bm{r}) 
&=&
g(r)
+
\frac{1}{Nn}
\frac{V}{2T}
\Pi_{\alpha\beta}
\left\langle
\sum_i \sum_{j\neq i}
\delta(\bm{r}-\bm{r}_{ij})
\tilde{\sigma}_{\alpha\beta}
\right\rangle_{\rm{eq}},
\end{eqnarray}
where $g(r)$ is the radial distribution function at equilibrium.
By substituting the expression for the stress
$\tilde{\sigma}_{\alpha\beta}$, Eq.~(\ref{eq:micro_stress}), and that
for the interparticle force $\bm{F}_{ij}^{(\rm{p})}$ therein,
Eq.~(\ref{eq:F_ij_p}), we obtain
\begin{eqnarray}
g_{\dot\gamma}(\bm{r}) 
&=&
g(r)
-
\frac{1}{Nn}
\frac{\zeta_e \dot\gamma d^2}{8T}
\Pi_{\alpha\beta}
\left\langle
\sum_i \sum_{j\neq i} \sum_{k\neq i}
\delta(\bm{r}-\bm{r}_{ij})
r_{ik,\beta} 
\delta(r_{ik}-d) \mathcal{P}(\hat{x}_{ik}, \hat{y}_{ik}) \hat{r}_{ik,\alpha}
\right\rangle_{\rm{eq}} 
\nonumber \\
&=&
g(r)
-
\frac{1}{Nn}
\frac{\zeta_e \dot\gamma d^3}{8T}
\Pi_{\alpha\beta}
\left\langle
\sum_i \sum_{j\neq i} \sum_{k\neq i}
\delta(\bm{r}-\bm{r}_{ij})
\delta(r_{ik}-d) \mathcal{P}(\hat{x}_{ik}, \hat{y}_{ik})
\hat{r}_{ik,\alpha} \hat{r}_{ik,\beta} 
\right\rangle_{\rm{eq}}. 
\hspace{2em}
\label{eq:g_gamma_1}
\end{eqnarray}
In the second equality, we have utilized $r_{ik,\beta}\delta(r_{ik}-d) =
d\, \hat{r}_{ik,\beta} \delta(r_{ik}-d)$. 
Equation~(\ref{eq:g_gamma_1}) is expressed in terms of the
triplet-correlation function $g^{(3)}(\bm{r},\bm{r}')$, Eq.~(\ref{eq:g3}), as
\begin{eqnarray}
g_{\dot\gamma}(\bm{r}) 
&=&
g(r)
-
n
\frac{\zeta_e \dot\gamma d^3}{8T}
\Pi_{\alpha\beta}
\int d^3 \bm{r}'
g^{(3)}(\bm{r},\bm{r}')
\delta(r'-d) \mathcal{P}(\hat{x}', \hat{y}') 
\hat{r}_{\alpha}' \hat{r}_{\beta}'.
\end{eqnarray}
We apply the factorization approximation to $g^{(3)}(\bm{r},\bm{r}')$,
\begin{eqnarray}
g^{(3)}(\bm{r},\bm{r}')
\approx
g(r)g(r')
\left[
1 + h(|\bm{r}-\bm{r}'|)
\right],
\label{eq:g3_angle}
\end{eqnarray}
\begin{figure*}
\centerline{
\includegraphics[width=6.5cm]{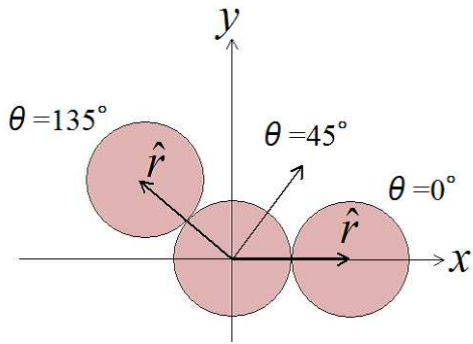}
}
\caption{Definition of the angle of the separation vector.
$\theta = 135$ and 45 degrees are the compressional and extensional directions of the shear, respectively.
$\theta = 0$ degree is the ``flow direction'', which is the direction of the flow of the solvent.
}
\label{Fig:angle}
\end{figure*}
\begin{figure*}
\centerline{
\includegraphics[width=8.5cm]{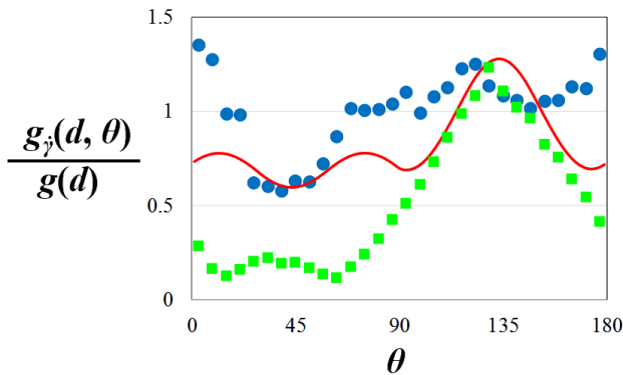}
}
\caption{Angular distribution of the contact value of the radial
distribution function at the steady state measured by MD simulation (circle).
The conditions are $\varphi=0.63$, $N=1000$, $\dot\gamma^{*} = 1\times
10^{-5}$.
The number of data averaged for each point is $10^4$. 
Also shown are the theoretical result of Grad's expansion (solid line) and the direct measurement of Grad's expansion (squares).}
\label{Fig:rdf_Grad}
\end{figure*}
which leads to
\begin{eqnarray}
g_{\dot\gamma}(\bm{r}) 
&\approx&
g(r)
\left\{
1
-
n
\frac{\zeta_e \dot\gamma d^3}{8T}
\Pi_{\alpha\beta}
\int d^3 \bm{r}'
g(r')
\left[
1 + h(|\bm{r}-\bm{r}'|)
\right]
\delta(r'-d) \mathcal{P}(\hat{x}', \hat{y}') 
\hat{r}_{\alpha}' \hat{r}_{\beta}'
\right\}
\nonumber \\
&=&
g(r)
\left\{
1
-
n
\frac{\zeta_e \dot\gamma d^5}{8T}
g(d)\,
\Pi_{\alpha\beta}
\int d\mathcal{S}'
\left[
1 + h(|\bm{r}-\bm{r}'|)
\right]
\mathcal{P}(\hat{x}', \hat{y}') 
\hat{r}_{\alpha}' \hat{r}_{\beta}'
\right\}.
\label{eq:g_gamma_2}
\end{eqnarray}
Here, $h(r) = g(r)-1$ is the pair-correlation function, and $\int
d\mathcal{S}'\cdots$ denotes angular integral with respect to
$\hat{\bm{r}}'$.

Let us consider the angular dependence of the contact value of the
radial distribution function in the $(x,y)$-plane.
From Eq.~(\ref{eq:g_gamma_2}), this is given by
\begin{eqnarray}
g_{\dot\gamma}(d, \theta) 
&\approx&
g(d)
\left\{
1
-
n
\frac{\zeta_e \dot\gamma d^5}{8T}
g(d)\,
\Pi_{\alpha\beta}
\int d\mathcal{S}'
\left[
1 + h(|\bm{r}-\bm{r}'|)
\right]
\mathcal{P}(\hat{x}', \hat{y}') 
\hat{r}_{\alpha}' \hat{r}_{\beta}'
\right\},
\label{eq:g_gamma_3}
\end{eqnarray}
where $\theta$ is the angle of $\hat{\bm{r}}$ with respect to the
$x$-axis, and $\theta=135^{\circ}$ and 45$^{\circ}$ are the directions
of compression and extension, respectively (cf. Fig.~\ref{Fig:angle}).
We adopt the approximate formula for the delta-function contribution of
$h(r)$ \citep{DTS2005},
\begin{eqnarray}
h(r)
\approx
\frac{1}{4\varphi\delta}
\left[
\frac{6A}{\left( \frac{r/d - 1}{\delta} + C\right)^4}
+
\frac{B}{\left( \frac{r/d - 1}{\delta} + C\right)^2}
\right],
\label{eq:h_Torquato}
\end{eqnarray}
where $\delta \approx (\varphi_J-\varphi)/(3\varphi_J)$ and $A=3.43, B=1.45,
C=2.25$.
For the evaluation of the r.h.s. of Eq.~(\ref{eq:g_gamma_3}), we perform
numerical integration.

The comparison of the both sides of Eq.~(\ref{eq:g_gamma_3}) is shown in
Fig.~\ref{Fig:rdf_Grad}.
The left-hand side (l.h.s.) is the radial distribution function at the
steady state and measured by MD simulation (circles).
The conditions for the simulation are described in the caption.
The angular integration on the r.h.s. has been evaluated as a double
integral with respect to the two angles $(\phi_1, \phi_2)$, where
$\cos\phi_1 = \hat{\bm{r}}\cdot\hat{\bm{r}}'$ and $\phi_2$ is the
azimuthal angle of $\hat{\bm{r}}'$ around $\hat{\bm{r}}$.
These angles are bounded in the range $\pi/3 < \phi_1 < \pi$ and $0 <
\phi_2 < 2\pi$, respectively.
The region $0< \phi_1 < \pi/3$ is excluded to avoid overlap.
The contact value of the equilibrium radial distribution function,
$g(d)$, is given by $g_0(\varphi)\approx 0.848/(\varphi_J-\varphi)$ for
$\varphi_J=0.639$, which is approximately 100 for $\varphi=0.63$.
The result for the r.h.s. shown in solid line in Fig.~\ref{Fig:rdf_Grad}
is reduced to 1/3 to fit the amplitude.
We see that the peak in the compression direction ($\theta =
135^{\circ}$) is captured by the theory, although the peak in the
direction $\theta = 0^{\circ}$ is not.
The reason why the theory still reasonably predicts the stress is that
the particles in contact with $\theta = 0^{\circ}$ does not contribute
to the stress, because they are driven with the same velocity by the
uniform shear.

For further validation, we have measured the r.h.s. of
Eq.~(\ref{eq:f_Grad}) directly by MD simulation.
We measure the distribution of the microscopic stress
$\tilde{\sigma}_{\alpha\beta}$ with respect to the angle of the
separation vector $\bm{r}_{ij}$, which resides in the $(x,y)$-plane.
To extract the contribution from the contacting pairs, we have assigned
1 to $f_{\rm{eq}}(\bm{r}_{ij})$ when the two particles are in contact,
and 0 otherwise.
The factor of $(\pi /180)^2 /Nn \times \Delta r \Delta \theta \Delta
\phi$ is multiplied to obtain values which correspond to the radial
distribution function.
Here $\phi$ is the angle of $\bm{r}_{ij}$ around the
$x$-axis and $\Delta r$, $\Delta \theta$, $\Delta \phi$ are the size of
the bins in each direction.
The result is shown in squares in Fig.~\ref{Fig:rdf_Grad}.
We see that the directly measured distribution has the same tendency
with the theoretically evaluated distribution, i.e. both exhibit a peak
in the compression direction, $\theta =135^{\circ}$, but the peak in
$\theta =0^{\circ}$ is not visible.

\begin{figure*}
\centerline{
\includegraphics[width=7.5cm]{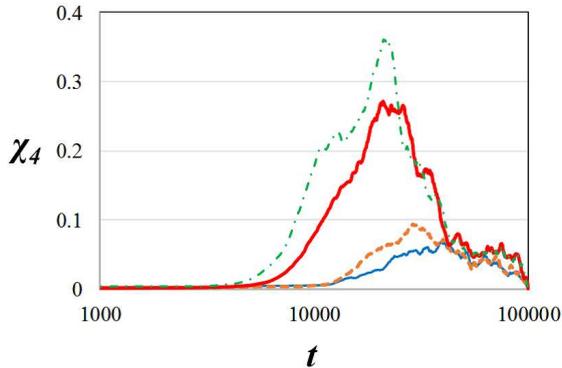}
}
\caption{Four-point susceptibility measured by MD simulation.
The conditions are $\varphi=0.60-0.635$, $N=1000$, $\dot\gamma^{*} = 1\times
10^{-5}$.
The number of data averaged for each point is 1000. 
The results are for $\varphi = 0.60$ (thin solid line), 0.62 (dashed
line), 0.63 (thick solid line), and 0.635 (dot-dashed line).
}
\label{Fig:chi4}
\end{figure*}

It should be noticed that the directly measured distribution does not
reflect the factorization approximation, Eq.~(\ref{eq:g3_angle}).
Hence, this result implies the validity of Eq.~(\ref{eq:g3_angle}).
As a check for the four-body correlation, we have measured the
four-point susceptibility, $\chi_4$, by MD simulation
(cf. Appendix~\ref{app:sec:chi4}).
The result is shown in Fig.~\ref{Fig:chi4}.
We see that no divergence is found in $\chi_4$ near the jamming point.
This implies that the divergence of the stress is determined by the
divergence of the radial distribution function at contact, $g_0$.
Thus, our theoretical treatment based on the factorization approximation
is sufficient to discuss the singularities of the stress at the jamming
point.

\section{Discussion}
First we discuss the hydrodynamic effects.
We have formulated the theory by the resistance matrix
$\olra{\zeta^{(N)}}$ to include the hydrodynamic effects.
In our formulation, which focuses on the proximity effects of the
particles, the far-field part of $\olra{\zeta^{(N)}}$ drops out, and the
lubrication part $\olra{\zeta^{(\rm{2})}_{\mathrm{lub}}}$ is taken into
account.
%
%
Although we have compared the theory and the MD simulation for the
simplified case where $\olra{\zeta^{(\rm{2})}_{\mathrm{lub}}}$ is
proportional to the unit tensor, it is possible to do so for a more
generic case.
This is left for future work.

Next we discuss the effect of the contact force of the particles.
In this work we have considered frictionless spheres, but it is reported
that the exponent of the divergence $\lambda$ depends on the friction
between the particles;
$\lambda = 1.6$ for frictionless spheres, while $\lambda = 2.4$ in the
strong friction limit~\citep{MSMD2014}.
Note that the jamming density $\varphi_J$ used in \cite{MSMD2014}
is 0.66, which reduces the exponent.
Extension of this work to frictional particles is necessary to address
this issue~\citep{SH2017}.

In the course of the derivation, we have assumed a separation between
$\mathcal{O}(g_0(\varphi))^2$ and $\mathcal{O}(g_0(\varphi))$ terms as
in Eqs.~(\ref{eq:ss3_order2}) and (\ref{eq:ss3_order1}).
This assumption is valid when the magnitudes of
$\mathcal{O}(g_0(\varphi))^2$ and $\mathcal{O}(g_0(\varphi))$ terms are
well separated, which holds for larger $\Lambda$ or $\varphi$ closer to
$\varphi_{J}$ ($\delta\varphi$ closer to zero).
For the case $\Lambda = 0.04$, this separation is reasonable only for
$\delta\varphi < 0.001$, so the applicability below the jamming point
must be perceived restricted than presented, as shown in
Appendix~\ref{app:sec:NumericalSolution}.

Finally, we compare our theory with \cite{GDLW2015}, which derives
constitutive laws $\delta\varphi(J)\sim J^{b_{\varphi}}$
i.e. $\eta_n\sim \delta\varphi^{-1/b_{\varphi}}$ and $\mu(J)-\mu_{0}\sim
J^{b_{\mu}}$ with $b_{\varphi}=b_{\mu}\approx 0.35$, under the
assumption that $\eta_s$ and $\eta_n$ diverges identically.
The equality $b_{\varphi}=b_{\mu}$ suggests $\delta\varphi \sim
\delta\mu$,
which is consistent with our theory, Eq.~(\ref{eq:mu_theory}), but the
exponent of the divergence $1/b_{\varphi}\approx 2.83$ differs from our
prediction of 2.
Their exponent relies on $g'(r)$ for $r > d$, so it might be applicable
to soft-core systems~\citep{KCIB2015}.
However, this is incompatible with hard spheres, because of
Eq.~(\ref{eq:F_ij_p_final}).
Note that their model does not consider the hydrodynamic interactions.

\section{Concluding Remarks}
In this paper, we have derived approximate analytic formulas for the
shear viscosity $\eta_s$, pressure viscosity $\eta_n$, and the stress
ratio $\mu$ for dense non-Brownian suspensions, valid at
$\varphi\lesssim\varphi_J$.
These formulas are derived from the microscopic overdamped equation of
motion for suspended frictionless hard spheres, taking into account the
proximity lubrication effect of the hydrodynamic interactions, and the
approximate distribution function which is an extension of Grad's
13-moment-like expansion.
We have performed MD simulations and confirmed that our theory
successfully derives the relations $\eta_s/\eta_0 \sim \eta_n/\eta_0
\sim \delta\varphi^{-2}$ and $\mu -\mu_{0} \sim \delta\varphi$, where
$\delta\varphi = \varphi_{J}-\varphi$ and $\mu_{0}\approx 0.16$.

%
\begin{acknowledgements}
\hspace{-1.5em}
{\bf Acknowledgements}

The authors are grateful to Satoshi Takada for providing the
prototype of the program for the event-driven MD simulation.
They are also grateful to Kuniyasu Saitoh, Atsushi Ikeda, Takeshi
Kawasaki, Yoshi Oono, Romain Mari, and Ryohei Seto for fruitful
discussions on the subject.
One of the authors (HH) appreciates useful comments by Prabhu R. Nott
and Corey O'Hern.
This work is partially supported by the Grant-in-Aid of MEXT for
Scientific Research (Grants No. 16H04025).
The MD simulations for this work have been carried out at the computer
facilities at the Yukawa Institute of Theoretical Physics, Kyoto
University.
\end{acknowledgements}
%

%
%
%

\appendix

\section{Force correlations}
\label{app:sec:ForceCorrelation}

We derive explicit expressions for the second term on the r.h.s. of
Eq.~(\ref{eq:dsdt}), which includes a derivative of the interparticle
forces.
Note that collisions take place only when the interparticle distance is
equal to the diameter in assemblies of hard spheres.
Hence, the derivative acts only on the angular coordinates, e.g.
\begin{eqnarray}
\frac{\partial F_{j,\alpha}^{(\mathrm{p})}}{\partial r_{i,\lambda}}
&=&
\frac{1}{2}
\zeta\dot\gamma d^2
\sum_{k\neq j} 
\delta(r_{jk}-d)
\frac{\partial}{\partial r_{i,\lambda}}
\left[
\mathcal{P}(\hat{x}_{jk},\hat{y}_{jk})
\hat{r}_{jk,\alpha}
\right]
\nonumber \\
&=&
\frac{1}{2}
\zeta\dot\gamma d^2
\sum_{k\neq j} 
\delta(r_{jk}-d)
\Theta(-\hat{x}_{jk}\hat{y}_{jk})
\frac{\partial}{\partial r_{i,\lambda}}
\left[
-\hat{x}_{jk}\hat{y}_{jk}\hat{r}_{jk,\alpha}
\right].
\label{eq:dFdr}
\end{eqnarray}
Here, the derivative of
$\Theta(-\hat{x}\hat{y})=\Theta(\hat{x})\Theta(-\hat{y}) +
\Theta(-\hat{x})\Theta(\hat{y})$ vanishes since it yields
$\hat{x}\delta(\hat{x})$ or $\hat{y}\delta(\hat{y})$.
This is explicitly confirmed by
\begin{eqnarray}
\hat{x}\hat{y}\hat{r}_{\alpha}
\frac{\partial}{\partial \hat{r}_{\lambda}} 
\Theta(-\hat{x}\hat{y})
=
\hat{x}\hat{y}\hat{r}_{\alpha}
\left[
\delta_{\lambda x}\delta(\hat{x})\Theta(-\hat{y})
-
\delta_{\lambda y}\Theta(\hat{x})\delta(\hat{y})
-
\delta_{\lambda x}\delta(\hat{x})\Theta(\hat{y})
+
\delta_{\lambda y}\Theta(\hat{x})\delta(\hat{y})
\right],
\hspace{2em}
\end{eqnarray}
where each term includes $\hat{x}\delta(\hat{x})$ or
$\hat{y}\delta(\hat{y})$.
From Eq.~(\ref{eq:dFdr}), the third term on the r.h.s. of
Eq.~(\ref{eq:dsdt}) is evaluated as
\begin{eqnarray}
&&
\hspace{-2em}
\sum_{i,j}
\left\langle
\!F_{i,\lambda}^{(\mathrm{p})}
r_{j,\beta}
\frac{\partial F_{j,\alpha}^{(\mathrm{p})}}{\partial r_{i,\lambda}}
\!\right\rangle
\!\!=
\frac{1}{2}
\zeta
\sum_{i,j}
\left\langle
\!F_{i,\lambda}^{(\mathrm{p})}
r_{j,\beta}\,
\dot\gamma d^2
\sum_{k\neq j}
\delta(r_{jk}-d) 
\Theta(-\hat{x}_{jk}\hat{y}_{jk})
\frac{\partial}{\partial r_{i,\lambda}}
\left[
-\hat{x}_{jk}\hat{y}_{jk}\hat{r}_{jk,\alpha}
\right]
\right\rangle
\nonumber \\
&=&
-\frac{1}{2}
\zeta
\sum_{i,j}
\left\langle
\!F_{i,\lambda}^{(\mathrm{p})}
r_{j,\beta}\,
\dot\gamma d
\sum_{k\neq j}
\tilde{\Delta}_{jk}
\Theta_{jk}
(\delta_{ij}-\delta_{ik})
\left( \delta_{\lambda x} \hat{y}_{jk}\hat{r}_{jk,\alpha}
+ \delta_{\lambda y} \hat{x}_{jk}\hat{r}_{jk,\alpha}
+ \delta_{\lambda\alpha} \hat{x}_{jk}\hat{y}_{jk}
\right)
\right\rangle
\nonumber \\
&=&
-\frac{1}{2}
\zeta\dot\gamma d^2
\sum_{i}
\sum_{j\neq i}
\left\langle
F_{i,\lambda}^{(\mathrm{p})}
\hat{r}_{ij,\beta}
\tilde{\Delta}_{ij}
\Theta_{ij}
\left( \delta_{\lambda x} \hat{y}_{ij}\hat{r}_{ij,\alpha}
+ \delta_{\lambda y} \hat{x}_{ij}\hat{r}_{ij,\alpha}
+ \delta_{\lambda\alpha} \hat{x}_{ij}\hat{y}_{ij}
\right)
\right\rangle
\nonumber \\
&=&
-\frac{1}{2}
\zeta\dot\gamma d^2
\sum_{i}
\sum_{j\neq i}
\left\langle
F_{i,\lambda}^{(\mathrm{p})}
\hat{r}_{ij,\beta}
\left( \delta_{\lambda x} \Delta_{ij}^{\alpha y}
+ \delta_{\lambda y} \Delta_{ij}^{\alpha x}
+ \delta_{\lambda\alpha} \Delta_{ij}^{xy}
\right)
\right\rangle,
\end{eqnarray}
where $r_{j,\beta}$ is cast into a relative coordinate $r_{ij,\beta}$
in the third equality, and $\tilde{\Delta}_{ij}:= \delta(r_{ij}-d)$,
$\Theta_{ij}:=\Theta(-\hat{x}_{ij}\hat{y}_{ij})$,
$\Delta_{ij}^{\alpha\beta}:=
\tilde{\Delta}_{ij}\hat{r}_{ij,\alpha}\hat{r}_{ij,\beta}\Theta_{ij}$
(cf. Eq.~(\ref{eq:Delta})).
Note that relative coordinates, e.g. $r_{ij,\beta}$, are expressed in
terms of normalized relative coordinates $\hat{r}_{ij,\beta}$ as $r_{ij}
= d\, \hat{r}_{ij,\beta}$, due to the delta function $\delta(r_{ij}-d)$.
If we further substitute $F_{i,\lambda}^{(\mathrm{p})}=\sum_{j\neq
i}F_{ij,\lambda}^{(\mathrm{p})}$, where $F_{ij,\lambda}^{(\mathrm{p})}$
is given by Eq.~(\ref{eq:F_ij_p_final}), we obtain
\begin{eqnarray}
&&
\hspace{-2em}
\sum_{i,j}
\left\langle
\!F_{i,\lambda}^{(\mathrm{p})}
r_{j,\beta}
\frac{\partial F_{j,\alpha}^{(\mathrm{p})}}{\partial r_{i,\lambda}}
\right\rangle
\nonumber \\
&=&
\frac{1}{4}
\zeta^2 \dot\gamma^2 d^4
\!\sum_{i}
\sum_{j\neq i}
\sum_{k\neq i}
\left\langle
\tilde{\Delta}_{ik} 
\hat{r}_{ik,\lambda}
\hat{x}_{ik}\hat{y}_{ik}
\Theta_{ik}
\hat{r}_{ij,\beta}
\left( \delta_{\lambda x} \Delta_{ij}^{\alpha y}
+ \delta_{\lambda y} \Delta_{ij}^{\alpha x}
+ \delta_{\lambda\alpha} \Delta_{ij}^{xy}
\right)
\right\rangle
\nonumber \\
&=&
\frac{1}{4}
\zeta^2 \dot\gamma^2 d^4
\!\sum_{i}
\sum_{j\neq i}
\sum_{k\neq i}
\left\langle
\Delta_{ik}^{xy} \hat{r}_{ik,\lambda}
\hat{r}_{ij,\beta}
\left( \delta_{\lambda x} \Delta_{ij}^{\alpha y}
+ \delta_{\lambda y} \Delta_{ij}^{\alpha x}
+ \delta_{\lambda\alpha} \Delta_{ij}^{xy}
\right)
\right\rangle
\nonumber \\
&=&
\frac{1}{4}
\zeta^2 \dot\gamma^2 d^4
\!\sum_{i}
\sum_{j\neq i}
\sum_{k\neq i}
\left\langle
\Delta_{ik}^{xy} 
\hat{r}_{ij,\beta}
\left( 
\hat{x}_{ik} \Delta_{ij}^{\alpha y}
+ \hat{y}_{ik} \Delta_{ij}^{\alpha x}
+ \hat{r}_{ik,\alpha} \Delta_{ij}^{xy}
\right)
\right\rangle.
\hspace{2.5em}
\end{eqnarray}
This can further be cast in the form
\begin{eqnarray}
&&
\hspace{-2em}
\sum_{i,j}
\left\langle
\!F_{i,\lambda}^{(\mathrm{p})}
r_{j,\beta}
\frac{\partial F_{j,\alpha}^{(\mathrm{p})}}{\partial r_{i,\lambda}}
\right\rangle
\nonumber \\
&=&
\frac{1}{4}
\zeta^2 \dot\gamma^2 d^4
\!\sum_{i}
\sum_{j\neq i}
\sum_{k\neq i}
\left\langle
\Delta_{ij}^{xy}
\Delta_{ik}^{xy} 
\hat{r}_{ij,\beta}
\hat{r}_{ik,\alpha}
+
\Delta_{ij}^{\alpha\beta}
\Delta_{ik}^{xy} 
\left( 
\hat{x}_{ij} \hat{y}_{ik}
+
\hat{y}_{ij} \hat{x}_{ik}
\right)
\right\rangle
\end{eqnarray}
by the identities $\hat{r}_{ij,\beta}\Delta_{ij}^{\alpha y} =
\hat{y}_{ij}\Delta_{ij}^{\alpha\beta}$ and
$\hat{r}_{ij,\beta}\Delta_{ij}^{\alpha x} =
\hat{x}_{ij}\Delta_{ij}^{\alpha\beta}$.

\section{Approximation of the force correlations}
\label{app:sec:ApproxForceCorrelations}

First we derive approximate expressions for the force correlations which
appear on the r.h.s. of Eq.~(\ref{eq:ss2}), by means of the approximate
formula Eq.~(\ref{eq:ss_formula}).
Then we apply the factorization approximation to the multi-body
correlations to obtain the final approximate expressions.

\subsection{Application of Grad's 13-moment-like expansion}
\label{app:sec:ApplyGrad}

\subsubsection{Canonical terms}

Let us consider the first term on the r.h.s. of Eq.~(\ref{eq:ss2}) for
illustration.
The second term can be evaluated in parallel.
The corresponding canonical term, i.e. the first term on the r.h.s. of
Eq.~(\ref{eq:ss_formula}), includes three indices $i,j,k$ for the
particles.
This can be decomposed into three- and two-body correlations as follows,
\begin{eqnarray}
&& \hspace{-1em}
\sum_{i=1}^{N}
\sum_{j\neq i}
\sum_{k\neq i}
\left\langle
\Delta_{ij}^{xy} \Delta_{ik}^{xy}
\left(
\hat{r}_{ij,\alpha} \hat{r}_{ik,\beta}
+
\hat{r}_{ij,\beta} \hat{r}_{ik,\alpha}
\right)
\right\rangle_{\mathrm{eq}}
\nonumber \\
&=&
\sum_{i,j,k}{}''
\left\langle
\Delta_{ij}^{xy} \Delta_{ik}^{xy}
\left(
\hat{r}_{ij,\alpha} \hat{r}_{ik,\beta}
+
\hat{r}_{ij,\beta} \hat{r}_{ik,\alpha}
\right)
\right\rangle_{\mathrm{eq}}
+
\frac{2}{d}\sum_{i,j}{}'
\left\langle
\tilde{\Delta}_{ij}
\hat{r}_{ij,\alpha}\hat{r}_{ij,\beta}
\hat{x}_{ij}^2 \hat{y}_{ij}^2\Theta_{ij}
\right\rangle_{\mathrm{eq}},
\hspace{1.5em}
\label{eq:can_1}
\end{eqnarray}
where $\tilde{\Delta}_{ij}:= \delta(r_{ij}-d)$, $\Theta_{ij}:=
\Theta(-\hat{x}_{ij}\hat{y}_{ij})$, and $\Delta_{ij}^{\alpha\beta}$ is
given by Eq.~(\ref{eq:Delta}).
Here, the summation $\sum_{i,j,k}{}''$ is performed over ($i,j,k$) where
all the indices are different, i.e. $i\neq j$, $j\neq k$, and $k\neq i$,
and the summation $\sum_{i,j}{}'$ is done for $i\neq j$.
The two-body correlation term (second term) is obtained by setting $j=k$
in the three-body correlation term (first term).
Equation~(\ref{eq:can_1}) can be further expressed as
\begin{eqnarray}
&& \hspace{-1em}
\sum_{i=1}^{N}
\sum_{j\neq i}
\sum_{k\neq i}
\left\langle
\Delta_{ij}^{xy} \Delta_{ik}^{xy}
\left(
\hat{r}_{ij,\alpha} \hat{r}_{ik,\beta}
+
\hat{r}_{ij,\beta} \hat{r}_{ik,\alpha}
\right)
\right\rangle_{\mathrm{eq}}
\nonumber \\
&=&
N n(\varphi)^2 
\int d^3 \boldsymbol{r} \int d^3 \boldsymbol{r}'\,
g^{(3)}(\boldsymbol{r},\boldsymbol{r}')
\delta(r-d) \delta(r'-d)
\left(
\hat{r}_{\alpha} \hat{r}_{\beta}'
+
\hat{r}_{\beta} \hat{r}_{\alpha}'
\right)
\hat{x}\hat{y} \Theta_{\hat{r}}
\hat{x}'\hat{y}' \Theta_{\hat{r}'}
\nonumber \\
&&
+
2N \frac{n(\varphi)}{d} 
\int d^3 \boldsymbol{r}\,
g(r)
\delta (r-d) 
\hat{r}_{\alpha} \hat{r}_{\beta} 
\hat{x}^2 \hat{y}^2 \Theta_{\hat{r}},
\label{eq:2nd_can}
\end{eqnarray}
where $\Theta_{\hat{r}}=\Theta(-\hat{x}\hat{y})$ and
$g^{(3)}(\boldsymbol{r},\boldsymbol{r}')$ and $g(r)$ are the
triplet- and pair-correlation functions, respectively (cf. Eq.~(\ref{eq:g3})).
Similarly, the canonical term for the second term on the r.h.s. of
Eq.~(\ref{eq:ss2}) is given by
\begin{eqnarray}
&& \hspace{-1em}
\sum_{i=1}^{N}
\sum_{j\neq i}
\sum_{k\neq i}
\left\langle
\Delta_{ij}^{\alpha\beta} \Delta_{ik}^{xy}
\left(
\hat{x}_{ij} \hat{y}_{ik}
+
\hat{y}_{ij} \hat{x}_{ik}
\right)
\right\rangle_{\mathrm{eq}}
\nonumber \\
&=&
N n(\varphi)^2 
\int d^3 \boldsymbol{r} \int d^3 \boldsymbol{r}'\,
g^{(3)}(\boldsymbol{r},\boldsymbol{r}')
\delta(r-d) \delta(r'-d)
\left(
\hat{x} \hat{y}'
+
\hat{y} \hat{x}'
\right)
\hat{r}_{\alpha}\hat{r}_{\beta} \Theta_{\hat{r}}
\hat{x}'\hat{y}' \Theta_{\hat{r}'}
\nonumber \\
&&
+
2N \frac{n(\varphi)}{d} 
\int d^3 \boldsymbol{r}\,
g(r)
\delta (r-d) 
\hat{r}_{\alpha} \hat{r}_{\beta} 
\hat{x}^2 \hat{y}^2 \Theta_{\hat{r}}.
\label{eq:3rd_can}
\end{eqnarray}

\subsubsection{Non-canonical terms}

Similarly to the canonical terms, let us consider the first term on the
r.h.s of Eq.~(\ref{eq:ss2}) for illustration.
The second term can be evaluated in parallel.
We consider the corresponding non-canonical term, i.e. the second term
on the r.h.s. of Eq.~(\ref{eq:ss_formula}).
This term includes five indices $i,j,k,l,m$ for the particles as
follows,
\begin{eqnarray}
&& \hspace{-2em}
\frac{V}{2T}
\Pi_{\rho\lambda}
\!\sum_{i=1}^{N}
\sum_{j\neq i}
\sum_{k\neq i}
\left\langle
\Delta_{ij}^{xy} \Delta_{ik}^{xy}
\hat{r}_{ij,\alpha} \hat{r}_{ik,\beta}
\tilde{\sigma}_{\rho\lambda}
\right\rangle_{\mathrm{eq}}
+
(\alpha \leftrightarrow \beta)
\nonumber \\
&=&
-\frac{1}{2T}
\Pi_{\rho\lambda}
\!\sum_{i=1}^{N}
\sum_{j\neq i}
\sum_{k\neq i}
\left\langle
\Delta_{ij}^{xy} \Delta_{ik}^{xy}
\hat{r}_{ij,\alpha}\hat{r}_{ik,\beta}
\sum_l
r_{l,\lambda} F_{l,\rho}^{(\mathrm{p})}
\right\rangle_{\mathrm{eq}}
+
(\alpha \leftrightarrow \beta)
\nonumber \\
&=&
-\frac{\zeta\dot\gamma d^2}{2T}
\Pi_{\rho\lambda}
\!\sum_{i=1}^{N}
\sum_{j\neq i}
\sum_{k\neq i}
\sum_{l,m}
\left\langle
\Delta_{ij}^{xy} \Delta_{ik}^{xy}
\Delta_{lm}^{xy}
\hat{r}_{ij,\alpha}\hat{r}_{ik,\beta}
r_{l,\lambda} \hat{r}_{lm,\rho} 
\right\rangle_{\mathrm{eq}}
+
(\alpha \leftrightarrow \beta),
\end{eqnarray}
where we have substituted $F_{l,\rho}^{(\mathrm{p})}=\sum_{m\neq
l}F_{lm,\rho}^{(\mathrm{p})}$ with $F_{lm,\rho}^{(\mathrm{p})}$ given by
Eq.~(\ref{eq:F_ij_p_final}) in the second equality and
$\Delta_{ij}^{\alpha\beta}:=\delta(r_{ij}-d)\hat{r}_{ij,\alpha}\hat{r}_{ij,\beta}\Theta(-\hat{x}_{ij}\hat{y}_{ij})$
(cf. Eq.~(\ref{eq:Delta})).
If the particles $l,m$ differ from any of the particle $i$, $j$, or $k$,
the correlation decouples and vanishes due to
$\langle\tilde{\sigma}_{\rho\lambda}\rangle_{\mathrm{eq}}=0$.
Hence, particle $l$ or $m$ must be equal to at least one of the
particles $i$, $j$, or $k$.
Let us choose $l$ to be equal to $i$, and redefine $m$ as $l$,
\begin{eqnarray}
&& \hspace{-2em}
\frac{V}{2T}
\Pi_{\rho\lambda}
\!\sum_{i=1}^{N}
\sum_{j\neq i}
\sum_{k\neq i}
\left\langle
\Delta_{ij}^{xy} \Delta_{ik}^{xy}
\hat{r}_{ij,\alpha} \hat{r}_{ik,\beta}
\tilde{\sigma}_{\rho\lambda}
\right\rangle_{\mathrm{eq}}
+
(\alpha \leftrightarrow \beta)
\nonumber \\
&=&
-\frac{\zeta\dot\gamma d^2}{2T}
\Pi_{\rho\lambda}
\!\sum_{i=1}^{N}
\sum_{j\neq i}
\sum_{k\neq i}
\sum_{l}
\left\langle
\Delta_{ij}^{xy} \Delta_{ik}^{xy}
\Delta_{il}^{xy} 
\hat{r}_{ij,\alpha}\hat{r}_{ik,\beta}
r_{i,\lambda} \hat{r}_{il,\rho} 
\right\rangle_{\mathrm{eq}}
+
(\alpha \leftrightarrow \beta).
\label{eq:nc_1}
\end{eqnarray}
This term includes four indices $i,j,k,l$ for the particles, and can be
decomposed into four-, three-, and two-body correlations.

The four-body correlation corresponds to the case where all the indices
$i,j,k,l$ are different from one another.
It can be expressed by the quadruplet-correlation function
$g^{(4)}(\boldsymbol{r},\boldsymbol{r}', \boldsymbol{r}'')$,
Eq.~(\ref{eq:g4}), as
\begin{eqnarray}
&& \hspace{-2em}
(\mbox{four-body correlation})
\nonumber \\
&=&
-\frac{\zeta\dot\gamma d^3}{4T}
\Pi_{\rho\lambda}
\!\sum_{i,j,k,l}{}'''
\left\langle
\Delta_{ij}^{xy} \Delta_{ik}^{xy}
\Delta_{il}^{xy}
\hat{r}_{ij,\alpha}\hat{r}_{ik,\beta}
\hat{r}_{il,\lambda} \hat{r}_{il,\rho} 
\right\rangle_{\mathrm{eq}}
+
(\alpha \leftrightarrow \beta)
\nonumber \\
&=&
-N \frac{\zeta\dot\gamma d^3}{4T}
\Pi_{\rho\lambda}
n(\varphi)^3 \int \!d^3 \boldsymbol{r} \!\int \!d^3 \boldsymbol{r}' \!\int \!d^3 \boldsymbol{r}'' \,
g^{(4)}(\boldsymbol{r},\boldsymbol{r}',\boldsymbol{r}'')
\delta(r-d)\delta(r'-d)\delta(r''-d)
\nonumber \\
&&
\times
\hat{r}_{\alpha} \hat{r}_{\beta}'
\hat{r}_{\lambda}{}'' \hat{r}_{\rho}{}''
\hat{x}\hat{y}  \Theta_{\hat{r}}
\hat{x}'\hat{y}'  \Theta_{\hat{r}'}
\hat{x}'' \hat{y}'' \Theta_{\hat{r}''}
+
(\alpha \leftrightarrow \beta),
\label{eq:nc_4body}
\end{eqnarray}
where the summation $\sum_{i,j,k,l}{}'''$ is performed over $(i,j,k,l)$
with all the indices different from one other.
Note that $r_{i,\lambda}$ is converted to $r_{il,\lambda}$ by virtue of
the odd parity with respect to the exchange $i \leftrightarrow l$.
This conversion allows us to express this term only in relative
coordinates.

The three-body correlation is obtained by setting $l$ equal to $j$ or
$k$ in Eq.~(\ref{eq:nc_4body}). 
Let us choose $l$ to be equal to $j$,
\begin{eqnarray}
&& \hspace{-2em}
(\mbox{three-body correlation})
\nonumber \\
&=&
-\frac{\zeta\dot\gamma d}{2T}
\Pi_{\rho\lambda}
\sum_{i,j,k}{}''
\left\langle
\tilde{\Delta}_{ij} \Delta_{ik}^{xy}
\hat{r}_{ij,\alpha}\hat{r}_{ik,\beta}
r_{i,\lambda} \hat{r}_{ij,\rho} 
\hat{x}_{ij}^2 \hat{y}_{ij}^2 \Theta_{ij}
\right\rangle_{\mathrm{eq}}
+
(\alpha \leftrightarrow \beta).
\end{eqnarray}
Here, $r_{i,\lambda}$ should be converted to $r_{ij,\lambda}$ or
$r_{ik,\lambda}$, in order for this term to be expressed only in
relative coordinates.
By noting that this term is {\it even} with respect to the exchange
$i\leftrightarrow j$ and {\it odd} with respect to $i \leftrightarrow
k$, $r_{ij,\lambda}$ vanishes and $r_{ik,\lambda}$ survives.
Hence, we obtain
\begin{eqnarray}
&& \hspace{-2em}
(\mbox{three-body correlation})
\nonumber \\
&=&
-\frac{\zeta\dot\gamma d^2}{4T}
\Pi_{\rho\lambda}
\sum_{i,j,k}{}''
\left\langle
\tilde{\Delta}_{ij} \Delta_{ik}^{xy}
\hat{r}_{ij,\alpha} \hat{r}_{ij,\rho} 
\hat{r}_{ik,\beta} \hat{r}_{ik,\lambda} 
\hat{x}_{ij}^2 \hat{y}_{ij}^2 \Theta_{ij}
\right\rangle_{\mathrm{eq}}
+
(\alpha \leftrightarrow \beta)
\nonumber \\
&=&
-N \frac{\zeta\dot\gamma d^2}{4T}
\Pi_{\rho\lambda} 
n(\varphi)^2 \int d^3 \boldsymbol{r} \int d^3 \boldsymbol{r}' \,
g^{(3)}(\boldsymbol{r},\boldsymbol{r}')
\delta(r-d)\delta(r'-d)
\hat{r}_{\alpha} \hat{r}_{\rho} 
\hat{r}_{\beta}' \hat{r}_{\lambda}' 
\hat{x}^2 \hat{y}^2 \Theta_{\hat{r}}
\hat{x}' \hat{y}' \Theta_{\hat{r}'}
\nonumber \\
&&
+
(\alpha \leftrightarrow \beta).
\label{eq:nc_3body}
\end{eqnarray}

Finally, the two-body correlation is obtained by setting $k=j$ in the
expression for the three-body correlation.
This is given by
\begin{eqnarray}
&& \hspace{-2em}
(\mbox{two-body correlation})
\nonumber \\
&=&
-\frac{\zeta\dot\gamma d}{4T}
\Pi_{\rho\lambda}
\sum_{i,j}{}'
\left\langle
\tilde{\Delta}_{ij}
\hat{r}_{ij,\alpha} \hat{r}_{ij,\beta} 
\hat{r}_{ij,\rho}  \hat{r}_{ij,\lambda} 
\hat{x}_{ij}^3 \hat{y}_{ij}^3 \Theta_{ij}
\right\rangle_{\mathrm{eq}}
+
(\alpha \leftrightarrow \beta)
\nonumber \\
&=& 
-N\frac{\zeta\dot\gamma d}{4T}
\Pi_{\rho\lambda}
n(\varphi)
\int d^{3}\boldsymbol{r} \,
g(r) \delta(r-d) 
\hat{r}_{\alpha} \hat{r}_{\beta}
\hat{r}_{\rho} \hat{r}_{\lambda}
\hat{x}^3 \hat{y}^3 \Theta_{\hat{r}}
+
(\alpha \leftrightarrow \beta)
\nonumber \\
&=& 
-N\frac{\zeta\dot\gamma d}{2T}
\Pi_{\rho\lambda}
n(\varphi)
\int d^{3}\boldsymbol{r} \,
g(r) \delta(r-d) 
\hat{r}_{\alpha} \hat{r}_{\beta}
\hat{r}_{\rho} \hat{r}_{\lambda}
\hat{x}^3 \hat{y}^3 \Theta_{\hat{r}}.
\label{eq:nc_2body}
\end{eqnarray}
From Eqs.~(\ref{eq:nc_4body}), (\ref{eq:nc_3body}), and
(\ref{eq:nc_2body}), we obtain
\begin{eqnarray}
&& \hspace{-2em}
\frac{V}{2T}
\Pi_{\rho\lambda}
\!\sum_{i=1}^{N}
\sum_{j\neq i}
\sum_{k\neq i}
\left\langle
\Delta_{ij}^{xy} \Delta_{ik}^{xy}
\hat{r}_{ij,\alpha} \hat{r}_{ik,\beta}
\tilde{\sigma}_{\rho\lambda}
\right\rangle_{\mathrm{eq}}
+
(\alpha \leftrightarrow \beta)
\nonumber \\
&&
\hspace{-1em}
\approx
-N \frac{\zeta\dot\gamma d^3}{4T}
\Pi_{\rho\lambda}
n(\varphi)^3 \int \!d^3 \boldsymbol{r} \!\int \!d^3 \boldsymbol{r}' \! \int \!d^3 \boldsymbol{r}'' \,
g^{(4)}(\boldsymbol{r},\boldsymbol{r}',\boldsymbol{r}'')
\delta(r-d)\delta(r'-d)\delta(r''-d)
\nonumber \\
&&
\times
\hat{r}_{\alpha} \hat{r}_{\beta}'
\hat{r}_{\lambda}{}'' \hat{r}_{\rho}{}''
\hat{x}\hat{y}  \Theta_{\hat{r}}
\hat{x}'\hat{y}'  \Theta_{\hat{r}'}
\hat{x}'' \hat{y}'' \Theta_{\hat{r}''}
+
(\alpha \leftrightarrow \beta)
\nonumber \\
&&
-
N \frac{\zeta\dot\gamma d^2}{4T}
\Pi_{\rho\lambda} 
n(\varphi)^2 \!\!\int \!d^3 \boldsymbol{r} \!\int \!d^3 \boldsymbol{r}' \,
g^{(3)}(\boldsymbol{r},\boldsymbol{r}')
\delta(r-d)\delta(r'-d)
\hat{r}_{\alpha} \hat{r}_{\rho} 
\hat{r}_{\beta}' \hat{r}_{\lambda}' 
\hat{x}^2 \hat{y}^2 \Theta_{\hat{r}}
\hat{x}' \hat{y}' \Theta_{\hat{r}'}
+
(\alpha \leftrightarrow \beta)
\nonumber \\
&&
-
N \frac{\zeta\dot\gamma d}{2T}
\Pi_{\rho\lambda} 
n(\varphi)
\int d^{3}\boldsymbol{r} \,
g(r) \delta(r-d) 
\hat{r}_{\alpha} \hat{r}_{\beta}
\hat{r}_{\rho} \hat{r}_{\lambda}
\hat{x}^3 \hat{y}^3 \Theta_{\hat{r}}.
\label{eq:2nd_noncan}
\end{eqnarray}
Similarly, the canonical term for the second term on the r.h.s. of
Eq.~(\ref{eq:ss2}) is given by
\begin{eqnarray}
&& \hspace{-2em}
\frac{V}{2T}
\Pi_{\rho\lambda}
\!\sum_{i=1}^{N}
\sum_{j\neq i}
\sum_{k\neq i}
\left\langle
\Delta_{ij}^{\alpha\beta} \Delta_{ik}^{xy}
\left(
\hat{x}_{ij} \hat{y}_{ik}
+
\hat{y}_{ij} \hat{x}_{ik} 
\right)
\tilde{\sigma}_{\rho\lambda}
\right\rangle_{\mathrm{eq}}
\nonumber \\
&&
\hspace{-1em}
\approx
-N \frac{\zeta\dot\gamma d^3}{4T}
\Pi_{\rho\lambda}
n(\varphi)^3 \int d^3 \boldsymbol{r} \int d^3 \boldsymbol{r}' \int d^3 \boldsymbol{r}'' \,
g^{(4)}(\boldsymbol{r},\boldsymbol{r}',\boldsymbol{r}'')
\delta(r-d)\delta(r'-d)\delta(r''-d)
\nonumber \\
&&
\times
\left(
\hat{x} \hat{y}'
+
\hat{y} \hat{x}' 
\right)
\hat{r}_{\lambda}{}'' \hat{r}_{\rho}{}''
\hat{r}_{\alpha}\hat{r}_{\beta}  \Theta_{\hat{r}}
\hat{x}'\hat{y}'  \Theta_{\hat{r}'}
\hat{x}'' \hat{y}'' \Theta_{\hat{r}''}
+
(\alpha \leftrightarrow \beta)
\nonumber \\
&&
-
N \frac{\zeta\dot\gamma d^2}{4T}
\Pi_{\rho\lambda} 
n(\varphi)^2 
\!\!\int \!d^3 \boldsymbol{r} \!\!\int \!d^3 \boldsymbol{r}' \,
g^{(3)}(\boldsymbol{r},\boldsymbol{r}')
\delta(r-d)\delta(r'-d)
\left(
\hat{x} \hat{y}'
+
\hat{y} \hat{x}'
\right)
\hat{r}_{\rho}  \hat{r}_{\lambda}' 
\hat{x} \hat{y} \hat{r}_{\alpha} \hat{r}_{\beta} \Theta_{\hat{r}}
\hat{x}' \hat{y}' \Theta_{\hat{r}'}
\nonumber \\
&&
-
N \frac{\zeta\dot\gamma d}{2T}
\Pi_{\rho\lambda} 
n(\varphi)
\int d^{3}\boldsymbol{r} \,
g(r) \delta(r-d) 
\hat{r}_{\alpha} \hat{r}_{\beta}
\hat{r}_{\rho} \hat{r}_{\lambda}
\hat{x}^3 \hat{y}^3 \Theta_{\hat{r}}.
\label{eq:3rd_noncan}
\end{eqnarray}

\subsection{Factorization approximation}
\label{app:sec:Factorization}

The first and second terms on the r.h.s. of Eq.~(\ref{eq:ss2}) are
expressed as spatial integrations of the correlation functions
$g^{(4)}(\bm{r},\bm{r}',\bm{r}'')$, $g^{(3)}(\bm{r},\bm{r}')$, and
$g(r)$ by means of the approximate formula Eq.~(\ref{eq:ss_formula}), as
is shown in Eqs.~(\ref{eq:2nd_can}), (\ref{eq:3rd_can}),
(\ref{eq:2nd_noncan}), and (\ref{eq:3rd_noncan}).
To proceed, it is necessary to adopt approximations for
$g^{(4)}(\bm{r},\bm{r}',\bm{r}'')$ and $g^{(3)}(\bm{r},\bm{r}')$, which
are difficult to evaluate.
As explained in Sec.~\ref{sec:ApproximateExpressions}, we adopt the
following approximations which are valid for the evaluation of
divergences in the vicinity of the jamming point,
\begin{eqnarray}
g^{(4)}(\bm{r},\bm{r}',\bm{r}'')
&\approx&
g(r)g(r')g(r'') 
\hspace{1em}(r,r',r'' \approx d),
\label{eq:g4_approx}
\\
g^{(3)}(\bm{r},\bm{r}')
&\approx&
g(r)g(r') 
\hspace{1em}(r,r' \approx d).
\label{eq:g3_approx}
\end{eqnarray}
We apply Eqs.~(\ref{eq:g4_approx}) and (\ref{eq:g3_approx}) to
Eqs.~(\ref{eq:2nd_can}), (\ref{eq:3rd_can}), (\ref{eq:2nd_noncan}), and
(\ref{eq:3rd_noncan}), and utilize
$g(r)\delta(r-d)=g_0(\varphi)\delta(r-d)$ in this section.

\subsubsection{Canonical terms}

By applying Eq.~(\ref{eq:g3_approx}) to Eq.~(\ref{eq:2nd_can}), the
radial integrations can be performed straightforwardly.
Thus we obtain
\begin{eqnarray}
&& \hspace{-1em}
\sum_{i=1}^{N}
\sum_{j\neq i}
\sum_{k\neq i}
\left\langle
\Delta_{ij}^{xy} \Delta_{ik}^{xy}
\left(
\hat{r}_{ij,\alpha} \hat{r}_{ik,\beta}
+
\hat{r}_{ij,\beta} \hat{r}_{ik,\alpha}
\right)
\right\rangle_{\mathrm{eq}}
\nonumber \\
&&
\approx
\frac{N}{d^2} \, 
\varphi^{*2} g_{0}(\varphi)^2
\int d\mathcal{S} \int d\mathcal{S}' \,
\left(
\hat{r}_{\alpha} \hat{r}_{\beta}'
+
\hat{r}_{\beta} \hat{r}_{\alpha}'
\right)
\hat{x}\hat{y} \Theta_{\hat{r}}
\hat{x}'\hat{y}' \Theta_{\hat{r}'}
+
2\frac{N}{d^2} 
\varphi^{*} g_{0}(\varphi) \int d\mathcal{S} \,
\hat{r}_{\alpha}\hat{r}_{\beta}
\hat{x}^2 \hat{y}^2 \Theta_{\hat{r}}
\nonumber \\
&&
=
\frac{N}{d^2} 
\left\{
\varphi^{*2}
g_{0}(\varphi)^2 \mathcal{S}_{\alpha\beta}^{(1:\mathrm{c2})}
+
\varphi^{*} g_{0}(\varphi) 
\mathcal{S}_{\alpha\beta}^{(1:\mathrm{c1})}
\right\},
\label{eq:FpalphaFpbeta_can}
\end{eqnarray}
where $\varphi^{*}=n(\varphi) d^3 = 6\varphi/\pi$ is the dimensionless
average number density, and the angular integrals
$\mathcal{S}_{\alpha\beta}^{(1:\mathrm{c2})}$ and
$\mathcal{S}_{\alpha\beta}^{(1:\mathrm{c1})}$ are given by
\begin{eqnarray}
\mathcal{S}_{\alpha\beta}^{(1:\mathrm{c2})}
&=&
\int d\mathcal{S} \int d\mathcal{S}' \,
\hat{x}\hat{y}
\hat{x}'\hat{y}'
\left(
\hat{r}_{\alpha} \hat{r}_{\beta}'
+
\hat{r}_{\beta} \hat{r}_{\alpha}'
\right)
\Theta_{\hat{r}}
\Theta_{\hat{r}'},
\label{eq:S_1c2}
\\
\mathcal{S}_{\alpha\beta}^{(1:\mathrm{c1})}
&=&
2\int d\mathcal{S} \,
\hat{x}^2 \hat{y}^2
\hat{r}_{\alpha} \hat{r}_{\beta}
\Theta_{\hat{r}}.
\label{eq:S_1c1}
\end{eqnarray}
Here, $\int d\mathcal{S}\cdots$ and $\int d\mathcal{S}'\cdots$ are
angular integrals with respect to $\hat{\boldsymbol{r}}$ and
$\hat{\boldsymbol{r}}'$, respectively.
Equation~(\ref{eq:3rd_can}) can be evaluated similarly.
The result is synthesized as follows,
\begin{eqnarray}
(\mbox{canonical})
&=&
-\frac{\zeta\dot\gamma}{4d}
\sum_{\ell =1}^{2}
\left\{
\varphi^{*3} g_{0}(\varphi)^2 \mathcal{S}_{\alpha\beta}^{(\ell :\mathrm{c2})} 
+
\varphi^{*2} g_{0}(\varphi) \mathcal{S}_{\alpha\beta}^{(\ell :\mathrm{c1})}
\right\},
\end{eqnarray}
where the angular integrals
$\mathcal{S}_{\alpha\beta}^{(2:\mathrm{c2})}$,
$\mathcal{S}_{\alpha\beta}^{(2:\mathrm{c1})}$ are given by
\begin{eqnarray}
\mathcal{S}_{\alpha\beta}^{(2:\mathrm{c2})}
&=&
\int d\mathcal{S} \int d\mathcal{S}' \,
\hat{r}_{\alpha} \hat{r}_{\beta}
\hat{x}'\hat{y}'
\left(
\hat{x}\hat{y}'
+
\hat{y}\hat{x}'
\right)
\Theta_{\hat{r}}
\Theta_{\hat{r}'},
\label{eq:S_2c2}
\\
\mathcal{S}_{\alpha\beta}^{(2:\mathrm{c1})}
&=&
\mathcal{S}_{\alpha\beta}^{(1:\mathrm{c1})}.
\label{eq:S_2c1}
\end{eqnarray}

\subsubsection{Non-canonical terms}

By applying Eqs.~(\ref{eq:g4_approx}) and (\ref{eq:g3_approx}) to
Eq.~(\ref{eq:2nd_noncan}), the radial integrations can be performed
straightforwardly.
Thus we obtain
\begin{eqnarray}
&& \hspace{-2em}
\frac{V}{2T}
\Pi_{\rho\lambda}
\!\sum_{i=1}^{N}
\sum_{j\neq i}
\sum_{k\neq i}
\left\langle
\Delta_{ij}^{xy} \Delta_{ik}^{xy}
\hat{r}_{ij,\alpha} \hat{r}_{ik,\beta}
\tilde{\sigma}_{\rho\lambda}
\right\rangle_{\mathrm{eq}}
+
(\alpha \leftrightarrow \beta)
\nonumber \\
&&
\hspace{-1em}
\approx
-N \frac{\zeta\dot\gamma}{4T}
\Pi_{\rho\lambda}
\varphi^{*3} g_{0}(\varphi)^3
\!\!\int\! d\mathcal{S} \!\int\! d\mathcal{S}' \!\!\int\! d\mathcal{S}''
\hat{r}_{\alpha}
\hat{r}_{\beta}'
\hat{r}_{\lambda}{}'' \hat{r}_{\rho}{}''
\hat{x}\hat{y}  \Theta_{\hat{r}}
\hat{x}'\hat{y}'  \Theta_{\hat{r}'}
\hat{x}'' \hat{y}'' \Theta_{\hat{r}''}
+
(\alpha \leftrightarrow \beta)
\nonumber \\
&&
-
N \frac{\zeta\dot\gamma}{4T}
\Pi_{\rho\lambda} 
\varphi^{*2} g_{0}(\varphi)^2
\!\int\! d\mathcal{S} \!\int\! d\mathcal{S}'
\hat{r}_{\alpha} \hat{r}_{\rho}  
\hat{r}_{\beta}' \hat{r}_{\lambda}' 
\hat{x}^2 \hat{y}^2  \Theta_{\hat{r}}
\hat{x}' \hat{y}' \Theta_{\hat{r}'}
+
(\alpha \leftrightarrow \beta)
\nonumber \\
&&
-
N \frac{\zeta\dot\gamma}{2T}
\Pi_{\rho\lambda} 
\varphi^{*} g_{0}(\varphi)
\int d\mathcal{S} \,
\hat{r}_{\alpha} \hat{r}_{\beta}
\hat{r}_{\lambda} \hat{r}_{\rho}
\hat{x}^3 \hat{y}^3 \Theta_{\hat{r}}
\nonumber \\
&=&
-N\frac{\zeta\dot\gamma}{4T}
\Pi_{\rho\lambda} 
\left[
\varphi^{*3} g_{0}(\varphi)^3 \mathcal{S}_{\alpha\beta\rho\lambda}^{(1:\mathrm{nc3})} 
+\varphi^{*2} g_{0}(\varphi)^2 \mathcal{S}_{\alpha\beta\rho\lambda}^{(1:\mathrm{nc2})} 
+\varphi^{*} g_{0}(\varphi) \mathcal{S}_{\alpha\beta\rho\lambda}^{(1:\mathrm{nc1})} 
\right].
\label{eq:FpalphaFpbeta_noncan}
\end{eqnarray}
Here, the angular integrals are given by
\begin{eqnarray}
\mathcal{S}_{\alpha\beta\rho\lambda}^{(1:\mathrm{nc3})}
&=&
\!\int\! d\mathcal{S}\! \int\! d\mathcal{S}' \!\int\! d\mathcal{S}'' \,
\hat{r}_{\alpha}
\hat{r}_{\beta}'
\hat{r}_{\lambda}{}'' \hat{r}_{\rho}{}''
\hat{x}\hat{y}  \Theta_{\hat{r}}
\hat{x}'\hat{y}'  \Theta_{\hat{r}'}
\hat{x}'' \hat{y}''  \Theta_{\hat{r}''}
+
(\alpha \leftrightarrow \beta),
\hspace{3em}
\label{eq:S_1nc3}
\\
\mathcal{S}_{\alpha\beta\rho\lambda}^{(1:\mathrm{nc2})}
&=&
\int d\mathcal{S} \int d\mathcal{S}' \,
\hat{r}_{\alpha} \hat{r}_{\rho}
\hat{r}_{\beta}' \hat{r}_{\lambda}' 
\hat{x}^2 \hat{y}^2  \Theta_{\hat{r}}
\hat{x}' \hat{y}' \Theta_{\hat{r}'}
+
(\alpha \leftrightarrow \beta),
\label{eq:S_1nc2}
\\
\mathcal{S}_{\alpha\beta\rho\lambda}^{(1:\mathrm{nc1})}
&=&
2\int d\mathcal{S} \,
\hat{r}_{\alpha} \hat{r}_{\beta}
\hat{r}_{\lambda} \hat{r}_{\rho}
\hat{x}^3 \hat{y}^3 \Theta_{\hat{r}}.
\label{eq:S_1nc1}
\end{eqnarray}
Equation~(\ref{eq:3rd_noncan}) can be evaluated similarly.
The result is synthesized as follows,
\begin{equation}
\hspace{-1em}
(\mbox{non-canonical})
=
\frac{\zeta\dot\gamma}{4d}
\frac{\zeta\dot\gamma d^2}{4T}
\Pi_{\rho\lambda} 
\!\sum_{\ell =1}^{2}
\left\{
\left[
\varphi^{*4}
g_{0}(\varphi)^3 \mathcal{S}_{\alpha\beta\rho\lambda}^{(\ell :\mathrm{nc3})} 
\!+\!
\varphi^{*3}
g_{0}(\varphi)^2 \mathcal{S}_{\alpha\beta\rho\lambda}^{(\ell :\mathrm{nc2})} 
\!+\!
\varphi^{*2}
g_{0}(\varphi) \mathcal{S}_{\alpha\beta\rho\lambda}^{(\ell :\mathrm{nc1})} 
\right]
\right\},
\end{equation}
where the angular integrals
$\mathcal{S}_{\alpha\beta\rho\lambda}^{(2:\mathrm{nc3})}$,
$\mathcal{S}_{\alpha\beta\rho\lambda}^{(2:\mathrm{nc2})}$,
$\mathcal{S}_{\alpha\beta\rho\lambda}^{(2:\mathrm{nc1})}$ are given by
\begin{eqnarray}
\mathcal{S}_{\alpha\beta\rho\lambda}^{(2:\mathrm{nc3})}
&=&
\!\int\! d\mathcal{S}\! \int\! d\mathcal{S}' \!\int\! d\mathcal{S}'' \,
\hat{r}_{\alpha} \hat{r}_{\beta}
\hat{x}'\hat{y}'
\left(
\hat{x}\hat{y}'
+
\hat{y}\hat{x}'
\right)
\hat{x}'' \hat{y}'' \hat{r}_{\lambda}{}'' \hat{r}_{\rho}{}''
\Theta_{\hat{r}}
\Theta_{\hat{r}'}
\Theta_{\hat{r}''},
\hspace{3em}
\label{eq:S_2nc3}
\\
\mathcal{S}_{\alpha\beta\rho\lambda}^{(2:\mathrm{nc2})}
&=&
\int d\mathcal{S} \int d\mathcal{S}' \,
\hat{x}\hat{y}
\hat{r}_{\alpha}\hat{r}_{\beta}\hat{r}_{\rho}
\left( \hat{x}\hat{y}' + \hat{y}\hat{x}' \right)
\hat{x}' \hat{y}' \hat{r}_{\lambda}'
\Theta_{\hat{r}}
\Theta_{\hat{r}'},
\label{eq:S_2nc2}
\\
\mathcal{S}_{\alpha\beta\rho\lambda}^{(2:\mathrm{nc1})}
&=&
\mathcal{S}_{\alpha\beta\rho\lambda}^{(1:\mathrm{nc1})}.
\label{eq:S_2nc1}
\end{eqnarray}

\section{Evaluation of the coefficients of the coupled equations}
\label{app:sec:CoeffEq}


We explicitly evaluate the angular integrals
$\mathcal{S}_{\alpha\beta\rho\lambda}^{(\ell :\mathrm{nc2})}$,
$\mathcal{S}_{\alpha\beta}^{(\ell :\mathrm{c1})}$, and
$\mathcal{S}_{\alpha\beta\rho\lambda}^{(\ell :\mathrm{nc1})}$, which appear in
Eqs.~(\ref{eq:ss3_order2}) and (\ref{eq:ss3_order1}).
These integrals enter as coefficients in the coupled equations for the
stress tensor components.
Eventually, Eqs.~(\ref{eq:ss3_order2}) and (\ref{eq:ss3_order1}) are cast in the form
\begin{eqnarray}
\frac{d}{dt}
\left(
\begin{array}{c}
\sigma_{xy}^{(2)} \\
\sigma_{\alpha\alpha}^{(2)} \\
\sigma_{xx}^{(2)} \\
\sigma_{yy}^{(2)} 
\end{array}
\right)
+
\dot\gamma
\!\left(
\begin{array}{c}
\frac{1}{2}\sigma_{yy}^{(2)} \\
\sigma_{xy}^{(2)} \\
\sigma_{xy}^{(2)} \\
0
\end{array}
\right)\!
&=&
\!\Lambda
\frac{\zeta\dot\gamma^2}{4d}
\varphi^{*3}
g_{0}(\varphi)^2 
\boldsymbol{\mathcal{A}}^{(2)}
\left(\!\!
\begin{array}{c}
\Pi_{xy}^{(2)} \\
\Pi_{xx}^{(2)} \\
\Pi_{yy}^{(2)} 
\end{array}
\!\!\right),
\label{eq:ss_coupled_2}
\\
\frac{d}{dt}
\left(
\begin{array}{c}
\sigma_{xy}^{(1)} \\
\sigma_{\alpha\alpha}^{(1)} \\
\sigma_{xx}^{(1)} \\
\sigma_{yy}^{(1)} 
\end{array}
\right)
+
\dot\gamma
\!\left(
\begin{array}{c}
\frac{1}{2}\sigma_{yy}^{(1)} \\
\sigma_{xy}^{(1)} \\
\sigma_{xy}^{(1)} \\
0
\end{array}
\right)\!
&=&
\!\frac{\zeta\dot\gamma^2}{4d}
\varphi^{*2}
g_{0}(\varphi)
\!\left\{ \!\!
- \boldsymbol{\mathcal{B}}
+
\Lambda
\boldsymbol{\mathcal{A}}^{(1)}
\left(\!\!
\begin{array}{c}
\Pi_{xy}^{(1)} \\
\Pi_{xx}^{(1)} \\
\Pi_{yy}^{(1)} 
\end{array}
\!\!\right)
\!\! \right\},
\label{eq:ss_coupled_1}
\end{eqnarray}
respectively.
Here, the elements of the vector $\boldsymbol{\mathcal{B}}:=\left(
\mathcal{B}_{xy}, \mathcal{B}_{\alpha\alpha}, \mathcal{B}_{xx},
\mathcal{B}_{yy}\right)^{T}$ and the matrices
\begin{eqnarray}
\boldsymbol{\mathcal{A}}^{(m)}
:=
\left(
\begin{array}{ccc}
\mathcal{A}_{xy;xy}^{(m)} & \mathcal{A}_{xy;xx}^{(m)} & \mathcal{A}_{xy;yy}^{(m)} \\
\mathcal{A}_{\alpha\alpha;xy}^{(m)} & \mathcal{A}_{\alpha\alpha;xx}^{(m)} & \mathcal{A}_{\alpha\alpha;yy}^{(m)} \\
\mathcal{A}_{xx;xy}^{(m)} & \mathcal{A}_{xx;xx}^{(m)} & \mathcal{A}_{xx;yy}^{(m)} \\
\mathcal{A}_{yy;xy}^{(m)} & \mathcal{A}_{yy;xx}^{(m)} & \mathcal{A}_{yy;yy}^{(m)} \\
\end{array}
\right)
\hspace{1em} (m=1,2)  
\end{eqnarray}
are given by the angular integrals.
In this section, we evaluate the elements of $\boldsymbol{\mathcal{B}}$ and
$\boldsymbol{\mathcal{A}}^{(m)}$ ($m=1,2$).
In the course of the evaluation, there appears integrals of the form
\begin{eqnarray}
\mathcal{A}_{pqr;stu}
&:=&
\int d\mathcal{S} \int d\mathcal{S}' \,
\Theta_{\hat{r}}
\Theta_{\hat{r}'}
\hat{x}^p \hat{y}^q \hat{z}^r
\hat{x}'{}^s \hat{y}'{}^t \hat{z}'{}^u,
\label{eq:A_pqr_stu}
\\
\mathcal{B}_{pqr}
&:=&
\int d\mathcal{S} \,
\Theta_{\hat{r}}
\hat{x}^p \hat{y}^q \hat{z}^r,
\label{eq:B_pqr}
\end{eqnarray}
where $\int d\mathcal{S}\cdots$ and $\int d\mathcal{S}'\cdots$ denote
angular integrals with respect to $\hat{\boldsymbol{r}}$ and $\hat{\boldsymbol{r}}'$,
respectively. 
The values for $\mathcal{A}_{pqr;stu}$ and $\mathcal{B}_{pqr}$ for
various combinations of $(p,q,r)$ and/or $(s,t,u)$ are collected in
Appendix~\ref{app:sec:angular_int}. 

\subsection{Equation for the shear stress}
\label{sec:shearstress}

We consider the case $(\alpha, \beta)=(x,y)$ in
Eqs.~(\ref{eq:ss3_order2}) and (\ref{eq:ss3_order1}).
Let us begin with the evaluation of the non-canonical term with $l=1$, 
\begin{eqnarray}
\Pi_{\rho\lambda}^{(2)}
\mathcal{S}_{xy\rho\lambda}^{(1:\mathrm{nc2})} 
&=&
\Pi_{\rho\lambda}^{(2)}
\int d\mathcal{S} \int d\mathcal{S}' \,
\hat{x}^2 \hat{y}^2 \hat{r}_{\rho} 
( \hat{x}\hat{y}' + \hat{y}\hat{x}')
\hat{x}' \hat{y}' \hat{r}_{\lambda}'
\Theta_{\hat{r}}
\Theta_{\hat{r}'},
\label{eq:S_yxrl_1nc2}
\\
\Pi_{\rho\lambda}^{(1)} 
\mathcal{S}_{xy\rho\lambda}^{(1:\mathrm{nc1})} 
&=&
2\Pi_{\rho\lambda}^{(1)} 
\int d\mathcal{S} \,
\hat{x}^4 \hat{y}^4
\hat{r}_{\rho} \hat{r}_{\lambda}
\Theta_{\hat{r}}.
\label{eq:S_yxrl_1nc1}
\end{eqnarray}
Recalling the implications of the symmetry, Eqs.~(\ref{eq:sym1}) and
(\ref{eq:sym2}), we obtain
\begin{eqnarray}
\Pi_{\rho\lambda}^{(2)} 
\mathcal{S}_{xy\rho\lambda}^{(1:\mathrm{nc2})} 
&=&
\left(\mathcal{A}_{420;220}+\mathcal{A}_{330;310}\right)\Pi_{xx}^{(2)}
+
\left(\mathcal{A}_{330;130}+\mathcal{A}_{240;220}\right)\Pi_{yy}^{(2)}
\nonumber \\
&&
+
\left(\mathcal{A}_{420;130}+2\mathcal{A}_{330;220}+\mathcal{A}_{240;310}
\right)\Pi_{xy}^{(2)},
\hspace{2em}
\label{eq:S_yx_rl_1nc2}
\\
\Pi_{\rho\lambda}^{(1)} 
\mathcal{S}_{xy\rho\lambda}^{(1:\mathrm{nc1})} 
&=&
2\mathcal{B}_{640}\Pi_{xx}^{(1)}
+
2\mathcal{B}_{460}\Pi_{yy}^{(1)}
+
2\mathcal{B}_{442}\Pi_{zz}^{(1)}
+
4\mathcal{B}_{550}\Pi_{xy}^{(1)}
\nonumber \\
&=&
2(\mathcal{B}_{640} - \mathcal{B}_{442})\Pi_{xx}^{(1)}
+
2(\mathcal{B}_{460} - \mathcal{B}_{442})\Pi_{yy}^{(1)}
+
4\mathcal{B}_{550}\Pi_{xy}^{(1)}.
\label{eq:S_yx_rl_1nc1}
\end{eqnarray}
Here, we have utilized the relation $\Pi_{zz}=-(\Pi_{xx}+\Pi_{yy})$.
Similarly, the term for $l=2$ can be evaluated as
\begin{eqnarray}
\Pi_{\rho\lambda}^{(2)} 
\mathcal{S}_{xy\rho\lambda}^{(2:\mathrm{nc2})}\!
&=&
\!\Pi_{\rho\lambda}^{(2)} 
\!\int\! d\mathcal{S} \!\int\! d\mathcal{S}' 
\hat{x}^2 \hat{y}^2 \hat{r}_{\rho}
( \hat{x}\hat{y}' + \hat{y}\hat{x}')
\hat{x}' \hat{y}' \hat{r}_{\lambda}'
\Theta_{\hat{r}}
\Theta_{\hat{r}'}
=
\Pi_{\rho\lambda}^{(2)} 
\mathcal{S}_{xy\rho\lambda}^{(1:\mathrm{nc2})}, 
\hspace{3em}
\label{eq:S_yx_rl_2nc2}
\\
\Pi_{\rho\lambda}^{(1)}
\mathcal{S}_{xy\rho\lambda}^{(2:\mathrm{nc1})}
&=&
\Pi_{\rho\lambda}^{(1)} 
\mathcal{S}_{xy\rho\lambda}^{(1:\mathrm{nc1})}.
\label{eq:S_yx_rl_2nc1}
\end{eqnarray}
The coefficients of the canonical term in Eq.~(\ref{eq:ss3_order1}) are
evaluated as
\begin{eqnarray}
\mathcal{S}_{xy}^{(1:\mathrm{c1})} 
&=&
2\int d\mathcal{S} \,
\hat{x}^3 \hat{y}^3
\Theta_{\hat{r}}
=
2\mathcal{B}_{330}
=
-\frac{16}{105},
\\
\mathcal{S}_{xy}^{(2:\mathrm{c1})} 
&=&
\mathcal{S}_{xy}^{(1:\mathrm{c1})} 
=
-\frac{16}{105}.
\end{eqnarray}
Hence, from Eqs.~(\ref{eq:S_yx_rl_1nc2}) and (\ref{eq:S_yx_rl_2nc2}), we
obtain
\begin{eqnarray}
\frac{d}{dt}
\sigma_{xy}^{(2)}
=
-
\frac{1}{2}
\sigma_{yy}^{(2)}
+
\frac{\zeta\dot\gamma}{4d}
\Lambda\,
\varphi^{*3}
g_{0}(\varphi)^2 
\left[
\mathcal{A}_{xy;xx}^{(2)} \Pi_{xx}^{(2)}
+
\mathcal{A}_{xy;yy}^{(2)} \Pi_{yy}^{(2)}
+
\mathcal{A}_{xy;xy}^{(2)} \Pi_{xy}^{(2)}
\right],
\label{eq:stress_2}
\end{eqnarray}
and from Eqs.~(\ref{eq:S_yx_rl_1nc1}) and (\ref{eq:S_yx_rl_2nc1}), we
obtain
\begin{eqnarray}
\frac{d}{dt}
\sigma_{xy}^{(1)}
=
-
\frac{1}{2}
\sigma_{yy}^{(1)}
+
\frac{\zeta\dot\gamma}{4d}
\varphi^{*2}
g_{0}(\varphi)
\left\{
-\frac{32}{105}
+
\Lambda
\left[
\mathcal{A}_{xy;xx}^{(1)} \Pi_{xx}^{(1)}
+
\mathcal{A}_{xy;yy}^{(1)} \Pi_{yy}^{(1)}
+
\mathcal{A}_{xy;xy}^{(1)} \Pi_{xy}^{(1)}
\right]
\right\},
\hspace{2.5em}
\label{eq:stress_1}
\end{eqnarray}
where the coefficients are evaluated as
\begin{eqnarray}
\mathcal{A}_{xy;xx}^{(2)}
&=&
2(\mathcal{A}_{420;220}+\mathcal{A}_{330;310}) 
=
\frac{12\pi^2 + 128}{1575},
\\
\mathcal{A}_{xy;yy}^{(2)}
&=&
\mathcal{A}_{xy;xx}^{(2)}
=
\frac{12\pi^2 + 128}{1575},
\\
\mathcal{A}_{xy;xy}^{(2)}
&=&
4(\mathcal{A}_{420;130}+\mathcal{A}_{330;220})
=
-\frac{32\pi}{315},
\end{eqnarray}
and
\begin{eqnarray}
\mathcal{A}_{xy;xx}^{(1)}
&=&
4(\mathcal{B}_{640} - \mathcal{B}_{442})
=
\frac{16\pi}{1155},
\\
\mathcal{A}_{xy;yy}^{(1)}
&=&
\mathcal{A}_{xy;xx}^{(1)}
=
\frac{16\pi}{1155},
\\
\mathcal{A}_{xy;xy}^{(1)}
&=&
8\mathcal{B}_{550}
=
-\frac{1024}{10395}.
\end{eqnarray}

\subsection{Equation for the pressure}

We take the trace with respect to the indices $(\alpha, \beta)$ in
Eqs.~(\ref{eq:ss3_order2}) and (\ref{eq:ss3_order1}).
The non-canonical term with $l=1$ is evaluated, by virtue of the
symmetry, Eqs.~(\ref{eq:sym1}) and (\ref{eq:sym2}), and with the aid of
$\Pi_{zz}=-(\Pi_{xx}+\Pi_{yy})$, as follows,
\begin{eqnarray}
\Pi_{\rho\lambda}^{(2)}
\mathcal{S}_{\alpha\alpha\rho\lambda}^{(1:\mathrm{nc2})} 
&=&
2\Pi_{\rho\lambda}^{(2)}
\int d\mathcal{S} \int d\mathcal{S}' \,
\hat{x}^2 \hat{y}^2 \hat{r}_{\alpha} \hat{r}_{\rho}
\hat{x}' \hat{y}' \hat{r}_{\alpha}' \hat{r}_{\lambda}'
\Theta_{\hat{r}}
\Theta_{\hat{r}'}
\nonumber \\
&=&
2\left( \mathcal{A}_{420;310}+\mathcal{A}_{330;220}\right)
\Pi_{xx}^{(2)}
+
2\left( \mathcal{A}_{330;220}+\mathcal{A}_{240;130}\right)
\Pi_{yy}^{(2)}
+
2\mathcal{A}_{222;112}
\Pi_{zz}^{(2)}
\nonumber \\
&&
+
2\left( \mathcal{A}_{420;220} + 2\mathcal{A}_{330;310}
+ \mathcal{A}_{240;220}\right)
\Pi_{xy}^{(2)}
\nonumber \\
&=&
2\left( \mathcal{A}_{420;310}+\mathcal{A}_{330;220}
-\mathcal{A}_{222;112}\right)
\Pi_{xx}^{(2)}
+
2\left( \mathcal{A}_{330;220}+\mathcal{A}_{240;130}
-\mathcal{A}_{222;112}\right)
\Pi_{yy}^{(2)}
\nonumber \\
&&
+
2\left( \mathcal{A}_{420;220} + 2\mathcal{A}_{330;310}
+ \mathcal{A}_{240;220}\right)
\Pi_{xy}^{(2)},
\label{eq:S_alphalpha_rl_1nc2}
\\
\Pi_{\rho\lambda}^{(1)}
\mathcal{S}_{\alpha\alpha\rho\lambda}^{(1:\mathrm{nc1})} 
&=&
2\Pi_{\rho\lambda}^{(1)}
\int d\mathcal{S}\,
\hat{x}^3 \hat{y}^3
\hat{r}_{\rho} \hat{r}_{\lambda}
\Theta_{\hat{r}}
\nonumber \\
&=&
2\mathcal{B}_{530}\Pi_{xx}^{(1)} 
+ 2\mathcal{B}_{350}\Pi_{yy}^{(1)}
+ 2\mathcal{B}_{332}\Pi_{zz}^{(1)}
+ 4\mathcal{B}_{440}\Pi_{xy}^{(1)}
\nonumber \\
&=&
2(\mathcal{B}_{530} - \mathcal{B}_{332})\Pi_{xx}^{(1)} 
+ 2(\mathcal{B}_{350} - \mathcal{B}_{332})\Pi_{yy}^{(1)}
+ 4\mathcal{B}_{440}\Pi_{xy}^{(1)}.
\label{eq:S_alphalpha_rl_1nc1}
\end{eqnarray}
Similarly, the term for $l=2$ are evaluated as
\begin{eqnarray}
\Pi_{\rho\lambda}^{(2)}
\mathcal{S}_{\alpha\alpha\rho\lambda}^{(2:\mathrm{nc2})}
&=&
\Pi_{\rho\lambda}^{(2)}
\int d\mathcal{S} \int d\mathcal{S}' \,
\hat{x}\hat{y}\hat{r}_{\rho}
\left(
\hat{x}\hat{y}' + \hat{y}\hat{x}'
\right)
\hat{x}' \hat{y}' \hat{r}_{\lambda}'
\Theta_{\hat{r}}
\Theta_{\hat{r}'}
\nonumber \\
&=&
\left( \mathcal{A}_{310;220}+\mathcal{A}_{220;310}\right)
\Pi_{xx}^{(2)}
+
\left( \mathcal{A}_{220;130}+\mathcal{A}_{130;220}\right)
\Pi_{yy}^{(2)}
\nonumber \\
&&
+
\left( \mathcal{A}_{310;130}+2\mathcal{A}_{220;220}
+ \mathcal{A}_{130;310}\right)
\Pi_{xy}^{(2)},
\label{eq:S_alphalpha_rl_2nc2}
\\
\Pi_{\rho\lambda}^{(1)}
\mathcal{S}_{\alpha\alpha\rho\lambda}^{(2:\mathrm{nc1})}
&=&
\Pi_{\rho\lambda}^{(1)}
\mathcal{S}_{\alpha\alpha\rho\lambda}^{(1:\mathrm{nc1})}.
\label{eq:S_alphalpha_rl_2nc1}
\end{eqnarray}
The coefficients of the canonical term in Eq.~(\ref{eq:ss3_order1}) are
evaluated as
\begin{eqnarray}
\mathcal{S}_{\alpha\alpha}^{(1:\mathrm{c1})}
&=&
2\int d\mathcal{S}\,
\hat{x}^2 \hat{y}^2 
\Theta_{\hat{r}}
=
2\mathcal{B}_{220}
=
\frac{2\pi}{15},
\\
\mathcal{S}_{\alpha\alpha}^{(2:\mathrm{c1})}
&=&
\mathcal{S}_{\alpha\alpha}^{(1:\mathrm{c1})}
=
\frac{2\pi}{15}.
\end{eqnarray}
Hence, from Eqs.~(\ref{eq:S_alphalpha_rl_1nc2}) and
(\ref{eq:S_alphalpha_rl_2nc2}), we obtain
\begin{eqnarray}
\frac{d}{dt}
\sigma_{\alpha\alpha}^{(2)}
=
-
\sigma_{xy}^{(2)}
+
\frac{\zeta\dot\gamma}{4d}
\Lambda\,
\varphi^{*3} g_{0}(\varphi)^2
\left[
\mathcal{A}_{\alpha\alpha ;xx}^{(2)}
\Pi_{xx}^{(2)}
+
\mathcal{A}_{\alpha\alpha ;yy}^{(2)}
\Pi_{yy}^{(2)}
+
\mathcal{A}_{\alpha\alpha ;xy}^{(2)}
\Pi_{xy}^{(2)}
\right],
\hspace{2em}
\label{eq:pressure_eq_ss_2}
\end{eqnarray}
and from Eqs.~(\ref{eq:S_alphalpha_rl_1nc1}) and
(\ref{eq:S_alphalpha_rl_2nc1}), we obtain
\begin{eqnarray}
\frac{d}{dt}
\sigma_{\alpha\alpha}^{(1)}
=
-
\sigma_{xy}^{(1)}
+
\frac{\zeta\dot\gamma}{4d}
\varphi^{*2} g_{0}(\varphi)
\left\{
\frac{4\pi}{15}
+
\Lambda
\left[
\mathcal{A}_{\alpha\alpha ;xx}^{(2)}
\Pi_{xx}^{(1)}
+
\mathcal{A}_{\alpha\alpha ;yy}^{(2)}
\Pi_{yy}^{(1)}
+
\mathcal{A}_{\alpha\alpha ;xy}^{(2)}
\Pi_{xy}^{(1)}
\right]
\right\},
\hspace{2.5em}
\label{eq:pressure_eq_ss_1}
\end{eqnarray}
where the coefficients are evaluated as
\begin{eqnarray}
\mathcal{A}_{\alpha\alpha ;xx}^{(2)}
&=&
2(\mathcal{A}_{420;310} + \mathcal{A}_{330;220}
- \mathcal{A}_{222;112} + \mathcal{A}_{310;220} )
=
-\frac{184\pi}{1575},
\\
\mathcal{A}_{\alpha\alpha ;yy}^{(2)}
&=&
\mathcal{A}_{\alpha\alpha ;xx}^{(2)}
=
-\frac{184\pi}{1575},
\\
\mathcal{A}_{\alpha\alpha ;xy}^{(2)}
&=&
4(\mathcal{A}_{420;220} + \mathcal{A}_{330;310})
+ 2(\mathcal{A}_{310;310}+\mathcal{A}_{220;220})
=
\frac{52\pi^2 + 604}{1575},
\end{eqnarray}
and
\begin{eqnarray}
\mathcal{A}_{\alpha\alpha ;xx}^{(1)}
&=&
4(\mathcal{B}_{530} - \mathcal{B}_{332}) 
=
-\frac{32}{315}
\\
\mathcal{A}_{\alpha\alpha ;yy}^{(1)}
&=&
\mathcal{A}_{\alpha\alpha ;xx}^{(1)}
=
-\frac{32}{315},
\\
\mathcal{A}_{\alpha\alpha ;xy}^{(1)}
&=&
8\mathcal{B}_{440}
=
\frac{8\pi}{105}.
\end{eqnarray}

\subsection{Equation for the $(x,x)$ component}

We consider the $(\alpha,\beta)=(x,x)$ components in
Eqs.~(\ref{eq:ss3_order2}) and (\ref{eq:ss3_order1}).
The non-canonical terms for $l=1$ are evaluated, by virtue of the
symmetry, Eqs.~(\ref{eq:sym1}) and (\ref{eq:sym2}), and with the aid of
$\Pi_{zz}=-(\Pi_{xx}+\Pi_{yy})$, as
\begin{eqnarray}
\Pi_{\rho\lambda}^{(2)}
\mathcal{S}_{xx\rho\lambda}^{(1:\mathrm{nc2})} 
&=&
2\Pi_{\rho\lambda}^{(2)}
\int d\mathcal{S} \int d\mathcal{S}' \,
\hat{x}^3 \hat{y}^2 \hat{r}_{\rho}
\hat{x}'{}^2 \hat{y}' \hat{r}_{\lambda}' 
\Theta_{\hat{r}}
\Theta_{\hat{r}'}
\nonumber \\
&=&
2\mathcal{A}_{420;310}\Pi_{xx}^{(2)}
+
2\mathcal{A}_{330;220}\Pi_{yy}^{(2)}
+
2\left( \mathcal{A}_{420;220}+\mathcal{A}_{330;310}\right)
\Pi_{xy}^{(2)},
\label{eq:S_xx_rl_1nc2}
\\
\Pi_{\rho\lambda}^{(1)}
\mathcal{S}_{xx\rho\lambda}^{(1:\mathrm{nc1})} 
&=&
2\Pi_{\rho\lambda}^{(1)}
\int d\mathcal{S} \,
\hat{x}^5 \hat{y}^3 \hat{r}_{\rho} \hat{r}_{\lambda}
\Theta_{\hat{r}}
\nonumber \\
&=&
2\mathcal{B}_{730}\Pi_{xx}^{(1)}
+
2\mathcal{B}_{550}\Pi_{yy}^{(1)}
+
2\mathcal{B}_{532}\Pi_{zz}^{(1)}
+
4\mathcal{B}_{640}\Pi_{xy}^{(1)}
\nonumber \\
&=&
2(\mathcal{B}_{730} - \mathcal{B}_{532})\Pi_{xx}^{(1)}
+
2(\mathcal{B}_{550} - \mathcal{B}_{532})\Pi_{yy}^{(1)}
+
4\mathcal{B}_{640}\Pi_{xy}^{(1)}.
\label{eq:S_xx_rl_1nc1}
\end{eqnarray}
Similarly, the terms for $l=2$ are evaluated as
\begin{eqnarray}
\Pi_{\rho\lambda}^{(2)}
\mathcal{S}_{xx\rho\lambda}^{(2:\mathrm{nc2})}
&=&
\Pi_{\rho\lambda}^{(2)}
\int d\mathcal{S} \int d\mathcal{S}' \,
\hat{x}^3 \hat{y} \hat{r}_{\rho}
\left(
\hat{x}\hat{y}' + \hat{y}\hat{x}'
\right)
\hat{x}' \hat{y}' \hat{r}_{\lambda}'
\Theta_{\hat{r}}
\Theta_{\hat{r}'}
\nonumber \\
&=&
\left( \mathcal{A}_{510;220}+\mathcal{A}_{420;310}\right)
\Pi_{xx}^{(2)}
+
\left( \mathcal{A}_{420;130}+\mathcal{A}_{330;220}\right)
\Pi_{yy}^{(2)}
\nonumber \\
&&
+
\left( \mathcal{A}_{510;130}+2\mathcal{A}_{420;220}
+\mathcal{A}_{330;310}
\right)
\Pi_{xy}^{(2)},
\label{eq:S_xx_rl_2nc2}
\\
\Pi_{\rho\lambda}^{(1)}
\mathcal{S}_{xx\rho\lambda}^{(2:\mathrm{nc1})}  
&=&
\Pi_{\rho\lambda}^{(1)}
\mathcal{S}_{xx\rho\lambda}^{(1:\mathrm{nc1})}.
\label{eq:S_xx_rl_2nc1}
\end{eqnarray}
The coefficients of the canonical terms in Eq.~(\ref{eq:ss3_order1}) are
evaluated as
\begin{eqnarray}
\mathcal{S}_{xx}^{(1:\mathrm{c1})} 
&=&
2\int d\mathcal{S}\,
\hat{x}^4 \hat{y}^2
\Theta_{\hat{r}}
=
2\mathcal{B}_{420}
=
\frac{2\pi}{35},
\\
\mathcal{S}_{xx}^{(2:\mathrm{c1})} 
&=&
\mathcal{S}_{xx}^{(1:\mathrm{c1})} 
=
\frac{2\pi}{35}.
\end{eqnarray}
Hence, from Eqs.~(\ref{eq:S_xx_rl_1nc2}) and (\ref{eq:S_xx_rl_2nc2}), we
obtain
\begin{eqnarray}
\frac{d}{dt}
\sigma_{xx}^{(2)}
=
-
\sigma_{xy}^{(2)}
+
\frac{\zeta\dot\gamma}{4d}
\Lambda\,
\varphi^{*3}
g_{0}(\varphi)^2
\left[
\mathcal{A}_{xx;xx}^{(2)} 
\Pi_{xx}^{(2)}
+
\mathcal{A}_{xx;yy}^{(2)} 
\Pi_{yy}^{(2)}
+
\mathcal{A}_{xx;xy}^{(2)} 
\Pi_{xy}^{(2)}
\right],
\label{eq:sxx_eq_ss_2}
\end{eqnarray}
and from Eqs.~(\ref{eq:S_xx_rl_1nc1}) and (\ref{eq:S_xx_rl_2nc1}), we
obtain
\begin{eqnarray}
\frac{d}{dt}
\sigma_{xx}^{(1)}
=
-
\sigma_{xy}^{(1)}
+
\frac{\zeta\dot\gamma}{4d}
\varphi^{*2}
g_{0}(\varphi)
\left\{
\frac{4\pi}{35}
+
\Lambda
\left[
\mathcal{A}_{xx;xx}^{(1)} 
\Pi_{xx}^{(1)}
+
\mathcal{A}_{xx;yy}^{(1)} 
\Pi_{yy}^{(1)}
+
\mathcal{A}_{xx;xy}^{(1)} 
\Pi_{xy}^{(1)}
\right]
\right\},
\hspace{2.5em}
\label{eq:sxx_eq_ss_1}
\end{eqnarray}
where the coefficients are evaluated as
\begin{eqnarray}
\mathcal{A}_{xx;xx}^{(2)}
&=&
3\mathcal{A}_{420;310}
+
\mathcal{A}_{510;220}
=
-\frac{104\pi}{1575},
\\
\mathcal{A}_{xx;yy}^{(2)}
&=&
3\mathcal{A}_{330;220}
+
\mathcal{A}_{420;130}
=
-\frac{8\pi}{175},
\\
\mathcal{A}_{xx;xy}^{(2)}
&=&
4\mathcal{A}_{420;220}
+
3\mathcal{A}_{330;310}
+
\mathcal{A}_{510;130}
=
\frac{24\pi^2 + 320}{1575},
\end{eqnarray}
and
\begin{eqnarray}
\mathcal{A}_{xx;xx}^{(1)}
&=&
4(\mathcal{B}_{730} - \mathcal{B}_{532}) 
=
-\frac{128}{2079},
\\
\mathcal{A}_{xx;yy}^{(1)}
&=&
4(\mathcal{B}_{550} - \mathcal{B}_{532}) 
=
-\frac{128}{3465},
\\
\mathcal{A}_{xx;xy}^{(1)}
&=&
8\mathcal{B}_{640}
=
\frac{8\pi}{231}.
\end{eqnarray}

\subsection{Equation for the $(y,y)$ component}
\label{sec:yy_component}

Finally, we consider the $(\alpha,\beta)=(y,y)$ components in
Eqs.~(\ref{eq:ss3_order2}) and (\ref{eq:ss3_order1}).
The non-canonical terms for $l=1$ are evaluated by virtue of the
symmetry, Eqs.~(\ref{eq:sym1}) and (\ref{eq:sym2}), and with the aid of
$\Pi_{zz}=-(\Pi_{xx}+\Pi_{yy})$, as
\begin{eqnarray}
\Pi_{\rho\lambda}^{(2)}
\mathcal{S}_{yy\rho\lambda}^{(1:\mathrm{nc2})} 
&=&
2\Pi_{\rho\lambda}^{(2)}
\int d\mathcal{S} \int d\mathcal{S}' \,
\hat{x}^2 \hat{y}^3 \hat{r}_{\rho}
\hat{x}' \hat{y}'{}^2 \hat{r}_{\lambda}' 
\Theta_{\hat{r}}
\Theta_{\hat{r}'}
\nonumber \\
&=&
2\mathcal{A}_{330;220}\Pi_{xx}^{(2)}
+
2\mathcal{A}_{240;130}\Pi_{yy}^{(2)}
+
2\left( \mathcal{A}_{330;130}+\mathcal{A}_{240;220}\right)
\Pi_{xy}^{(2)},
\label{eq:S_yy_rl_1nc2}
\\
\Pi_{\rho\lambda}^{(1)}
\mathcal{S}_{yy\rho\lambda}^{(1:\mathrm{nc1})} 
&=&
2\Pi_{\rho\lambda}^{(1)}
\int d\mathcal{S} \,
\hat{x}^3 \hat{y}^5 \hat{r}_{\rho} \hat{r}_{\lambda}
\Theta_{\hat{r}}
\nonumber \\
&=&
2\mathcal{B}_{550}\Pi_{xx}^{(1)}
+
2\mathcal{B}_{370}\Pi_{yy}^{(1)}
+
2\mathcal{B}_{352}\Pi_{zz}^{(1)}
+
4\mathcal{B}_{460}\Pi_{xy}^{(1)}
\nonumber \\
&=&
2(\mathcal{B}_{550} - \mathcal{B}_{352})\Pi_{xx}^{(1)}
+
2(\mathcal{B}_{370} - \mathcal{B}_{352})\Pi_{yy}^{(1)}
+
4\mathcal{B}_{460}\Pi_{xy}^{(1)}.
\label{eq:S_yy_rl_1nc1}
\end{eqnarray}
Similarly, the terms for $l=2$ are evaluated as
\begin{eqnarray}
\Pi_{\rho\lambda}^{(2)}
\mathcal{S}_{yy\rho\lambda}^{(2:\mathrm{nc2})}
&=&
\Pi_{\rho\lambda}^{(2)}
\int d\mathcal{S} \int d\mathcal{S}' \,
\hat{x} \hat{y}^3 \hat{r}_{\rho}
\left(
\hat{x}\hat{y}' + \hat{y}\hat{x}'
\right)
\hat{x}' \hat{y}' \hat{r}_{\lambda}'
\Theta_{\hat{r}}
\Theta_{\hat{r}'}
\nonumber \\
&=&
\left( \mathcal{A}_{330;220}+\mathcal{A}_{240;310}\right)
\Pi_{xx}^{(2)}
+
\left( \mathcal{A}_{240;130}+\mathcal{A}_{150;220}\right)
\Pi_{yy}^{(2)}
\nonumber \\
&&
+
\left( \mathcal{A}_{330;130}+2\mathcal{A}_{240;220}
+\mathcal{A}_{150;310}
\right)
\Pi_{xy}^{(2)},
\label{eq:S_yy_rl_2nc2}
\\
\Pi_{\rho\lambda}^{(1)}
\mathcal{S}_{yy\rho\lambda}^{(2:\mathrm{nc1})}  
&=&
\Pi_{\rho\lambda}^{(1)}
\mathcal{S}_{yy\rho\lambda}^{(1:\mathrm{nc1})}.
\label{eq:S_yy_rl_2nc1}
\end{eqnarray}
The coefficients of the canonical terms in Eq.~(\ref{eq:ss3_order1}) are
evaluated as
\begin{eqnarray}
\mathcal{S}_{yy}^{(1:\mathrm{c1})} 
&=&
2\int d\mathcal{S}\,
\hat{x}^2 \hat{y}^4
\Theta_{\hat{r}}
=
2\mathcal{B}_{240}
=
\frac{2\pi}{35},
\\
\mathcal{S}_{yy}^{(2:\mathrm{c1})} 
&=&
\mathcal{S}_{yy}^{(1:\mathrm{c1})} 
=
\frac{2\pi}{35}.
\end{eqnarray}
Hence, from Eqs.~(\ref{eq:S_yy_rl_1nc2}) and (\ref{eq:S_yy_rl_2nc2}), we
obtain
\begin{eqnarray}
\frac{d}{dt}
\sigma_{yy}^{(2)}
=
\frac{\zeta\dot\gamma}{4d}
\Lambda\,
\varphi^{*3}
g_{0}(\varphi)^2
\left[
\mathcal{A}_{yy;xx}^{(2)} 
\Pi_{xx}^{(2)}
+
\mathcal{A}_{yy;yy}^{(2)} 
\Pi_{yy}^{(2)}
+
\mathcal{A}_{yy;xy}^{(2)} 
\Pi_{xy}^{(2)}
\right],
\label{eq:nsd_eq_ss_2}
\end{eqnarray}
and from Eqs.~(\ref{eq:S_yy_rl_1nc1}) and (\ref{eq:S_yy_rl_2nc1}), we
obtain
\begin{eqnarray}
\frac{d}{dt}
\sigma_{yy}^{(1)}
=
\frac{\zeta\dot\gamma}{4d}
\varphi^{*2}
g_{0}(\varphi)
\left\{
\frac{4\pi}{35}
+
\Lambda
\left[
\mathcal{A}_{yy;xx}^{(1)} 
\Pi_{xx}^{(1)}
+
\mathcal{A}_{yy;yy}^{(1)} 
\Pi_{yy}^{(1)}
+
\mathcal{A}_{yy;xy}^{(1)} 
\Pi_{xy}^{(1)}
\right]
\right\},
\label{eq:nsd_eq_ss_1}
\end{eqnarray}
where the coefficients are evaluated as
\begin{eqnarray}
\mathcal{A}_{yy;xx}^{(2)}
&=&
3\mathcal{A}_{330;220}
+
\mathcal{A}_{240;310}
=
-\frac{8\pi}{175},
\\
\mathcal{A}_{yy;yy}^{(2)}
&=&
3\mathcal{A}_{240;130}
+
\mathcal{A}_{150;220}
=
-\frac{104\pi}{1575},
\\
\mathcal{A}_{yy;xy}^{(2)}
&=&
3\mathcal{A}_{330;130}
+
4\mathcal{A}_{240;220}
+
\mathcal{A}_{150;310}
=
\frac{24\pi^2 + 320}{1575},
\end{eqnarray}
and
\begin{eqnarray}
\mathcal{A}_{yy;xx}^{(1)}
&=&
4(\mathcal{B}_{550} - \mathcal{B}_{352}) 
=
-\frac{128}{3465},
\\
\mathcal{A}_{yy;yy}^{(1)}
&=&
4(\mathcal{B}_{370} - \mathcal{B}_{352}) 
=
-\frac{128}{2079},
\\
\mathcal{A}_{yy;xy}^{(1)}
&=&
8\mathcal{B}_{460}
=
\frac{8\pi}{231}.
\end{eqnarray}

\subsection{Summary}

From Secs.~\ref{sec:shearstress}--\ref{sec:yy_component}, the elements
of the matrices $\boldsymbol{\mathcal{A}}^{(m)}$ ($m=1,2$) and the vector
$\boldsymbol{\mathcal{B}}$ are evaluated as follows,
\begin{eqnarray}
\boldsymbol{\mathcal{A}}^{(2)}
&=&
\left(
\begin{array}{ccc}
-\frac{32\pi}{315} & \frac{12\pi^2 + 128}{1575} & \frac{12\pi^2 + 128}{1575} \\
\frac{52\pi^2 + 604}{1575} & -\frac{184\pi}{1575} & -\frac{184\pi}{1575} \\
\frac{24\pi^2 + 320}{1575} & -\frac{104\pi}{1575} & -\frac{8\pi}{175} \\
\frac{24\pi^2 + 320}{1575} & -\frac{8\pi}{175} & -\frac{104\pi}{1575} \\
\end{array}
\right)
\approx
\left(
\begin{array}{ccc}
-0.319 & 0.156  & 0.156 \\
0.709 & -0.367 & -0.367 \\
0.354 & -0.207 & -0.144 \\
0.354 & -0.144 & -0.207 
\end{array}
\right),
\hspace{2.0em}
\label{eq:A2_final}
\\
\boldsymbol{\mathcal{A}}^{(1)}
&=&
\left(
\begin{array}{ccc}
-\frac{1024}{10395} &  \frac{16\pi}{1155} & \frac{16\pi}{1155} \\
\frac{8\pi}{105} & -\frac{32}{315} & -\frac{32}{315} \\
\frac{8\pi}{231} & -\frac{128}{2079} & -\frac{128}{3465} \\
\frac{8\pi}{231} & -\frac{128}{3465} & -\frac{128}{2079} 
\end{array}
\right)
\approx
\left(
\begin{array}{ccc}
0.0435 & 0.0435 & -0.0985 \\
-0.102 & -0.102 & 0.239 \\
-0.0616 & -0.0369 & 0.109 \\
-0.0369 & -0.0616 & 0.109
\end{array}
\right),
\label{eq:A1_final}
\\
\boldsymbol{\mathcal{B}}
&=&
\left(
\begin{array}{c}
-\frac{32}{105} \\ 
\frac{4\pi}{15} \\ 
\frac{4\pi}{35} \\ 
\frac{4\pi}{35} 
\end{array}
\right)
\approx
\left(
\begin{array}{c}
-0.305 \\ 
0.838 \\ 
0.359 \\ 
0.359
\end{array}
\right).
\label{eq:B_final}
\end{eqnarray}

\section{Solution of the coupled equations}
\label{app:sec:solution}

\subsection{Analytic solutions of the steady equations}
\label{app:sec:analytic_solution}

We show the analytic solutions of the steady equations for
Eqs.~(\ref{eq:ss3_order2}) and (\ref{eq:ss3_order1}), or
Eqs.~(\ref{eq:ss_coupled_2}) and (\ref{eq:ss_coupled_1}), in this
section.
The solution of Eq.~(\ref{eq:ss3_order2}) or (\ref{eq:ss_coupled_2}) is
given by
\begin{eqnarray}
\left(
\begin{array}{c}
P^{(2)} \\
\sigma_{xy}^{(2)} \\
\sigma_{xx}^{(2)} \\
\sigma_{yy}^{(2)} \\
\sigma_{zz}^{(2)} 
\end{array}
\right) 
=
P^{(2)}
\left(
\begin{array}{c}
1 \\
\Pi_{xy}^{(2)} \\
\Pi_{xx}^{(2)}-1 \\
\Pi_{yy}^{(2)}-1 \\
-(\Pi_{xx}^{(2)}+\Pi_{yy}^{(2)}+1)
\end{array}
\right)
\approx
\left(
\begin{array}{c}
0.659 \\
0.107 \\
-1.21 \\
-0.097 \\
-0.673
\end{array}
\right)
\Lambda
\frac{\zeta\dot\gamma}{4d}
\varphi^{*3}
g_0(\varphi)^2,
\label{eq:ss_final_2}
\end{eqnarray}
which coincides with the asymptotic numerical solution of the transient
equation, supplemented with a relaxation term of the pressure for
numerical stability (see Fig.~\ref{Fig:transient} and
Appendix~\ref{app:sec:NumericalSolution}).
The solution of Eq.~(\ref{eq:ss3_order1}) or (\ref{eq:ss_coupled_1}) is
given by
\begin{eqnarray}
\left(
\begin{array}{c}
P^{(1)} \\
\sigma_{xy}^{(1)} \\
\sigma_{xx}^{(1)} \\
\sigma_{yy}^{(1)} \\
\sigma_{zz}^{(1)} 
\end{array}
\right) 
&=&
P^{(1)*}
\left(
\begin{array}{c}
1 \\
\Pi_{xy}^{(1)} \\
\Pi_{xx}^{(1)}-1 \\
\Pi_{yy}^{(1)}-1 \\
-(\Pi_{xx}^{(1)}+\Pi_{yy}^{(1)}+1)
\end{array}
\right)
\frac{\zeta\dot\gamma}{4d}
\varphi^{*2}
g_0(\varphi),
\label{eq:ss_final_1}
\end{eqnarray}
where $P^{(1)*}$, $\Pi_{xx}^{(1)}$, $\Pi_{yy}^{(1)}$, and
$\Pi_{xy}^{(1)}$ are given in terms of $\Lambda$ by
\begin{eqnarray}
P^{(1)*}
&\approx&
\frac{0.0203}{3.38 - 0.363\Lambda},
\label{eq:ss_final_12}
\\
\hspace{-1em}
\left(\!\!
\begin{array}{c}
\Pi_{xx}^{(1)} \\
\Pi_{yy}^{(1)} \\
\Pi_{xy}^{(1)}
\end{array}
\!\!\right)\!
&\approx&
\!\frac{1}{\Lambda (0.000469 P^{(1)*}\! + 3.17\!\times\! 10^{-5})}
\!\left(\!\!
\begin{array}{c}
(0.00521 - 0.00100\Lambda)P^{(1)*}\! + 0.000168 \\
0.000469 \Lambda P^{(1)*} + 8.59\!\times\! 10^{-5} \\
(0.00332 - 7.58\times 10^{-5}\Lambda) P^{(1)*}\! + 0.000210
\end{array}
\!\!\!\right)\!.
\hspace{2.5em}
\label{eq:ss_final_13}
\end{eqnarray}
The derivation of the solution is shown in
Appendices~\ref{app:sec:order2} and \ref{app:sec:order1}.
%
Note that $\Lambda$ introduced in Eq.~(\ref{eq:Lambda}) is merely a
constant, as discussed in Sec.~\ref{sec:muJ}.
In Eqs.~(\ref{eq:ss_final_2}), (\ref{eq:ss_final_12}), and
(\ref{eq:ss_final_13}), the numerical coefficients are determined
analytically in rational forms, but only their approximate values in
decimals are shown for brevity.

\subsubsection{Order $\mathcal{O}(g_0(\varphi)^2)$}
\label{app:sec:order2}

Let us solve the set of equations of
$\mathcal{O}(g_0(\varphi)^2)$, Eq.~(\ref{eq:ss_coupled_2}), with time
derivatives set to zero.
We cast these equations into the form
\begin{eqnarray}
&&
\boldsymbol{A}^{(2)}(P^{(2)*})
\left(
\begin{array}{c}
\Pi_{xy}^{(2)} \\
\Pi_{xx}^{(2)} \\
\Pi_{yy}^{(2)} 
\end{array}
\right)
=
\left(
\begin{array}{c}
-P^{(2)*} \\
0 \\
0 
\end{array}
\right),
\\
\boldsymbol{A}^{(2)}(P^{(2)*})
&=&
\left(\!
\begin{array}{ccc}
2\mathcal{A}_{xy;xy}^{(2)} &
2\mathcal{A}_{xy;xx}^{(2)} &
2\mathcal{A}_{xy;yy}^{(2)} - P^{(2)*} \\
\mathcal{A}_{\alpha\alpha;xy}^{(2)}-\mathcal{A}_{xx;xy}^{(2)} &
\mathcal{A}_{\alpha\alpha;xx}^{(2)}-\mathcal{A}_{xx;xx}^{(2)} & 
\mathcal{A}_{\alpha\alpha;yy}^{(2)}-\mathcal{A}_{xx;yy}^{(2)} \\
\mathcal{A}_{yy;xy}^{(2)} &
\mathcal{A}_{yy;xx}^{(2)} &
\mathcal{A}_{yy;yy}^{(2)} 
\end{array}
\!\right)
\nonumber \\
&=&
\left(\!
\begin{array}{ccc}
-0.638 &
0.313 & 
0.313 - P^{(2)*} \\
0.356 &
-0.160 & 
-0.223 \\
0.354 &
-0.144 &
-0.207 
\end{array}
\!\right),
\hspace{1em}
\label{eq:A2_P2}
\end{eqnarray}
where we have defined
\begin{eqnarray}
P^{(2)*}
&:=&
\frac{P^{(2)}}{\Lambda \Sigma^{(2)}},
\\
\Sigma^{(2)}
&:=&
\frac{\zeta\dot\gamma}{4d}
\varphi^{*3}
g_{0}(\varphi)^2.
\end{eqnarray}
This can be solved as
\begin{eqnarray}
\left(
\begin{array}{c}
\Pi_{xy}^{(2)} \\
\Pi_{xx}^{(2)} \\
\Pi_{yy}^{(2)} 
\end{array}
\right)
&=&
\boldsymbol{A}^{(2)}(P^{(2)*})^{-1}
\left(
\begin{array}{c}
-P^{(2)*} \\
0 \\
0 
\end{array}
\right) 
=
-\frac{P^{(2)*}}{\det\boldsymbol{A}^{(2)}(P^{(2)*})}
\left(
\begin{array}{c}
A_{xy}^{(2)} \\
A_{xx}^{(2)} \\
A_{yy}^{(2)} 
\end{array}
\right)
\nonumber \\
&=&
-\frac{P^{(2)*}}{A^{(2)} P^{(2)*}+B^{(2)}}
\left(
\begin{array}{c}
A_{xy}^{(2)} \\
A_{xx}^{(2)} \\
A_{yy}^{(2)} 
\end{array}
\right),
\label{eq:Pi_2}
\end{eqnarray}
where the numerical factors are given by
\begin{eqnarray}
A_{xy}^{(2)}
&=&
0.00102,
\hspace{1em}
A_{yy}^{(2)}
=
0.00533,
\hspace{1em}
A_{xx}^{(2)}
=
-0.00518,
\label{eq:A2_xx_yy_zz}
\\
A^{(2)}
&=&
-0.00532,
\hspace{1em}
B^{(2)}
=
-0.00061.
\label{eq:A2_B2}
\end{eqnarray}
From Eqs.~(\ref{eq:A2_P2}) and (\ref{eq:Pi_2}), the numerical factors in
Eqs.~(\ref{eq:A2_xx_yy_zz}) and (\ref{eq:A2_B2}) are determined in
rational forms.
However, the expressions are extremely complicated, so we only show
their approximate values in decimals.
All the decimals which appear in the remainder are also approximations
of rational forms.
From Eqs.~(\ref{eq:ss_coupled_2}) and (\ref{eq:Pi_2}), we obtain
\begin{eqnarray}
P^{(2)}
&=&
\Lambda
\Sigma^{(2)}
\left(
\mathcal{A}_{xx;xy}^{(2)}
+
\frac{\Pi_{xx}^{(2)}}{\Pi_{xy}^{(2)}} \mathcal{A}_{xx;xx}^{(2)}
+
\frac{\Pi_{yy}^{(2)}}{\Pi_{xy}^{(2)}} \mathcal{A}_{xx;yy}^{(2)}
\right)
\nonumber \\
&=&
\Lambda
\Sigma^{(2)}
\left(
\mathcal{A}_{xx;xy}^{(2)}
+
\frac{A_{xx}^{(2)}}{A_{xy}^{(2)}} \mathcal{A}_{xx;xx}^{(2)}
+
\frac{A_{yy}^{(2)}}{A_{xy}^{(2)}} \mathcal{A}_{xx;yy}^{(2)}
\right)
\nonumber \\
&\approx&
0.659\Lambda\, 
\Sigma^{(2)},
\label{eq:P_2}
\end{eqnarray}
which in turn determines $\Pi_{xx}^{(2)}$, $\Pi_{yy}^{(2)}$, and
$\Pi_{xy}^{(2)}$ by Eq.~(\ref{eq:Pi_2}).
The stress components are determined as
\begin{eqnarray}
\sigma_{xy}^{(2)} 
&=&
P^{(2)}\Pi_{xy}^{(2)}
\approx
0.107\Lambda\, 
\Sigma^{(2)},
\label{eq:sxy_2}
\\
\sigma_{xx}^{(2)} 
&=&
P^{(2)}(\Pi_{xx}^{(2)}-1)
\approx
-1.21\Lambda\, 
\Sigma^{(2)},
\label{eq:sxx_2}
\\
\sigma_{yy}^{(2)} 
&=&
P^{(2)}(\Pi_{yy}^{(2)}-1)
\approx
-0.097\Lambda\, 
\Sigma^{(2)},
\label{eq:syy_2}
\\
\sigma_{zz}^{(2)} 
&=&
P^{(2)}(\Pi_{zz}^{(2)}-1)
=
-P^{(2)}(\Pi_{xx}^{(2)}+\Pi_{yy}^{(2)}+1)
\approx
-0.673\Lambda\, 
\Sigma^{(2)},
\label{eq:szz_2}
\\
\mu_{0}
&=&
\mu(\delta\varphi\to 0)
=
\frac{\sigma_{xy}^{(2)}}{P^{(2)}}
=
0.163.
\end{eqnarray}
Note that the magnitudes of $\sigma_{\alpha\beta}^{(2)}$ with
$(\alpha,\beta)=(x,y),(x,x),(y,y),(z,z)$ are proportional to
$\Lambda$, and the stress ratio $\mu_0$ is independent of $\Lambda$.

\subsubsection{Order $\mathcal{O}(g_0(\varphi))$}
\label{app:sec:order1}

Next we solve the set of equations of $\mathcal{O}(g_0(\varphi))$,
Eq.~(\ref{eq:ss_coupled_1}), with time derivatives set to zero.
This can be cast into the form
\begin{eqnarray}
&&
\boldsymbol{A}^{(1)}(P^{(1)*})
\left(
\begin{array}{c}
\Pi_{xy}^{(1)} \\
\Pi_{xx}^{(1)} \\
\Pi_{yy}^{(1)} 
\end{array}
\right)
=
\left(
\begin{array}{c}
2\mathcal{B}_{xy}^{*} - P^{(1)*} \\
\mathcal{B}_{\alpha\alpha}^{*} - \mathcal{B}_{xx}^{*} \\
\mathcal{B}_{yy}^{*} 
\end{array}
\right),
\\
\boldsymbol{A}^{(1)}(P^{(1)*})
&=&
\left(\!\!
\begin{array}{ccc}
2\mathcal{A}_{xy;xy}^{(1)} &
2\mathcal{A}_{xy;xx}^{(1)} &
2\mathcal{A}_{xy;yy}^{(1)} - P^{(1)*} \\
\mathcal{A}_{\alpha\alpha;xy}^{(1)}-\mathcal{A}_{xx;xy}^{(1)} &
\mathcal{A}_{\alpha\alpha;xx}^{(1)}-\mathcal{A}_{xx;xx}^{(1)} & 
\mathcal{A}_{\alpha\alpha;yy}^{(1)}-\mathcal{A}_{xx;yy}^{(1)} \\
\mathcal{A}_{yy;xy}^{(1)} &
\mathcal{A}_{yy;xx}^{(1)} &
\mathcal{A}_{yy;yy}^{(1)} \\
\end{array}
\!\!\right)
\nonumber \\
&=&
\left(\!\!
\begin{array}{ccc}
-0.197 &
0.0870 & 
0.0870 - P^{(1)*} \\
0.131 &
-0.0400 & 
-0.0646 \\
0.109 &
-0.0369 & 
-0.0616 
\end{array}
\!\!\right),
\hspace{1.5em}
\end{eqnarray}
where we have defined
\begin{eqnarray}
P^{(1)*}
&:=&
\frac{P^{(1)}}{\Lambda\Sigma^{(1)}},
\\
\Sigma^{(1)}
&:=&
\frac{\zeta\dot\gamma}{4d}
\varphi^{*2}
g_{0}(\varphi),
\\
\mathcal{B}_{\alpha\alpha}^{*}
&:=&
\frac{\mathcal{B}_{\alpha\alpha}}{\Lambda},
\hspace{0.5em}
\mathcal{B}_{xx}^{*}
:=
\frac{\mathcal{B}_{xx}}{\Lambda},
\hspace{0.5em}
\mathcal{B}_{yy}^{*}
:=
\frac{\mathcal{B}_{yy}}{\Lambda},
\hspace{0.5em}
\mathcal{B}_{xy}^{*}
:=
\frac{\mathcal{B}_{xy}}{\Lambda}.
\end{eqnarray}
This can be solved in parallel to Eq.~(\ref{eq:Pi_2}) as
\begin{eqnarray}
\left(
\begin{array}{c}
\Pi_{xy}^{(1)} \\
\Pi_{xx}^{(1)} \\
\Pi_{yy}^{(1)} 
\end{array}
\right)
&=&
\boldsymbol{A}^{(1)}(P^{(1)*})^{-1}
\left(
\begin{array}{c}
2\mathcal{B}_{xy}^{*} - P^{(1)*} \\
\mathcal{B}_{\alpha\alpha}^{*} - \mathcal{B}_{xx}^{*} \\
\mathcal{B}_{yy}^{*} 
\end{array}
\right) 
\nonumber \\
&=&
\frac{1}{\det\boldsymbol{A}^{(1)}(P^{(1)*})}
\left(
\begin{array}{c}
\Lambda^{-1}(A_{xy}^{(1)}P^{(1)*} + B_{xy}^{(1)}) - C_{xy}^{(1)}P^{(1)*} \\
\Lambda^{-1}(A_{xx}^{(1)}P^{(1)*} + B_{xx}^{(1)}) - C_{xx}^{(1)}P^{(1)*} \\
\Lambda^{-1}(A_{yy}^{(1)}P^{(1)*} + B_{yy}^{(1)}) - C_{yy}^{(1)}P^{(1)*}
\end{array}
\right)
\nonumber \\
&=&
\frac{1}{A^{(1)} P^{(1)*}+B^{(1)}}
\left(
\begin{array}{c}
\Lambda^{-1}(A_{xy}^{(1)}P^{(1)*} + B_{xy}^{(1)}) - C_{xy}^{(1)}P^{(1)*}\\
\Lambda^{-1}(A_{xx}^{(1)}P^{(1)*} + B_{xx}^{(1)}) - C_{xx}^{(1)}P^{(1)*}\\
\Lambda^{-1}(A_{yy}^{(1)}P^{(1)*} + B_{yy}^{(1)}) - C_{yy}^{(1)}P^{(1)*}
\end{array}
\right),
\label{eq:Pi_1}
\end{eqnarray}
where the numerical factors are given by
\begin{eqnarray}
A_{xy}^{(1)}
&=&
0.00332,
\hspace{1em}
B_{xy}^{(1)}
=
0.000210,
\hspace{1em}
C_{xy}^{(1)}
=
7.58\times 10^{-5},
\\
A_{xx}^{(1)}
&=&
0.00521,
\hspace{1em}
B_{xx}^{(1)}
=
0.000168,
\hspace{1em}
C_{xx}^{(1)}
=
0.00100,
\\
A_{yy}^{(1)}
&=&
0,
\hspace{1em}
B_{yy}^{(1)}
=
8.59\times 10^{-5},
\hspace{1em}
C_{yy}^{(1)}
=
-0.000469,
\\
A^{(1)}
&=&
0.000469,
\hspace{1em}
B^{(1)}
=
3.17\times 10^{-5}.
\end{eqnarray}
From Eqs.~(\ref{eq:ss_coupled_1}) and (\ref{eq:Pi_1}), we obtain
\begin{eqnarray}
P^{(1)*}
&=&
-\frac{\mathcal{B}_{xx}^{*}}{\Pi_{xy}^{(1)}}
+
\mathcal{A}_{xx;xy}^{(1)}
+
\frac{\Pi_{xx}^{(1)}}{\Pi_{xy}^{(1)}} \mathcal{A}_{xx;xx}^{(1)}
+
\frac{\Pi_{yy}^{(1)}}{\Pi_{xy}^{(1)}} \mathcal{A}_{xx;yy}^{(1)}
\nonumber \\
&=&
-\frac{A^{(1)}P^{(1)*} + B^{(1)}}
{(A_{xy}^{(1)}-\Lambda C_{xy}^{(1)})P^{(1)*} + B_{xy}^{(1)}} 
\mathcal{B}_{xx}
+
\mathcal{A}_{xx;xy}^{(1)}
\nonumber \\
&&
+
\frac{(A_{xx}^{(1)}-\Lambda C_{xx}^{(1)})P^{(1)*}+B_{xx}^{(1)}}
{(A_{xy}^{(1)}-\Lambda C_{xy}^{(1)})P^{(1)*} + B_{xy}^{(1)}} 
\mathcal{A}_{xx;xx}^{(1)}
+
\frac{(A_{yy}^{(1)}-\Lambda C_{yy}^{(1)})P^{(1)*}+B_{yy}^{(1)}}
{(A_{xy}^{(1)}-\Lambda C_{xy}^{(1)})P^{(1)*} + B_{xy}^{(1)}} 
\mathcal{A}_{xx;yy}^{(1)},
\hspace{2.5em}
\end{eqnarray}
from which $P^{(1)*}$ is determined as
\begin{eqnarray}
&&
\hspace{-2em}
P^{(1)*}
=
\frac{X + \sqrt{ X^2 + 4Y}}{2},
\\
&&
\hspace{-2em}
X
:=
\frac{
\mathcal{A}_{xx;xx}^{(1)}(A_{xx}^{(1)}\!-\!\Lambda C_{xx}^{(1)})
\!+\!\mathcal{A}_{xx;yy}^{(1)}(A_{yy}^{(1)}\!-\!\Lambda C_{yy}^{(1)})
\!+\!\mathcal{A}_{xx;xy}^{(1)}(A_{xy}^{(1)}\!-\!\Lambda C_{xy}^{(1)})
\!-\!
A^{(1)}\mathcal{B}_{xx} 
\!-\! B_{xy}^{(1)} 
}
{A_{xy}^{(1)} - \Lambda C_{xy}^{(1)}},
\hspace{2.5em}
\\
&&
\hspace{-2em}
Y
:=
\frac{
\mathcal{A}_{xx;xx}^{(1)}B_{xx}^{(1)}
+\mathcal{A}_{xx;yy}^{(1)}B_{yy}^{(1)}
+\mathcal{A}_{xx;xy}^{(1)}B_{xy}^{(1)} 
- B^{(1)}\mathcal{B}_{xx} 
}
{A_{xy}^{(1)} - \Lambda C_{xy}^{(1)}}.
\end{eqnarray}
Evaluation of the numerators of $X$ and $Y$ yields
\begin{eqnarray}
X
&\approx&
\frac{3.63\times 10^{-5}\Lambda - 3.38\times10^{-4}}{A_{xy}^{(1)} -
\Lambda C_{xy}^{(1)}},
\\
Y&\approx&
\frac{2.03\times10^{-6}}{A_{xy}^{(1)} -
\Lambda C_{xy}^{(1)}},
\end{eqnarray}
which implies $X<0$, $Y>0$, and $|X|\gg |Y|$.
Hence, we obtain
\begin{eqnarray}
P^{(1)*}
&\approx&
\frac{|Y|}{|X|}
=
\frac{0.0203}{3.38 - 0.363\Lambda},
\\
P^{(1)}
&=&
P^{(1)*}
\Sigma^{(1)}
=
\frac{0.0203}{3.38 - 0.363\Lambda}
\Sigma^{(1)},
\label{eq:P1}
\end{eqnarray}
which in turn determines $\Pi_{xx}^{(1)}$, $\Pi_{yy}^{(1)}$, and
$\Pi_{xy}^{(1)}$ by Eq.~(\ref{eq:Pi_1}).
The stress components are determined as in
Eqs.~(\ref{eq:sxy_2})--(\ref{eq:szz_2}).
For instance, for $\Lambda = 0.04$, we obtain
\begin{eqnarray}
P^{(1)}
&\approx&
1.20\times 10^{-4}\,
\Sigma^{(1)},
\\
\sigma_{xy}^{(1)}
&\approx&
0.0399\,
\Sigma^{(1)},
\\
\sigma_{xx}^{(1)}
&\approx&
0.0345\,
\Sigma^{(1)},
\\
\sigma_{yy}^{(1)}
&\approx&
0.0148\,
\Sigma^{(1)},
\\
\sigma_{zz}^{(1)}
&\approx&
-0.0497\,
\Sigma^{(1)}.
\end{eqnarray}

\subsection{Numerical solutions of the transient equations}
\label{app:sec:NumericalSolution}

Let us solve the transient equations of $\mathcal{O}(g_0(\varphi)^2)$,
Eq.~(\ref{eq:ss3_order2}), numerically.
For this purpose, we choose the stress components
$\sigma_{\alpha\beta}^{(2)}$ rather than the deviatoric stress
components $\Pi_{\alpha\beta}^{(2)}$ as independent variables.
Equation~(\ref{eq:ss3_order2}) is expressed solely in terms of
$\sigma_{\alpha\beta}^{(2)}$ as
\begin{eqnarray}
\frac{d}{dt} \sigma_{xy}^{(2)}
&=&
-\frac{1}{2}\dot\gamma \sigma_{yy}^{(2)}
+
\dot\gamma \Lambda \frac{\Sigma^{(2)}}{P^{(2)}}
\left[
\mathcal{A}^{(2)}_{xy;xx} \sigma_{xx}^{(2)}
+
\mathcal{A}^{(2)}_{xy;yy} \sigma_{yy}^{(2)}
+
\mathcal{A}^{(2)}_{xy;xy} \sigma_{xy}^{(2)}
\right]
\nonumber \\
&&
+
\dot\gamma \Lambda \Sigma^{(2)}
(\mathcal{A}^{(2)}_{xy;xx} + \mathcal{A}^{(2)}_{xy;yy}),
\label{eq:dsdt_xy}
\hspace{2em}
\\
\frac{d}{dt}P^{(2)}
&=&
\frac{1}{3}\dot\gamma \sigma_{xy}^{(2)}
-\frac{1}{3}\dot\gamma \Lambda \frac{\Sigma^{(2)}}{P^{(2)}}
\left[
\mathcal{A}^{(2)}_{\alpha\alpha;xx} \sigma_{xx}^{(2)}
+
\mathcal{A}^{(2)}_{\alpha\alpha;yy} \sigma_{yy}^{(2)}
+
\mathcal{A}^{(2)}_{\alpha\alpha;xy} \sigma_{xy}^{(2)}
\right]
\nonumber \\
&&
-\frac{1}{3} \dot\gamma \Lambda \Sigma^{(2)}
(\mathcal{A}^{(2)}_{\alpha\alpha;xx} +
\mathcal{A}^{(2)}_{\alpha\alpha;yy}),
\label{eq:dsdt_P}
\hspace{2em}
\\
\frac{d}{dt}\sigma_{xx}^{(2)}
&=&
-\dot\gamma \sigma_{xy}^{(2)}
+\dot\gamma \Lambda \frac{\Sigma^{(2)}}{P^{(2)}}
\left[
\mathcal{A}^{(2)}_{xx;xx} \sigma_{xx}^{(2)}
+
\mathcal{A}^{(2)}_{xx;yy} \sigma_{yy}^{(2)}
+
\mathcal{A}^{(2)}_{xx;xy} \sigma_{xy}^{(2)}
\right]
\nonumber \\
&&
+
\dot\gamma \Lambda \Sigma^{(2)}
(\mathcal{A}^{(2)}_{xx;xx} + \mathcal{A}^{(2)}_{xx;yy}),
\label{eq:dsdt_xx}
\\
\frac{d}{dt}\sigma_{yy}^{(2)}
&=&
\dot\gamma \Lambda \frac{\Sigma^{(2)}}{P^{(2)}}
\left[
\mathcal{A}^{(2)}_{yy;xx} \sigma_{xx}^{(2)}
+
\mathcal{A}^{(2)}_{yy;yy} \sigma_{yy}^{(2)}
+
\mathcal{A}^{(2)}_{yy;xy} \sigma_{xy}^{(2)}
\right]
\nonumber \\
&&
+
\dot\gamma \Lambda \Sigma^{(2)}
(\mathcal{A}^{(2)}_{yy;xx} + \mathcal{A}^{(2)}_{yy;yy}),
\label{eq:dsdt_yy}
\end{eqnarray}
where $\Sigma^{(2)}$ is defined as
\begin{eqnarray}
\Sigma^{(2)}
:=
\frac{\zeta\dot\gamma}{4d}
\varphi^{*3} g_{0}(\varphi)^2
> 0.
\end{eqnarray}
The first terms on the r.h.s. of
Eqs.~(\ref{eq:dsdt_xy})--(\ref{eq:dsdt_xx}) correspond to the heating
due to shear, and the terms proportional to $\Sigma^{(2)}$ are the
relaxation terms which originate from dissipation.

It should be noted that Eq.~(\ref{eq:dsdt_P}) is singular in the sense
that the self-relaxation term proportional to $P^{(2)}$ is absent from
the right-hand side.
In contrast, the other equations for $\sigma_{xy}^{(2)}$,
$\sigma_{xx}^{(2)}$, and $\sigma_{yy}^{(2)}$ include self-relaxation
terms proportional to themselves.
%
%
\begin{figure}
\begin{center}
\includegraphics[width=7.5cm]{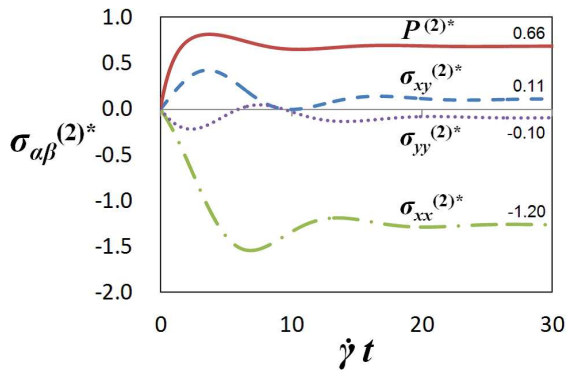} 
\end{center} 
\caption {Transient numerical solution for
 $\sigma_{xy}^{(2)*}:=\sigma_{xy}^{(2)}/(\Lambda\Sigma^{(2)})$,
 $\sigma_{xx}^{(2)*}:=\sigma_{xx}^{(2)}/(\Lambda\Sigma^{(2)})$,
 $\sigma_{yy}^{(2)*}:=\sigma_{yy}^{(2)}/(\Lambda\Sigma^{(2)})$, and
 $P^{(2)*}:=P^{(2)}/(\Lambda\Sigma^{(2)})$ with $\Sigma^{(2)} :=
 \frac{\zeta\dot\gamma}{4d} \varphi^{*3} g_{0}(\varphi)^2$, for the case
 $\xi = 0.001$.} 
\label{Fig:transient}
\end{figure}
%
%
This singularity in Eq.~(\ref{eq:dsdt_P}) causes instability in
the numerical integration.
To avoid this problem, we add a self-relaxation term to the r.h.s. of
Eq.~(\ref{eq:dsdt_P}), which vanishes in the steady state, as
\begin{eqnarray}
\frac{d}{dt}P^{(2)}
&=&
\frac{1}{3}\dot\gamma \sigma_{xy}^{(2)}
+
\frac{1}{3}
\dot\gamma \Lambda \frac{\Sigma^{(2)}}{P^{(2)}}
\left[
0.367 \sigma_{xx}^{(2)}
+
0.367 \sigma_{yy}^{(2)}
-
0.709 \sigma_{xy}^{(2)}
\right]
\nonumber \\
&&
-
\xi (P^{(2)}-P_{\mathrm{ss}}^{(2)})
+
0.245 \dot\gamma \Lambda \Sigma^{(2)},
\label{eq:dsdt_P_reg}
\end{eqnarray}
where $P_{\mathrm{ss}}^{(2)} = 0.659\Lambda \Sigma^{(2)}$ is the steady
solution and $\xi >0$ is a viscous constant.
The numerical solution of the coupled equations Eqs.~(\ref{eq:dsdt_xy}),
(\ref{eq:dsdt_xx}), (\ref{eq:dsdt_yy}), and (\ref{eq:dsdt_P_reg}) is
shown in Fig.~\ref{Fig:transient}, for the case $\xi = 0.001$ and
initial conditions
$\sigma_{xy}^{(2)}(t=0)=\sigma_{xx}^{(2)}(t=0)=\sigma_{yy}^{(2)}(t=0)=0$,
$P^{(2)}(t=0) = 0.01\Lambda\Sigma^{(2)}$.
We confirm that the asymptotic steady values coincide with the
analytical solutions, Eq.~(\ref{eq:ss_final_2}).

\begin{figure}
\begin{center}
\includegraphics[width=7.5cm]{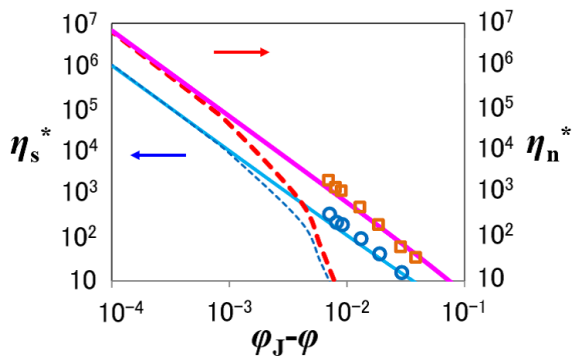} 
\end{center} 
\caption {Shear and pressure viscosities for the case
$\Lambda = 0.04$.
The thick-solid line, thick-dashed line, open squares are the
 results of the pressure viscosity with splitting, without splitting,
 and of MD.
The thin-solid line, thin-dashed line, open circles are the results of
 the shear viscosity with splitting, without splitting, and of MD.  }
\label{Fig:Nosplitting}
\end{figure}

Next, we numerically solve the transient equations without splitting
into $\mathcal{O}(g_0(\varphi)^2)$ and $\mathcal{O}(g_0(\varphi))$,
i.e. Eq.~(\ref{eq:ss4}) or (\ref{eq:ss4_components}).
We choose $\sigma_{\alpha\beta}$ as independent variables as in
Eqs.~(\ref{eq:dsdt_xy})--(\ref{eq:dsdt_yy}), and attach a
self-relaxation term to the pressure equation as in
Eq.~(\ref{eq:dsdt_P_reg}).
We will not explicitly write down the equations and present the result
in Fig.~\ref{Fig:Nosplitting}.
We see that the result with splitting deviates from that without
splitting for $\delta\varphi = \varphi_J-\varphi > 10^{-3}$.
%
%
%
%

\section{Comparison with empirical relations}
\label{app:sec:MorrisBoulay}

\begin{figure}
\centerline{
\includegraphics[width=8.5cm]{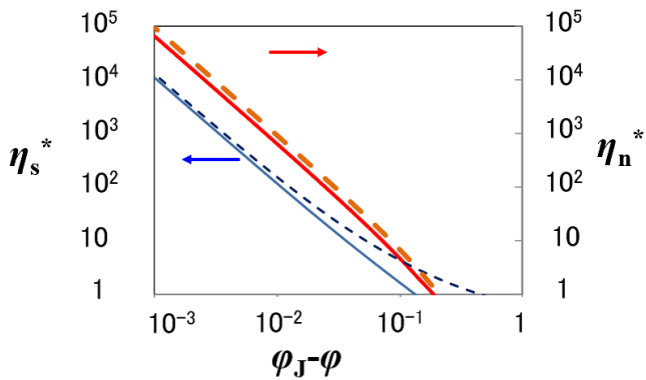}
} 
\caption {Comparison with the empirical relation~\citep{MB1999}.
The results of the normalized shear and pressure viscosities are shown
 in thin-dashed line and thick-dashed line, respectively.
The results are multiplied by 0.3 for comparison to the theoretical
results for $\Lambda = 0.04$ shown in Fig.~\ref{Fig:viscosities}, which
are also displayed.}
\label{Fig:compare_MB}
\end{figure}

We compare our theory with empirical relations which describe
experimental results well.
The empirical equations for the normalized pressure and shear
viscosities are, respectively, given by
\begin{eqnarray}
\eta_n^{*(\mathrm{MB})}
&=&
K_n
\left( \frac{\varphi/\varphi_J}{1 - \varphi/\varphi_J} \right)^{2},
\label{eq:eta_n_MB}
\\
\eta_s^{*(\mathrm{MB})}
&=&
K_s
\left( \frac{\varphi/\varphi_J}{1 - \varphi/\varphi_J} \right)^{2}
+
2.5 \left( \frac{\varphi_J}{1-\varphi/\varphi_J} \right)
+
1,
\label{eq:eta_s_MB}
\end{eqnarray}
where the upper script (MB) is named after the authors, and the
coefficients are given by $K_n=0.75$ and $K_s=0.1$~\citep{MB1999}.
In Fig.~\ref{Fig:compare_MB}, we present the results of
Eqs.~(\ref{eq:eta_n_MB}) and (\ref{eq:eta_s_MB}), together with our
theoretical results.
For comparison, the results of Eqs.~(\ref{eq:eta_n_MB}) and
(\ref{eq:eta_s_MB}) are multiplied by 0.3 for comparison.
In the dense region, where the first term on the r.h.s. of
Eq.~(\ref{eq:eta_s_MB}) is dominant, we obtain
\begin{eqnarray}
\mu^{(\mathrm{MB})}(\delta\varphi\to 0)  
= 
\frac{\eta_s^{*(\mathrm{MB})}(\delta\varphi\to 0)}{\eta_n^{*(\mathrm{MB})}(\delta\varphi\to 0)}
=
\frac{K_s}{K_n} 
= 0.133.
\end{eqnarray}
This is in a relatively good agreement with the result of the present
theory, 0.163.
For lower densities, from Eq.~(\ref{eq:eta_n_MB}) and
(\ref{eq:eta_s_MB}), we obtain
\begin{eqnarray}
\mu  ^{(\mathrm{MB})}
= 
\frac{\eta_s^{*(\mathrm{MB})}}{\eta_n^{*(\mathrm{MB})}}
=
\frac{K_s}{K_n}
+
\frac{2.5}{K_n (\varphi/\varphi_J)^2} \delta\varphi
+
\frac{1}{K_n \varphi^2} \delta\varphi^2
\approx
0.133
+
\frac{3.33}{(\varphi/\varphi_J)^2} \delta\varphi,
\end{eqnarray}
where we have neglected the $\mathcal{O}(\delta\varphi^2)$ term in the last
equality.
Specifically, for $10^{-2}<\delta\varphi <10^{-1}$ and
$\varphi_J=0.639$, we obtain
\begin{eqnarray}
0.133 + 3.44\, \delta\varphi 
< \mu^{(\mathrm{MB})}
< 0.133 + 4.68\, \delta\varphi.
\end{eqnarray}
This is also in a relatively good agreement with our theory, $\mu\approx
0.163+1.78\varphi^{*-1}\delta\varphi$ (cf. Eq.(\ref{eq:mu_theory})).
%
%
%

%
\section{Details of the event-driven MD simulation}
\label{app:sec:EDMD}

The scheme of the event-driven MD simulation is discussed.
It is based on the scheme for Brownian hard spheres~\citep{SVM2007},
which solves the overdamped equation of motion,
\begin{eqnarray}
\zeta \dot{\boldsymbol{r}}_{i}
=
\boldsymbol{F}_{i}^{(\mathrm{p})}
+
\zeta\dot\gamma y_i \boldsymbol{e}_x
+
\boldsymbol{F}_{i}^{(\mathrm{r})}.
\label{eq:eom_Brownian}
\end{eqnarray}
Here, $\zeta = 3\pi d\,\eta_0$ is the scalar resistance and
$\boldsymbol{F}_{i}^{(\mathrm{r})}$ is the random fluctuation force
exerted by the solvent, which is assumed to be Gaussian,
\begin{eqnarray}
\left\langle
\boldsymbol{F}_{i}^{(\mathrm{r})}(t)
\boldsymbol{F}_{i}^{(\mathrm{r})}(t') 
\right\rangle
=
2D_0 \zeta^2 \delta(t-t').
\end{eqnarray}
The diffusion constant is related to $\zeta$ as $D_0 =
T_{\mathrm{eq}}/\zeta$ via the fluctuation-dissipation theorem, where
$T_{\mathrm{eq}}$ is the equilibrium temperature of the solvent.
The units of length and time are chosen as $d$ and $d^2/D_0$.
If we require that the physical mass of the particle $m$ does not appear
explicitly in the overdamped dynamics of Eq.~(\ref{eq:eom_Brownian}),
the only combination with the dimension of mass is $\zeta d^2/D_0$, and
hence we choose this as the unit of mass.
The algorithm consists of three steps:
\begin{enumerate}
\item for each time step $t_n = n\Delta t$ ($n=0,1,\cdots$), random
    velocities $\boldsymbol{v}_i^{(\mathrm{r})} = (\Delta
    \boldsymbol{r}_i - \langle \Delta\boldsymbol{r}\rangle)/\Delta t$
    are sampled for $i=1,\cdots,N$ according to the Maxwellian
    distribution $f(\boldsymbol{v}) = (m_{\mathrm{v}}/(2\pi
    T_{\mathrm{eq}}))^{-3/2}\exp\left( -m_{\mathrm{v}}\boldsymbol{v}^2/(2 T_{\mathrm{eq}})\right)$,
    where $m_{\mathrm{v}}$ is the ``virtual mass'' related to $D_0$ as $D_0 =
    T_{\mathrm{eq}}\Delta t/(2m_{\mathrm{v}})$;
\item add uniform shear velocity $\dot\gamma y_i \boldsymbol{e}_x$ to the random
    velocity; 
\item evolve between $t_n$ and $t_{n+1}$ by the event-driven MD.
\end{enumerate}
Note that $m_{\mathrm{v}}$ is a virtual quantity introduced in the
Maxwellian distribution, so as to ensure the diffusive motion of the
particles.
In fact, from $D_0 = T_{\mathrm{eq}}/\zeta$ and $D_0 =
T_{\mathrm{eq}}\Delta t/(2m_{\mathrm{v}})$, $m_\mathrm{v}$ is determined
as 
\begin{eqnarray}
m_{\mathrm{v}}=\zeta\Delta t/2
=
\zeta d^2 \Delta t^*/(2D_0),
\label{eq:mv_ud}
\end{eqnarray}
which depends on $\Delta t = \Delta t^* d^2 /D_0$ and
hence is not physical.
(We attach $*$ to dimensionless quantities.)

We modify the above scheme to adapt to non-Brownian hard spheres.
We can eliminate the Brownian motion by taking the limit $D_0 \to 0$, or
$T_{\mathrm{eq}} \to 0$, with $m_{\mathrm{v}}=\zeta\Delta t/2$ fixed.
By this choice, for each time step $t_n = n\Delta t$, the velocity of
the particles is set to the uniform shear velocity $\dot\gamma y_i
\boldsymbol{e}_x$, and the dynamics is evolved between $t_n$ and $t_{n+1}$ by
the event-driven MD.
However, taking the limit $D_0 \to 0$ requires us to choose other units
for the time and mass, which are uniquely determined to be
$\dot\gamma^{-1}$ and $\zeta/\dot\gamma$, respectively.
This implies that $\Delta t$ and $m_{\mathrm{v}}$ should be scaled as
\begin{eqnarray}
\Delta t = \dot\gamma^{-1}\Delta t^{*}
\end{eqnarray}
and
\begin{eqnarray}
m_{\mathrm{v}} =
(\zeta/\dot\gamma )(\Delta t^{*}/2).
\label{eq:mv_od}
\end{eqnarray}
%
%
%
Although the physical mass $m$ exists in real suspensions, the
overdamped approximation, Eq.~(\ref{eq:eom_Brownian}), superficially
replaces $m$ with the virtual mass $m_{\mathrm{v}}$, which is given by
Eq.~(\ref{eq:mv_ud}) for the Brownian and (\ref{eq:mv_od}) for the
non-Brownian suspensions, respectively.

\begin{figure}
\centerline{
\includegraphics[width=6.0cm]{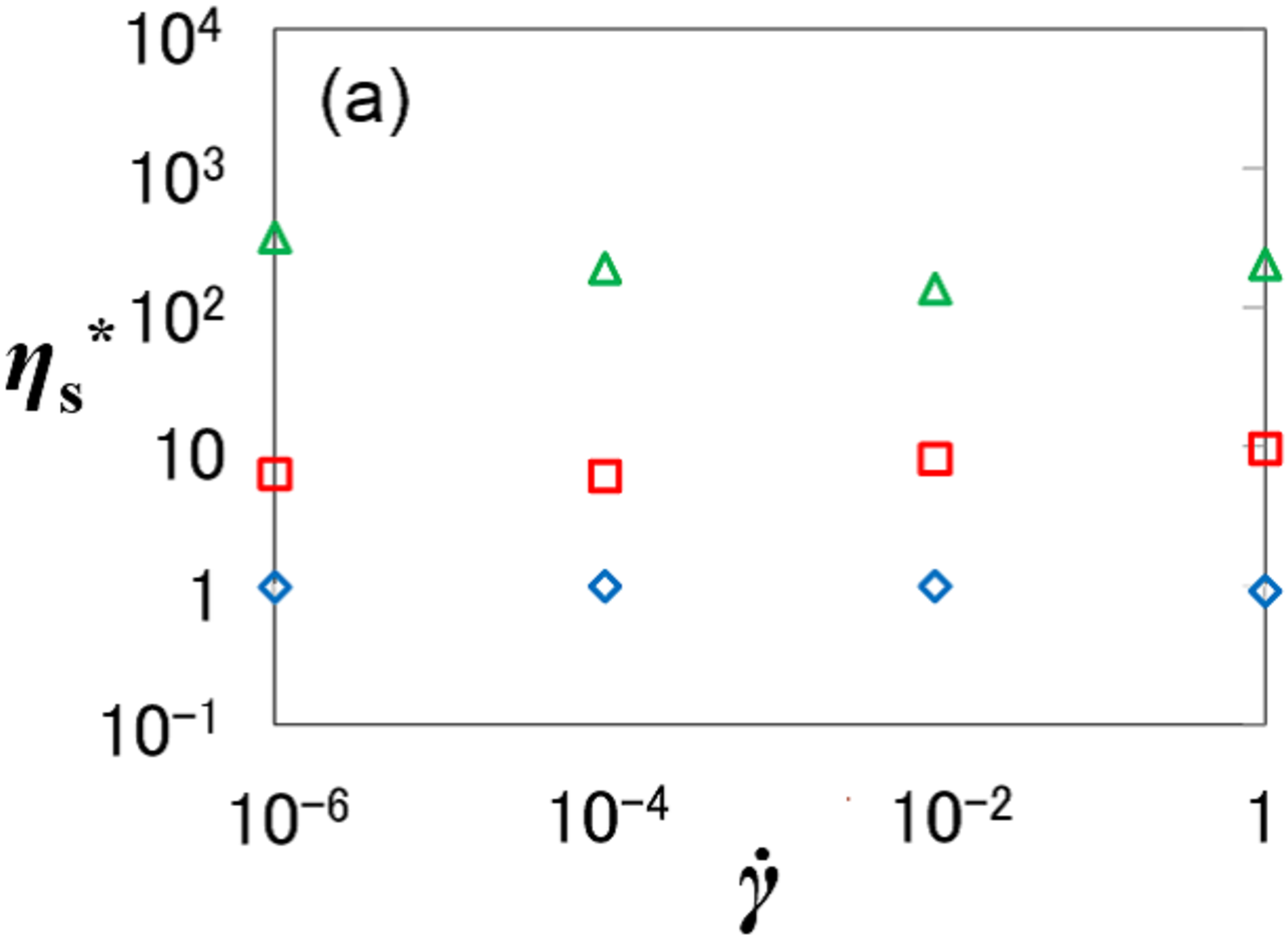}
\hspace{2em}
\includegraphics[width=6.0cm]{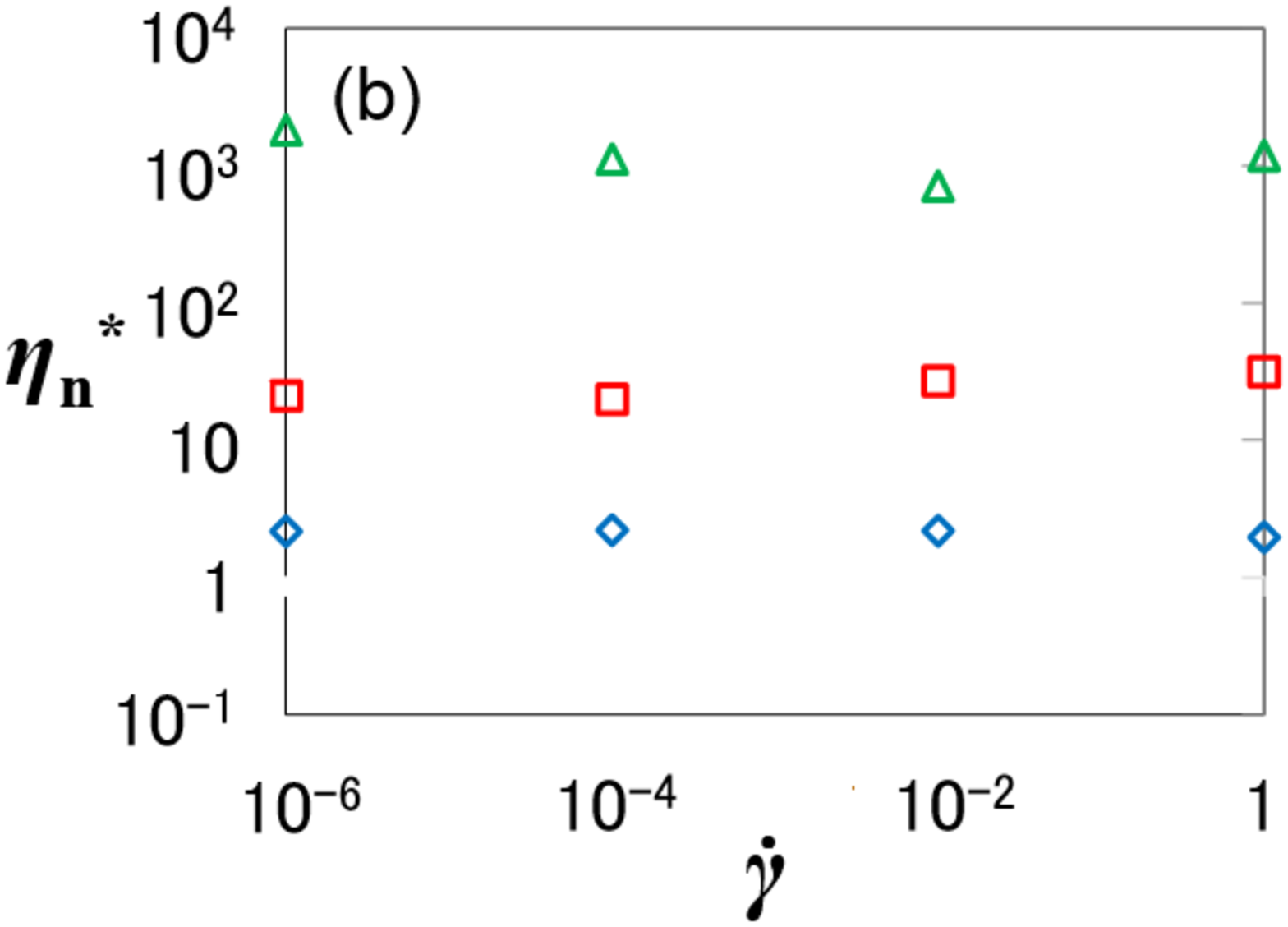}
}
\caption {(a) Shear-rate dependence of the shear viscosity and (b) the
 pressure viscosity. The results for $\varphi = 0.50$, 0.60,
 0.63 are shown in open diamonds, squares, and triangles,
 respectively. }
\label{Fig:Newtonian}
\end{figure}

\begin{figure}
\centerline{
\includegraphics[width=7.0cm]{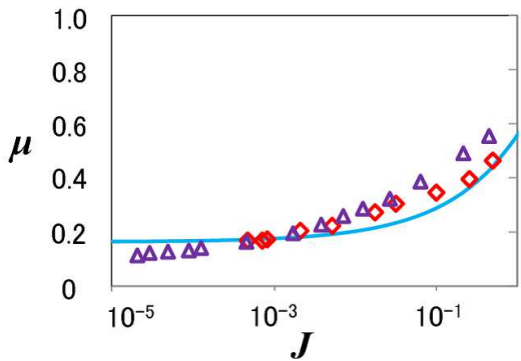}
}
\caption{Comparison of the theoretical result with those of hard-sphere
and soft-sphere MD simulations. The results of the theory, hard-sphere
MD, and soft-sphere MD are shown in solid line, open diamonds, and
crosses, respectively.}  \label{Fig:ratio_soft}
\end{figure}

After equilibration from an initial configuration without overlapping of
the spheres, we start the sampling.
The average of the stress tensor is evaluated by
\begin{eqnarray}
\sigma_{\alpha\beta}
=
\sigma_{\alpha\beta}^{(K)}
+
\sigma_{\alpha\beta}^{(C)},
\end{eqnarray}
where
\begin{eqnarray}
\sigma_{\alpha\beta}^{(K)}
=
-\frac{1}{V}
\left\langle
\sum_{i=1}^{N}
\frac{1}{m_{\mathrm{v}}} p_{i,\alpha} p_{i,\beta}
\right\rangle
=
-\frac{1}{V}
\left\langle
\sum_{i=1}^{N}
m_{\mathrm{v}} v_{i,\alpha} v_{i,\beta}
\right\rangle
\label{eq:stress_K_ave_MD}
\end{eqnarray}
is the average kinetic stress and
\begin{eqnarray}
\sigma_{\alpha\beta}^{(C)}
=
-\frac{1}{V}
\left\langle
\sum_{i=1}^{N}
\frac{1}{2t_m}
\sum_{\mathrm{coll}}
r_{ij,\alpha} p_{ij,\beta}
\right\rangle
=
-\frac{1}{V}
\left\langle
\sum_{i=1}^{N}
\frac{m_{\mathrm{v}}}{2t_m}
\sum_{\mathrm{coll}}
r_{ij,\alpha} v_{ij,\beta}
\right\rangle
\label{eq:stress_C_ave_MD}
\end{eqnarray}
is the average contact stress, with
\begin{eqnarray}
\bm{p}_{i}
=
m_{\mathrm{v}} 
\bm{v}_{i}
=
m_{\mathrm{v}} 
( \dot{\bm{r}}_{i} - \dot\gamma y_{i} \bm{e}_{x})
\end{eqnarray}
the peculiar momentum.
%
%
%
In Eq.~(\ref{eq:stress_C_ave_MD}), $t_m$ is a time interval which is
introduced to evaluate the force from the momentum transfer as
$\boldsymbol{F}_{ij}=\boldsymbol{p}_{ij}/t_m$, and the summation
$\sum_{\mathrm{coll}}$ is performed over all the colliding pairs ($i$,
$j$).
The parameters are set as $N=1000$, $\Delta t^{*}=0.001$, and $t_m^{*} =
10\Delta t^{*}=0.01$.
By the choice $t_m^{*}\propto \Delta t^{*}$,
Eq.~(\ref{eq:stress_C_ave_MD}) is insensitive to $\Delta t^{*}$.
Furthermore, we have verified that $\sigma_{\alpha\beta}^{(C)}$ is
insensitive to the choice of $t_m^*$.
We have confirmed that the results are almost insensitive to the shear
rate, which exemplifies the Newtonian behavior
(cf. Fig~\ref{Fig:Newtonian}).
We have also confirmed that, although the magnitude of
$\sigma_{\alpha\beta}^{(K)}$ is proportional to $\Delta t^*$, 
it does not exhibit any divergence in approaching the jamming point,
which validates Eq.~(\ref{eq:micro_stress}).
We have compared the result of the event-driven MD simulation with that
of soft-sphere MD simulation ($\dot\gamma^{*} = 10^{-7}$) and found
reasonable agreement between them (cf. Fig.~\ref{Fig:ratio_soft}).
%
%
%
%
%
%
%

\section{Implication of the symmetry}
\label{app:sec:Implication_symmetry}

Here we discuss another implication of the symmetry.
From Eq.~(\ref{eq:eom}), the velocity distribution of the particles
deviates from the uniform profile of that of the solvent, $\dot\gamma
y_{i}\boldsymbol{e}_{x}$, if the average force is non-vanishing,
$\langle \boldsymbol{F}_{i}^{(\mathrm{p})}\rangle \neq 0$.
However, this is actually not the case.
From Eqs.~(\ref{eq:F_ij_p_final}) and (\ref{eq:ss_formula}), we have
\begin{eqnarray}
\left\langle
\boldsymbol{F}_{i}^{(\mathrm{p})}
\right\rangle
&=&
\sum_{j\neq i}
\left\langle
\boldsymbol{F}_{ij}^{(\mathrm{p})}
\right\rangle
=
-\frac{1}{2}
\zeta_{e}\dot\gamma d^2
\sum_{j\neq i}
\left\langle
\hat{\boldsymbol{r}}_{ij}\delta(r_{ij}-d)
\hat{x}_{ij}\hat{y}_{ij}\Theta(-\hat{x}_{ij}\hat{y}_{ij})
\right\rangle
\nonumber \\
&=&
-\frac{1}{2}
\zeta_{e}
\dot\gamma d^2
\!\sum_{j\neq i}
\!\left\{
\left\langle
\Delta_{ij}\hat{\boldsymbol{r}}_{ij}
\hat{x}_{ij}\hat{y}_{ij}
\Theta_{ij}
\right\rangle_{\mathrm{eq}}
+
\frac{V}{2T}
\Pi_{\alpha\beta}
\!\left\langle
\Delta_{ij}\hat{\boldsymbol{r}}_{ij}
\hat{x}_{ij}\hat{y}_{ij}
\Theta_{ij}
\tilde{\sigma}_{\alpha\beta}
\right\rangle_{\mathrm{eq}}
\right\},
\hspace{2.5em}
\end{eqnarray}
where $\Delta_{ij}:= \delta(r_{ij}-d)$ and
$\Theta_{ij}:=\Theta(-\hat{x}_{ij}\hat{y}_{ij})$. 
From Eq.~(\ref{eq:sym2}) it is obvious that the first canonical term
vanishes (cubic in $\hat{\boldsymbol{r}}_{ij}$).
By noting that $\tilde{\sigma}_{\alpha\beta}$ is quartic in
$\hat{\boldsymbol{r}}$, the second non-canonical term also vanishes.
Hence, the linear profile of the velocity of the particles is preserved
in the steady state.

\section{Angular integrals}
\label{app:sec:angular_int}

\begin{figure}
\begin{center}
\includegraphics[width=10.0cm]{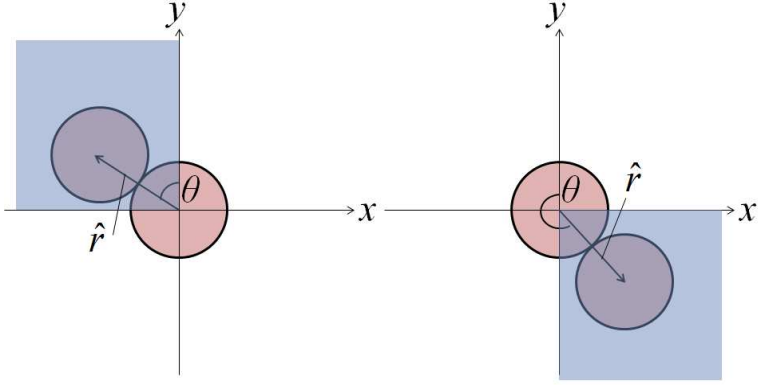} 
\end{center} 
\caption {Spherical coordinate for the angular integral.
}
\label{Fig:coord}
\end{figure}
\begin{figure}
\begin{center}
\includegraphics[width=10.0cm]{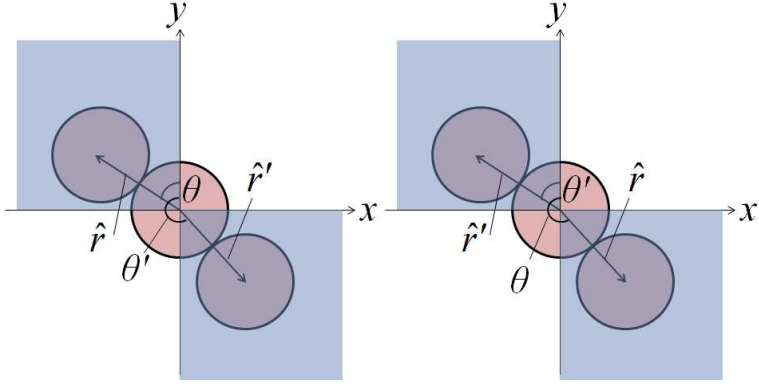} 
\end{center} 
\caption {Possible configurations for the double angular integrals.
}
\label{Fig:coord2}
\end{figure}

We collect the results of the angular integrals $\mathcal{A}_{pqr;stu}$
and $\mathcal{B}_{pqr}$, defined in Eqs.~(\ref{eq:A_pqr_stu}) and
(\ref{eq:B_pqr}). 
Let us introduce the following spherical coordinate
(cf. Fig.~\ref{Fig:coord}),
\begin{eqnarray}
\hat{z}
&=&
\sin\theta \cos\phi,
\\
\hat{x}
&=&
\sin\theta \sin\phi,
\\ 
\hat{y}
&=&
\cos\theta.
\end{eqnarray}
Then, Eq.~(\ref{eq:B_pqr}) is parametrized as
\begin{eqnarray}
\mathcal{B}_{pqr}
&=&
\int d\mathcal{S} \,
\left(
\Theta(\hat{y})\Theta(-\hat{x})
+
\Theta(-\hat{y})\Theta(\hat{x})
\right) 
\hat{x}^p \hat{y}^q \hat{z}^r
\nonumber \\
&=&
=
\left\{
\int_{0}^{1} d(\cos\theta) \int_{-\pi}^{0} d\phi 
+
\int_{-1}^{0} d(\cos\theta) \int_{0}^{\pi} d\phi 
\right\}
\hat{x}^p \hat{y}^q \hat{z}^r. 
\end{eqnarray}
Equation~(\ref{eq:A_pqr_stu}) is a double angular integral with respect
to $\hat{\boldsymbol{r}}$ and $\hat{\boldsymbol{r}}'$, which can be classified into the
two cases depicted in Fig.~\ref{Fig:coord2}.
Accordingly, it is parametrized as follows,
\begin{eqnarray}
\mathcal{A}_{pqr;stu}
&=&
\left\{
\int_0^1 \!\!d(\cos\theta) \int_{-\pi}^0 \!\!d\phi
\int_{-1}^0 \!\!d(\cos\theta') \int_0^{\pi} \!\!d\phi'
+
\int_{-1}^0 \!\!d(\cos\theta) \int_0^{\pi} \!\!d\phi
\int_0^1 \!\!d(\cos\theta') \int_{-\pi}^0 \!\!d\phi'
\right\}
\nonumber \\
&\times&
\hat{x}^p \hat{y}^q \hat{z}^r
\hat{x}'{}^s \hat{y}'{}^t \hat{z}'{}^u.
\label{eq:A_pqr_stu_param}
\end{eqnarray}
Note that only the integrands which are even with respect to the parity
transformation ``$\hat{x}\to -\hat{x}$ and $\hat{y}\to -\hat{y}$''
survive (cf. Eq.~(\ref{eq:ss3})).
For parity-even terms, Eq.~(\ref{eq:A_pqr_stu_param}) reduces to
\begin{eqnarray}
\mathcal{A}_{pqr;stu}
=
2\int_0^1 d(\cos\theta) \int_{-\pi}^0 d\phi
\int_{-1}^0 d(\cos\theta') \int_0^{\pi} d\phi'
\hat{x}^p \hat{y}^q \hat{z}^r
\hat{x}'{}^s \hat{y}'{}^t \hat{z}'{}^u.
\label{eq:A_pqr_stu_param2}
\end{eqnarray}
%
%
%
%
%
%
%
%
Hence, Eq.~(\ref{eq:A_pqr_stu}) is expressed as
\begin{eqnarray}
\hspace{-2em}
\mathcal{A}_{pqr;stu}
&=&
2\mathcal{B}_{pqr} \mathcal{C}_{stu},
\\
\mathcal{B}_{pqr}
&=&
\int_0^1 d(\cos\theta) \int_{-\pi}^0 d\phi \,
\hat{x}^p \hat{y}^q \hat{z}^r
=
\int_0^1 d(\cos\theta) \int_{-\pi}^0 d\phi \,
\cos^q \theta \sin^{p+r} \theta \sin^p \phi \cos^r \phi,
\\
\mathcal{C}_{stu}
&=&
\int_{-1}^0 \!\!d(\cos\theta') \int_0^{\pi} \!\!d\phi'
\hat{x}'{}^s \hat{y}'{}^t \hat{z}'{}^u 
=
\int_{-1}^0 \!\!d(\cos\theta') \int_0^{\pi} \!\!d\phi'
\cos^t \theta \sin^{s+u} \theta \sin^s \phi \cos^u \phi.
\hspace{2em}
\end{eqnarray}
Furthermore, $\mathcal{B}_{pqr}=\mathcal{C}_{pqr}$ holds, so we finally
obtain
\begin{eqnarray}
\mathcal{A}_{pqr;stu}
=
2\mathcal{B}_{pqr} \mathcal{B}_{stu}.
\label{eq:A_2BB}
\end{eqnarray}
It should be noted that $\mathcal{B}_{pqr}=\mathcal{B}_{qpr}$ holds from
symmetry.
We collect the results of $\mathcal{B}_{pqr}$ necessary for our purpose
below:
\begin{eqnarray}
\mathcal{B}_{730}
&=&
\mathcal{B}_{370}
=
\int_{0}^{1} d(\cos\theta) \int_{-\pi}^{0} d\phi\,
\cos^3\theta \sin^7\theta \sin^7\phi
=
-\frac{64}{3465},
\\
\mathcal{B}_{710}
&=&
\mathcal{B}_{170}
=
\int_{0}^{1} d(\cos\theta) \int_{-\pi}^{0} d\phi\,
\cos\theta \sin^7\theta \sin^7\phi
=
-\frac{32}{315},
\end{eqnarray}
\begin{eqnarray}
\mathcal{B}_{640}
&=&
\mathcal{B}_{460}
=
\int_{0}^{1} d(\cos\theta) \int_{-\pi}^{0} d\phi\,
\cos^4\theta \sin^6\theta \sin^6\phi
=
\frac{\pi}{231},
\\
\mathcal{B}_{620}
&=&
\mathcal{B}_{260}
=
\int_{0}^{1} d(\cos\theta) \int_{-\pi}^{0} d\phi\,
\cos^2\theta \sin^6\theta \sin^6\phi
=
\frac{\pi}{63},
\end{eqnarray}
\begin{eqnarray}
\mathcal{B}_{550}
&=&
\int_{0}^{1} d(\cos\theta) \int_{-\pi}^{0} d\phi\,
\cos^5\theta \sin^5\theta \sin^5\phi
=
-\frac{128}{10395},
\\
\mathcal{B}_{530}
&=&
\mathcal{B}_{350}
=
\int_{0}^{1} d(\cos\theta) \int_{-\pi}^{0} d\phi\,
\cos^3\theta \sin^5\theta \sin^5\phi
=
-\frac{32}{945},
\\
\mathcal{B}_{510}
&=&
\mathcal{B}_{150}
=
\int_{0}^{1} d(\cos\theta) \int_{-\pi}^{0} d\phi\,
\cos\theta \sin^5\theta \sin^5\phi
=
-\frac{16}{105},
\end{eqnarray}
\begin{eqnarray}
\mathcal{B}_{440}
&=&
\int_{0}^{1} d(\cos\theta) \int_{-\pi}^{0} d\phi\,
\cos^4\theta \sin^4\theta \sin^4\phi
=
\frac{\pi}{105},
\\
\mathcal{B}_{420}
&=&
\mathcal{B}_{240}
=
\int_{0}^{1} d(\cos\theta) \int_{-\pi}^{0} d\phi\,
\cos^2\theta \sin^4\theta \sin^4\phi
=
\frac{\pi}{35},
\end{eqnarray}
\begin{eqnarray}
\mathcal{B}_{330}
&=&
\int_{0}^{1} d(\cos\theta) \int_{-\pi}^{0} d\phi\,
\cos^3\theta \sin^3\theta \sin^3\phi
=
-\frac{8}{105},
\\
\mathcal{B}_{310}
&=&
\mathcal{B}_{130}
=
\int_{0}^{1} d(\cos\theta) \int_{-\pi}^{0} d\phi\,
\cos\theta \sin^3\theta \sin^3\phi
=
-\frac{4}{15},
\end{eqnarray}
\begin{eqnarray}
\mathcal{B}_{220}
&=&
\int_{0}^{1} d(\cos\theta) \int_{-\pi}^{0} d\phi\,
\cos^2\theta \sin^2\theta \sin^2\phi
=
\frac{\pi}{15},
\end{eqnarray}
\begin{eqnarray}
\mathcal{B}_{040}
&=&
\int_{0}^{1} d(\cos\theta) \int_{-\pi}^{0} d\phi\,
\cos^4\theta
=
\frac{\pi}{5},
\\
\mathcal{B}_{020}
&=&
\int_{0}^{1} d(\cos\theta) \int_{-\pi}^{0} d\phi\,
\cos^2\theta
=
\frac{\pi}{3},
\end{eqnarray}
\begin{eqnarray}
\mathcal{B}_{442}
&=&
\int_{0}^{1} d(\cos\theta) \int_{-\pi}^{0} d\phi\,
\cos^4\theta \sin^{6}\theta
\sin^4\phi \cos^2\phi
=
\frac{\pi}{1155},
\\
\mathcal{B}_{242}
&=&
\mathcal{B}_{422}
=
\int_{0}^{1} d(\cos\theta) \int_{-\pi}^{0} d\phi\,
\cos^4\theta \sin^{4}\theta
\sin^2\phi \cos^2\phi
=
\frac{\pi}{315},
\end{eqnarray}
\begin{eqnarray}
\mathcal{B}_{532}
&=&
\mathcal{B}_{352}
=
\int_{0}^{1} d(\cos\theta) \int_{-\pi}^{0} d\phi\,
\cos^3\theta \sin^{7}\theta
\sin^5\phi \cos^2\phi
=
-\frac{32}{10395},
\\
\mathcal{B}_{152}
&=&
\mathcal{B}_{512}
=
\int_{0}^{1} d(\cos\theta) \int_{-\pi}^{0} d\phi\,
\cos^5\theta \sin^{3}\theta
\sin\phi \cos^2\phi
=
-\frac{16}{945},
\end{eqnarray}
\begin{eqnarray}
\mathcal{B}_{332}
&=&
\int_{0}^{1} d(\cos\theta) \int_{-\pi}^{0} d\phi\,
\cos^3\theta \sin^{5}\theta
\sin^3\phi \cos^2\phi
=
-\frac{8}{945},
\\
\mathcal{B}_{132}
&=&
\mathcal{B}_{312}
=
\int_{0}^{1} d(\cos\theta) \int_{-\pi}^{0} d\phi\,
\cos^3\theta \sin^{3}\theta
\sin\phi \cos^2\phi
=
-\frac{4}{105},
\end{eqnarray}
\begin{eqnarray}
\mathcal{B}_{222}
&=&
\int_{0}^{1} d(\cos\theta) \int_{-\pi}^{0} d\phi\,
\cos^2\theta \sin^4\theta \cos^2\phi \sin^2\phi
=
\frac{\pi}{105},
\\
\mathcal{B}_{112}
&=&
\int_{0}^{1} d(\cos\theta) \int_{-\pi}^{0} d\phi\,
\cos\theta \sin^3\theta \sin\phi \cos^2\phi
=
-\frac{2}{15}.
\end{eqnarray}
In the above calculation, we have utilized the following formulas:
\begin{eqnarray}
\int_{0}^{1}d(\cos\theta)\,
\cos^n \theta \sin^m \theta
&=&
\left\{
\begin{array}{c}
\int_0^1 dx\, x^n (1-x^2)^{m/2}
\hspace{1em}
(\mbox{$n,m$: even}) \\
\int_0^{\pi/2}
d\theta \, \cos^n\theta (1-\cos^2\theta)^{(m+1)/2}
\hspace{1em}
(\mbox{$n,m$: odd})
\end{array}
\right.
,
\\
\int_{-\pi}^{0}d\phi\,
\sin^n \phi
&=&
\left\{
\begin{array}{c}
2 \cdot \frac{\pi}{2}\frac{(n-1)!!}{n!!}
\hspace{1em}
(\mbox{$n$: even}) \\
-2 \cdot \frac{(n-1)!!}{n!!}
\hspace{1em}
(\mbox{$n$: odd})
\end{array}
\right.
.
\end{eqnarray}

From the above results for $\mathcal{B}_{pqr}$, we obtain
$\mathcal{A}_{pqr;stu}$ via Eq.~(\ref{eq:A_2BB}) as follows,
\begin{eqnarray}
\mathcal{A}_{420;220}
&=&
\mathcal{A}_{240;220} 
=
2\frac{\pi}{35}\frac{\pi}{15} = \frac{2\pi^2}{525},
\\
\mathcal{A}_{420;130}
&=&
\mathcal{A}_{240;310} 
=
2\frac{\pi}{35}\left(-\frac{4}{15}\right)=-\frac{8\pi}{525},
\end{eqnarray}
\begin{eqnarray}
\mathcal{A}_{330;310}
&=&
\mathcal{A}_{330;130} 
=
-2\frac{8}{105}\left(-\frac{4}{15}\right)=\frac{64}{1575},
\\
\mathcal{A}_{330;220}
&=&
-2\frac{8}{105}\frac{\pi}{15}=-\frac{16\pi}{1575},
\end{eqnarray}
\begin{eqnarray}
\mathcal{A}_{310;130}
&=&
\mathcal{A}_{130;130} 
=
2\left(-\frac{4}{15}\right)^2 = \frac{32}{225},
\\
\mathcal{A}_{310;220}
&=&
\mathcal{A}_{130;220} 
=
2\left(-\frac{4}{15}\right)\frac{\pi}{15}=-\frac{8\pi}{225},
\\
\mathcal{A}_{310;112}
&=&
\mathcal{A}_{130;112} 
=
2\left(-\frac{4}{15}\right)
\left(-\frac{2}{15}\right)=\frac{16}{225},
\end{eqnarray}
\begin{eqnarray}
\mathcal{A}_{220;220}
&=&
2\left(\frac{\pi}{15}\right)^2=\frac{2\pi^2}{225},
\\
\mathcal{A}_{220;112}
&=&
2\frac{\pi}{15} \left( -\frac{2}{15}\right)
=
-\frac{4\pi}{225},
\\
\mathcal{A}_{222;112}
&=&
2\frac{\pi}{105}\left(-\frac{2}{15}\right)=-\frac{4\pi}{1575},
\end{eqnarray}
\begin{eqnarray}
\mathcal{A}_{040;310}
&=&
\mathcal{A}_{040;130} 
=
2\frac{\pi}{5}\left(-\frac{4}{15}\right)=-\frac{8\pi}{75},
\\
\mathcal{A}_{040;220}
&=&
2\frac{\pi}{5}\frac{\pi}{15}=\frac{2\pi^2}{75},
\\
\mathcal{A}_{040;112}
&=&
2\frac{\pi}{5} \left( -\frac{2}{15}\right)
=
-\frac{4\pi}{75},
\end{eqnarray}
\begin{eqnarray}
\mathcal{A}_{020;310}
&=&
\mathcal{A}_{020;130} 
=
2\frac{\pi}{3}\left(-\frac{4}{15}\right)=-\frac{8\pi}{45},
\\
\mathcal{A}_{020;220}
&=&
2\frac{\pi}{3}\frac{\pi}{15}=\frac{2\pi^2}{45},
\\
\mathcal{A}_{020;112}
&=&
2\frac{\pi}{3} \left( -\frac{2}{15}\right)
=
-\frac{4\pi}{45},
\end{eqnarray}
\begin{eqnarray}
\mathcal{A}_{510;220}
&=&
\mathcal{A}_{150;220} 
=
-2\frac{16}{105}\frac{\pi}{15}=-\frac{32\pi}{1575},
\\
\mathcal{A}_{510;130}
&=&
\mathcal{A}_{150;130} 
=
-2\frac{16}{105}\left(-\frac{4}{15}\right)=\frac{128}{1575}.
\end{eqnarray}

\section{Four-point susceptibility}
\label{app:sec:chi4}

The simulation method for the four-point susceptibility is shown.
The four-point density correlation function is given by
\begin{eqnarray}
g^{(4)} (\bm{r}_1, \bm{r}_2, t)
=
\left\langle
\rho(\bm{r}_1, 0)\rho(\bm{r}_1, t)
\rho(\bm{r}_2, 0)\rho(\bm{r}_2, t)
\right\rangle
-
\left\langle
\rho(\bm{r}_1, 0)\rho(\bm{r}_1, t)
\right\rangle
\left\langle
\rho(\bm{r}_2, 0)\rho(\bm{r}_2, t)
\right\rangle,
\hspace{1.5em}
\end{eqnarray}
where $\rho(\bm{r},t) := \sum_{i=1}^N \delta( \bm{r}-\bm{r}_{i}(t))$ is
the density.
The four-point susceptibility is obtained by integrating the spatial
degrees of freedom in $g^{(4)}(\bm{r}_1, \bm{r}_2, t)$,
\begin{eqnarray}
\chi_4^0(t)
=
\frac{\beta V}{N^2} 
\int d^3 \bm{r}_1 \int d^3 \bm{r}_2\,
g^{(4)}(\bm{r}_1, \bm{r}_2, t),
\end{eqnarray}
where $\beta$ is the inverse temperature.
This can be expressed by the order parameter
\begin{eqnarray}
Q_0(t) 
:=
\int d^3 \bm{r} \, 
\rho(\bm{r},0)\rho(\bm{r},t)
=
\sum_{i=1}^N
\sum_{j=1}^N
\delta(\bm{r}_{i}(0)-\bm{r}_{j}(t))
\end{eqnarray}
as
\begin{eqnarray}
\chi_4^0(t) 
=
\frac{\beta V}{N^2}
\left[
\left\langle
Q_0(t)^2
\right\rangle
-
\left\langle
Q_0(t)
\right\rangle^2
\right].
\end{eqnarray}
There is a problem in evaluating $\chi_4^0(t)$ by a simulation, because
$Q_0(t)$ is ill defined for a finite system.
We follow the method of \cite{GNS2000} and modify $Q_0(t)$ by an
``overlap'' function $w(r)$ \citep{Parisi1997} that is unity inside a
region of size $a$ and zero otherwise, where $a$ is chosen to be of the
order of the particle diameter:
\begin{eqnarray}
Q(t) 
&:=&
\int d^3 \bm{r}_1 \int d^3 \bm{r}_2 \, 
\rho(\bm{r}_1,0)\rho(\bm{r}_2,t) w(|\bm{r}_1-\bm{r}_2|)
\nonumber \\
&=&
\sum_{i=1}^N \sum_{j=1}^N
\int d^3 \bm{r} \, w(|\bm{r}|)\delta(\bm{r}+\bm{r}_{i}(0)-\bm{r}_{j}(t)) 
=
\sum_{i=1}^N \sum_{j=1}^N
w(|\bm{r}_{ij}-\bm{\mu}_j|).
\end{eqnarray}
Here $\bm{r}_{ij} := \bm{r}_i(0)-\bm{r}_j(0)$, $\bm{\mu}_i :=
\bm{r}_i(t)-\bm{r}_i(0)$ and $a$ is chosen as $0.3 \, d$.
Replacing $Q_0(t)$ by $Q(t)$ yields
\begin{eqnarray}
\chi_4(t) 
=
\frac{\beta V}{N^2}
\left[
\left\langle
Q(t)^2
\right\rangle
-
\left\langle
Q(t)
\right\rangle^2
\right],
\end{eqnarray}
which we have measured by MD simulation (cf. Fig.~\ref{Fig:chi4}).

\bibliographystyle{jfm}
\bibliography{Refs_20190218}
\end{document}